\documentclass[preprint,english,showpacs,preprintnumbers,amsmath,amssymb,floatfix]{revtex4}
\usepackage[T1]{fontenc}
\usepackage[latin9]{inputenc}
\usepackage{color}
\usepackage{array}
\usepackage{amstext}
\usepackage{graphicx}
\usepackage{esint}

\makeatletter

\@ifundefined{textcolor}{}
{%
 \definecolor{BLACK}{gray}{0}
 \definecolor{WHITE}{gray}{1}
 \definecolor{RED}{rgb}{1,0,0}
 \definecolor{GREEN}{rgb}{0,1,0}
 \definecolor{BLUE}{rgb}{0,0,1}
 \definecolor{CYAN}{cmyk}{1,0,0,0}
 \definecolor{MAGENTA}{cmyk}{0,1,0,0}
 \definecolor{YELLOW}{cmyk}{0,0,1,0}
 }

\@ifundefined{definecolor}
 {\usepackage{color}}{}
\@ifundefined{definecolor}
 {\usepackage{color}}{}
\makeatother

\makeatother

\usepackage{babel}

\begin{document}

\title{Impact of Charm H1-ZEUS Combined data and Determination of the Strong Coupling in Two Different Schemes}

\author{A.~Vafaee}
\email[]{a.vafaee@semnan.ac.ir}
\author{A.~N.~Khorramian}
\email[]{Khorramiana@semnan.ac.ir}
\affiliation{Faculty of Physics, Semnan University, P. O. Box 35131-19111, Semnan, Iran}

\date{\today}

\begin{abstract}
We study the impact of recent measurements of charm cross section H1-ZEUS combined data on simultaneous determination of parton distribution functions (PDFs) and the strong coupling, $\alpha_s(M^2_Z)$, in two different schemes. We perform several fits based on Thorne-Roberts (RT) and Thorne-Roberts optimal (RTOPT) schemes at next-to-leading order (NLO). We show that adding charm cross section H1-ZEUS combined data reduces the uncertainty of the gluon distribution and improves the fit quality up to $\sim 0.4$~\% and $\sim 0.9$~\%~, without and with the charm contribution, from the RT scheme to the RTOPT scheme, respectively. We also emphasise the central role of the strong coupling, $\alpha_s(M^2_Z)$, in revealing the impact of charm flavour contribution, when it is considered as an extra free parameter. We show that in going from the RT scheme to the RT OPT scheme, we get $\sim 0.9$~\% and $\sim 2.0$~\%~improvement in the value of $\alpha_s(M^2_Z)$, without and with the charm flavour contribution respectively.  
\end{abstract}


\maketitle

\section{\label{introduction}Introduction}
When the mass of a quark is significantly larger than the quantum chromodynamics (QCD) scale parameter, $\Lambda_{QCD} \sim 250$~MeV, we categorize it as a heavy quark \cite{Behnke:2015qja}. The production of heavy quarks in photoproduction ($\gamma p$) and deep inelastic scattering (DIS) of ${e^\pm}p$ was one of the main tasks at HERA. The only heavy quarks kinematically accessible at HERA were beauty and charm quarks, and investigation of the impact of charm quark cross section H1-ZEUS combined data \cite{Abramowicz:1900rp}  on simultaneous determination of parton distribution functions and the strong coupling, $\alpha_s(M^2_Z)$, is the main topic of this analysis. In deep inelastic ${e^\pm}p$ scattering we can approximate the ratio of photon couplings corresponding to a heavy quark, $Q_h, h=b, c$, by
\begin{equation} \label{eq:hfrac}
f(h) \sim \frac{Q_h^2}{\Sigma{Q_q^2}} ,
\end{equation}
where $Q_h=\frac{1}{3},~\frac{2}{3}$ are the beauty and charm electric charges, respectively, and $Q_q$~, with $q=u,d,s,c,b$~, represent the kinematically accessible quark flavours. 

 Now, for the charm quark we have
\begin{equation} \label{eq:cfrac}
f(c) \sim \frac{Q_c^2}{Q_d^2+Q_u^2+Q_s^2+Q_c^2+Q_b^2}
        = \frac{4}{11} \simeq 0.36~.
\end{equation}
From Eq. \ref{eq:cfrac}~ we see that up to 36 percent of the cross sections at HERA originate from processes with charm quarks in the final state. This is our main motivation to investigate the impact of only charm quarks on simultaneous determination of parton distribution functions or their uncertainties and the strong coupling, $\alpha_s(M^2_Z)$.

The ratio $f(c) \simeq 0.36~$ implies that charm quarks are an integral part of the quark-antiquark sea within the proton. On the other hand, the proton has no net charm flavour number, which in turn implies that the charm quarks within the proton can only arise in pairs of $c\overline{c}$. Since the charm-quark mass is about $1.5$ GeV~, at the low-energy limit the $c\overline{c}$ pairs are considerably heavier than that to have a contribution within the proton. 

 Although consideration of so-called intrinsic charm (IC) \cite{Brodsky:1980pb} may alter this simple view of the heavy flavour content of the proton, at present there is no evidence for the existence of such a contribution from HERA data. Therefore, in this analysis the charm quarks within the proton are as usual considered  as virtual quarks, which in turn arise as fluctuations of probing the gluon content of the proton.
 
 The charm PDFs play an important role in hadronic collisions and cause  photons to emerge from hard parton-parton interactions in association with one or more charm quark jets. Clearly, to study and analyse these processes, we need the charm PDFs, which in turn have sizeable uncertainties. A series of experimental measurements involving charm (or beauty) and photon final states have recently been published by the CDF and D0 Collaborations \cite{Abazov:2012ea,Abazov:2009de,D0:2012gw,Abazov:2014hoa,Aaltonen:2009wc,Aaltonen:2013ama}.

 As we noted, the charm quark mass is about $1.5$ GeV~, whereas the QCD scale is about $\Lambda_{QCD} \sim 0.25$ GeV~, so it is reasonable to treat the charm quark mass as a hard scale in perturbative quantum chromodynamics (pQCD) and investigate the charm mass effect in pQCD. Accordingly, in this study we use the full HERA run I and II combined data \cite{Abramowicz:2015mha} as a new measurements of inclusive deep inelastic scattering cross sections for our base data set and then we investigate, simultaneously, the impact of charm quark cross section H1-ZEUS combined data \cite{Abramowicz:1900rp} on the central value of the PDFs and determination of the strong coupling, $\alpha_s(M^2_Z)$.  
  
  Although the charm quark mass is large compared to the QCD scale, it is small with respect to other pQCD scales, such as the transverse momentum of a quark or a jet, $p_T$, and the virtuality of the photon, $Q^2$. This smallness leads to the logarithmic correction terms, $\sim [\alpha_s \ln(p^2_{T}/m_c^2)]^n$ and $\sim [\alpha_s \ln(Q^2/m_c^2)]^n$~, corresponding to $p_T$ and $Q^2$, respectively.
At present, the order of magnitude and treatment of these correction terms are open questions.

 The outline of this  paper is as follows. In Section~2, we describe the theoretical framework of our study and discuss the reduced cross sections. We introduce the data set which we use in this QCD-analysis and discuss our methodology in Section~3. In Section~4, the impact of charm quark cross section H1-ZEUS combined data on QCD fit quality is discussed. We explain the impact of charm production data on PDFs and $\alpha_s(M^2_Z)$ in Section~5. We present our results in Section~6 and conclude with a summary in Section~7.

\section{\label{dis}Cross sections and parton distributions}
In perturbative quantum chromodynamics, the deep inelastic scattering of ${e^\pm}p$, at the centre-of-mass energies up to $\sqrt{s} \simeq 320\,$GeV at HERA, plays a central role in probing the structure of the proton, as a sea of strongly interacting quarks and gluons. For neutral current (NC) interactions, the reduced cross sections can be expressed in terms of the generalized structure functions as:
\begin{eqnarray}
   \sigma_{r,NC}^{{\pm}}&=&   \frac{d^2\sigma_{NC}^{e^{\pm} p}}{d{x_{\rm Bj}}dQ^2} \frac{Q^4 x_{\rm Bj}}{2\pi \alpha^2 Y_+}                                               \nonumber\\
    &=&  \tilde{F_2} \mp \frac{Y_-}{Y_+} x\tilde{F_3} -\frac{y^2}{Y_+} \tilde{F_{\rm L}}~,
    \label{eq:NC}
\end{eqnarray} 
where  $Y_{\pm} = 1 \pm (1-y)^2$ and $\alpha$ is the fine-structure constant which is defined 
at zero momentum transfer. The generalized structure functions $\tilde F_2$, $\tilde F_L$ and $\tilde F_3$ can be expressed as linear combinations of the proton structure functions $F^{\gamma}_2, F^{\gamma Z}_2, F^{\gamma Z}_3, F^Z_2$ and $F^Z_3$ as follows:
\begin{eqnarray} \label{strf}                                                   
 \tilde F_2 &=& F_2 - \kappa_Z v_e  \cdot F_2^{\gamma Z} +                      
  \kappa_Z^2 (v_e^2 + a_e^2 ) \cdot F_2^Z~, \nonumber \\   
 \tilde F_L &=& F_{\rm L} - \kappa_Z v_e  \cdot F_{\rm L}^{\gamma Z} +                      
  \kappa_Z^2 (v_e^2 + a_e^2 ) \cdot F_{\rm L}^Z~, \nonumber \\                     
 x\tilde F_3 &=&  - \kappa_Z a_e  \cdot xF_3^{\gamma Z} +                    
  \kappa_Z^2 \cdot 2 v_e a_e  \cdot xF_3^Z~,                                   
\end{eqnarray}
where $v_e$ and $a_e$ are the vector and axial-vector weak couplings of the electron to the $Z$ boson, and $\kappa_Z(Q^2)$ is defined as
\begin{eqnarray}
\kappa_Z(Q^2) &=&   \frac{Q^2}{(Q^2+M_Z^2)(4\sin^2 \theta_W \cos^2 \theta_W)}.
\end{eqnarray}
This analysis is based on xFitter, an open source QCD framework 
\cite{xFitter} which is an update of the former HERAFitter package \cite{ Sapronov}. The values of the $Z$-boson mass and the electroweak mixing angle are $M_Z=91.1876$~GeV and $\sin^2 \theta_W=0.23127$, respectively, and electroweak effects have been treated only at leading order (LO).

 In the range of low values of $Q^2$, $Q^2\ll M_Z^2$, the $Z$ boson exchange contribution may be ignored and then the reduced NC DIS cross sections can be expressed by 
\begin{eqnarray}
   \sigma_{r,NC}^{{\pm}}&= &   F_2  - \frac{y^2}{Y_{+}}  F_L~.
    \label{eq:LOWNC}
\end{eqnarray}

 Similarly, the reduced charged current (CC) deep inelastic $e^{\pm}p$ scattering cross sections may be expressed as follows:
 
\begin{eqnarray}
\sigma_{r,CC}^{\pm} &=&\frac{2\pi x_{\rm Bj}}{G^2_F} \left[\frac{M^2_W+Q^2}{M^2_W}\right]^2 \frac{d^2\sigma_{CC}^{e^{\pm} p}}{d{x_{\rm Bj}}dQ^2} \nonumber \\
&= &  \frac{Y_{+}}{2}  W_2^{\pm} \mp \frac{Y_{-}}{2}x  W_3^{\pm} - \frac{y^2}{2} W_L^{\pm}~~,
\end{eqnarray}
where $\tilde W_2^{\pm}$, $\tilde W_3^{\pm}$ and $\tilde W_L^{\pm}$  are  another set of structure functions and $G_F$ is the Fermi constant, which is related to the weak coupling constant $g$ and electromagnetic coupling constant $e$ by:
\begin{eqnarray}
G^2_F = \frac{e^2}{4\sqrt{2}{\sin ^2\theta_W}M^2_W} = \frac{g^2}{4M_W}.
\end{eqnarray}
The values of the Fermi constant and $W$-boson mass in the xFitter QCD framework \cite{xFitter} are: $G_F=1.16638\times 10^{-5} $~GeV$^{-2}$
and $M_W=80.385$~GeV.

 In the quark parton model (QPM), the sums and differences
of quark and anti-quark distributions, depending on the 
charge of the lepton beam, can be represented by $W_2^\pm$, $xW_3^\pm$ structure functions, respectively, and $W_{\rm L}^\pm = 0$~:
\begin{eqnarray}
 \label{WSF}
    W_2^{+}  \approx  x\overline{U}+xD\,,\hspace{0.05cm} ~~~~~~~
    W_2^{-}  \approx  xU+x\overline{D}\,,\hspace{0.05cm} \nonumber \\   
   xW_3^{+}  \approx   xD-x\overline{U}\,,\hspace{0.05cm} ~~~~~~~
   xW_3^{-}  \approx  xU-x\overline{D}\,.
\end{eqnarray}
According to Eq.~\ref{WSF}~ we have:

\begin{eqnarray}
\sigma_{r,CC}^{+} \approx  x\overline{U}+ (1-y)^2xD \nonumber \\
\sigma_{r,CC}^{-} \approx  xU +(1-y)^2 x\overline{D}.
\end{eqnarray}
Now it is possible to determine both the valence-quark distributions, $xu_v$ and $xd_v$, and the combined sea-quark distributions, $x\overline{U}$ and $x\overline{D}$, by combination of NC and CC measurements.

 In analogy to the inclusive NC deep inelastic $e^{\pm}p$ cross section, the reduced cross sections for charm-quark production, $\sigma_{red}^{C\bar{C}}$,  can be expressed by
\begin{eqnarray}
	\sigma_{red}^{C\bar{C}} &=& 
	          \frac{d\sigma^{C\bar{C}}(e^{\pm} p)}{d{x_{\rm Bj}} \, dQ^2} \cdot \frac{x_{\rm Bj} \, Q^4}{2 \pi \alpha^2 Y_{+}} \nonumber \\ 
	          &=&F_2^{C\bar C} \mp \frac{Y_{-}}{Y_{+}}x  F_3^{C\bar C} - \frac{y^2}{Y_{+}}  F_L^{C\bar C}~, 
    \label{eq:NCheavy}
\end{eqnarray}	
where $Y_{\pm} = (1 \pm (1-y)^2)$, $\alpha$ is the electromagnetic coupling constant, and $F_2^{C\bar{C}}$, $xF_3^{C\bar{C}}$ and $F_L^{C\bar{C}}$ are charm-quark contributions to the 
inclusive structure functions $F_2$, $xF_3$ and $F_L$, respectively.

 In the kinematic region at HERA, the $F_2^{C\bar{C}}$ structure function makes a dominant contribution. The $xF_3^{C\bar{C}}$ structure function contributes only from $Z^0$ exchange and $\gamma Z^0$, which implies that for the $Q^{2} \ll M_{Z}^{2}$ region, this contribution may be ignored. Finally, the contribution of longitudinal charm-quark structure function, $F_L^{C\bar{C}}$, is suppressed only for the $y^2 \ll 1$ region which can be a few percent in the kinematic region accessible at HERA and therefore cannot be ignored.
 
 Therefore, neglecting the $xF_3^{C\bar C}$ structure function contribution, the reduced charm-quark cross section, $\sigma_{red}^{ C\bar C}$, for both positron and electron beams, can be expressed by  
\begin{eqnarray}
   \sigma_{red}^{ C\bar C}&=&   \frac{d^2\sigma^{C\bar C}(e^{\pm}p)}{dxdQ^2} \frac{xQ^4}{2\pi\alpha^2 Y_{+}} \nonumber\\
  &= &  F_2^{C\bar C}  - \frac{y^2}{Y_{+}}  F_L^{C\bar C}~. 
    \label{eq:NCheavylast}
\end{eqnarray}
Accordingly, at high $y$, the reduced charm-quark cross section, $\sigma_{red}^{ C\bar C}$, and the $F_2^{C\bar C}$ structure function only differ by a small $F_L^{C\bar C}$ contribution \cite{Daum:1996ec}.

 In the QPM, the structure functions depend only on the $Q^2$ 
variable and then they can be directly related to the PDFs. In QCD, however, and especially when heavy flavour           production is included, the structure functions depend on both $x$ and $Q^2$ variables, \cite{Sjostrand:2001yu,Aktas:2006hy,DeRoeck:2011na,Ball:2010de,Aaron:2011gp,Tung:2006tb,Aaron:2009aa,Engelen:1998rf,CooperSarkar:2012tx,Frixione:1993yw,Marchesini:1991ch,Jung:1993gf,Sjostrand:1985ys,Sjostrand:1986hx,Lonnblad:1992tz,Kuraev:1976ge,Ciafaloni:1987ur,Jung:2000hk,Beneke:1994rs,Agashe:2014kda,Schmidt:2012az,Alekhin:2010sv,Gao:2013wwa,Martin:1998sq,Pumplin:2002vw,Chekanov:2002pv}. In Section \ref{methodology}, based on our methodology, we extract the PDFs as functions of $x$ and $Q^2$ variables, using full HERA run I and II combined data, with and without the charm cross section H1-ZEUS combined measurements data set included.

\section{\label{methodology}Data Set and Methodology}         
In this paper, we use two different data sets: the full HERA run I and II combined NC and CC DIS $e^{\pm}p$ scattering cross sections \cite{Abramowicz:2015mha}, and the charm production reduced cross section measurements data \cite{Abramowicz:1900rp}. In our analysis, we choose the full HERA run I and II combined data as our base data set, and then we investigate the impact of charm production reduced cross section data on simultaneous determination of PDFs and the strong coupling, $\alpha_s(M^2_Z)$~ in the Thorne-Roberts (RT) and Thorne-Roberts optimal (RTOPT) schemes. The kinematic ranges for these two data sets have been reported in Ref.~\cite{Vafaee:2017nze}.

 We use xFitter \cite{xFitter}, version 1.2.2, as our QCD fit framework. Using the QCDNUM package \cite{Botje:2010ay}, version 17-01/12, we evolved the parton distribution functions and $\alpha_s(M^2_Z)$. In the evolution of PDFs and $\alpha_s(M^2_Z)$, we set our theory type based on the DGLAP collinear evolution equations \cite{DGLAP} and make several fits at leading order and next-to-leading order in the RT and RTOPT schemes.
 
 The RT scheme is a General Mass-Variable Flavour Number Scheme (GM-VFNS). Really, the RT scheme was designed to provide a smooth transition from the massive FFN scheme at low scales $Q^2 \sim m_h^2$ to the massless ZM-VFNS scheme at high scales $Q^2 \gg m_h^2$. However, the connection is not unique. A GM-VFNS can be defined by demanding equivalence of the $n_f = n$ (FFNS) and $n_f = n + 1$ flavour (ZM-VFNS) descriptions above the transition
point for the new parton distributions. Of course they are by definition identical below this point, at all orders. The RT scheme has two different variants: RT standard and RT optimal, with a smoother transition across the heavy flavour threshold region. A review of the two different schemes has been given in Ref.~\cite{Vafaee:2017nze}.
 
  To investigate the impact of charm production reduced cross section data,  we need to use the heavy flavour scheme in our analysis. Different theoretical groups use various heavy flavour schemes. For example, some theory groups such as CT10 \cite{Lai:2010vv}, ABKM09 \cite{Alekhin:2009vn}, and NNPDF2.1 \cite{Ball:2008by,Mironov:2009uv} used  
S-ACOT \cite{Collins:1998rz}, FFNS \cite{Martin:2006qz} and  FONLL \cite{Forte:2010ta}, respectively and some other groups such as MSTW08 \cite{Martin:2009iq} and HERAPDF1.5/2.0 \cite{Aaron:2009aa} used the RT (also referred to as TR) standard and optimal heavy flavour schemes \cite{Thorne:2006qt,Thorne:2012az} to calculate the reduced charm cross sections in DIS. On the other hand, to include heavy flavour contributions, the perturbative QCD scales $\mu_f^2$ and $\mu_r^2$  play a central rule. Some groups such as CT10 \cite{Lai:2010vv} and ABKM09  \cite{Alekhin:2009vn}  choose  $\mu_f^2 = \mu_r^2=Q^2+m_C^2$ and $\mu_f^2 = \mu_r^2=Q^2+4m_C^2$  respectively, where $m_C$ denotes the pole mass of the charm quark, whereas the NNPDF2.1 \cite{Ball:2008by,Mironov:2009uv}, HERAPDF1.5 \cite{Aaron:2009aa} and  MSTW08 \cite{Martin:2009iq} groups use $\mu_f = \mu_r=Q$ in their heavy quark QCD approach.
  
  To include the heavy-flavor contributions, we use both RT and RTOPT schemes,  and choose $\mu_f = \mu_r=Q$~ as the perturbative quantum chromodynamics scale for the pole mass of the charm quark $m_c=1.5 \pm 0.15$ GeV.
  
 The last step in our QCD analysis is the minimization procedure. In this regard, we use the standard MINUIT minimization package~\cite{James:1975dr}, as a powerful program for minimization, correlations and parameter errors. 
   
   In order to minimize the influence of higher twist contributions we use kinematic cuts. In the various DIS analyses, different kinds of kinematic cuts should be applied. In this analysis we imposed a kinematic cut $Q^2$=3.5 GeV$^2$  to omit all data with $Q^2$ less than this value. The cuts on the kinematic coverage of the DIS data have been made for values of $Q^2$ between $Q^2=0.045$\,GeV$^2$ and $Q^2=50000$\,GeV$^2$ and values of $x_{\rm Bj}$ 
between $x_{\rm Bj}=6\times10^{-7}$ and $x_{\rm Bj}=0.65$. The cuts on $Q^2$  not only significantly increase the number of data points available to constrain PDFs, but also allow access to a greater range of kinematics, which in turn lead to reduced PDF uncertainties, especially at higher values of $x$.

 In this analysis, based on the HERAPDF approach \cite{Abramowicz:2015mha}, we generically parameterized the PDFs of the proton, $xf(x)$, at the initial scale of the QCD evolution $Q^2_0= 1.9$ GeV$^2$ as
\begin{equation}
 xf(x) = A x^{B} (1-x)^{C} (1 + D x + E x^2)~~,
\label{eqn:pdf}
\end{equation}
where in the infinite momentum frame, $x$ is the fraction of the proton's momentum taken by the struck parton.
  
To determine the normalization constants $A$ for the valence and gluon distributions, we use the QCD number and momentum sum rules. Using this functional form, Eq. \ref{eqn:pdf} leads to the following independent combinations of parton distribution functions:
\begin{eqnarray}
\label{eq:xgpar}
xg(x) &=   & A_g x^{B_g} (1-x)^{C_g} - A_g' x^{B_g'} (1-x)^{C_g'}  ,  \\
\label{eq:xuvpar}
xu_v(x) &=  & A_{u_v} x^{B_{u_v}}  (1-x)^{C_{u_v}}\left(1+E_{u_v}x^2 \right) , \\
\label{eq:xdvpar}
xd_v(x) &=  & A_{d_v} x^{B_{d_v}}  (1-x)^{C_{d_v}} , \\
\label{eq:xubarpar}
x\bar{U}(x) &=  & A_{\bar{U}} x^{B_{\bar{U}}} (1-x)^{C_{\bar{U}}}\left(1+D_{\bar{U}}x\right) , \\
\label{eq:xdbarpar}
x\bar{D}(x) &= & A_{\bar{D}} x^{B_{\bar{D}}} (1-x)^{C_{\bar{D}}} .
\end{eqnarray}
where $xg(x)$ is the gluon distribution, $xu_{{v}}(x)$ and $xd_{{v}}(x)$ are the valence-quark distributions, and $x\bar{U}(x)$ and $x\bar{D}(x)$ are the $u$-type and $d$-type anti-quark distributions, which are identical to the sea-quark distributions. A review of HERAPDF functional form, including some more details, can be found in Ref.~\cite{Vafaee:2017nze}.

\section{\label{qcdfitquality}Impact of Charm Production Data on the QCD fit quality }
We now investigate the impact of the charm cross section H1-ZEUS combined measurements on simultaneous determination of PDFs and $\alpha_s(M^2_Z)$. We also explain how adding these data improve the uncertainty of PDFs, reducing the error bars of some parton distributions, especially gluon distributions and some of their ratios, when the HERA run I and II combined inclusive DIS $e^{\pm}p$ scattering cross sections data are chosen as a ``BASE''.
To investigate the fit quality, we use the $\chi^2$ definition as reported in Ref.~\cite{Vafaee:2017nze}.

 For HERA run I and II combined inclusive DIS $e^{\pm}p$ scattering cross sections and the charm cross section H1-ZEUS combined measurements, the number of data points are 1307 and 42, respectively. Accordingly, the total number of data points for BASE and BASE plus charm, which we refer to sometimes as ``TOTAL'', are 1307 and 1349, respectively.  In various configurations, the $Q^2 \ge 1.5$\,GeV$^2$ range was covered by the HERA run I and II combined data \cite{Abramowicz:2015mha}. The MINUIT parameters are sensitive to the $Q^2_{min}$ value, so to get a convergent fit result we set $Q^2_{min}=3.5$ GeV$^2$, as suggested in~Ref \cite{Abramowicz:2015mha}. Clearly, this cut on $Q^2$  omits all data with $Q^2$ less than $Q^2_{min}=3.5$~GeV$^2$ and therefore, reduces the total number of data points from 1307 and 1349 to 1145 and 1192 for the BASE and TOTAL data sets, respectively. Now, based on Table~\ref{tab:data}, we can present our QCD fit quality as follows:

for the RT scheme:
\begin{eqnarray}
\noindent\centerline{ $\chi^2_{TOTAL}$ / dof = $\frac{1335}{1131}=1.180~~$for BASE~,} \label{lobase}\\
\noindent\centerline{ $\chi^2_{TOTAL}$ / dof = $\bf \frac{1389}{1178}=1.179 ~~$for TOTAL~,} \label{lototal}
\end{eqnarray} 
and for the RTOPT scheme:
\begin{eqnarray}
\noindent\centerline{ $\chi^2_{TOTAL}$ / dof = $\frac{1331}{1131}=1.176~~$for BASE~,} \label{nlobase}\\
\noindent\centerline{ $\chi^2_{TOTAL}$ / dof = $\bf \frac{1378}{1178}=1.169 ~~$for TOTAL~,} \label{nlototal}
\end{eqnarray}
where dof refers to the $\chi^2$ per degrees of freedom and is defined as the number of data points minus the number of free parameters.
As we can  see from Eqs.~(19-22), we obtain four different values of $\chi^2_{\rm TOTAL}$/dof, corresponding to four different fits, which in turn imply four different fit-qualities in some PDFs. Now, according to the relative change of $\chi^2$, which is defined by $\frac{\chi^2_{\rm RT}-\chi^2_{\rm RT\;OPT}}{\chi^2_{\rm RT}}$ and the numerical results of Eqs.~(19-22), we see that in going from the RT scheme to the RT OPT scheme, we get $\sim 0.4$~\% and $\sim 0.9$~\%~improvement in the fit quality, without and with the charm flavour contribution, respectively. Clearly these differences in fit quality imply a significant reduction of some PDF uncertainties, especially for gluon distributions, as we will explain in the next section.

\begin{table}[h]
\begin{center}
\begin{tabular}{|l|c|c|c|c|}
\hline
\hline
{ \bf Order} & \multicolumn{4}{c|}{ {\bf NLO} }    \\ \hline
 { \bf Experiment} & {$~~~~$RT BASE$~~~~$} & {RT TOTAL} & { $~~~$RTOPT BASE$~~~$} & { RTOPT TOTAL} \\ \hline
  HERA I+II CC $e^{+}p$ \cite{Abramowicz:2015mha} & 44 / 39& 45 / 39& 44 / 39& 44 / 39 \\ 
  HERA I+II CC $e^{-}p$ \cite{Abramowicz:2015mha} & 49 / 42& 49 / 42& 50 / 42& 49 / 42 \\ 
  HERA I+II NC $e^{-}p$ \cite{Abramowicz:2015mha} & 221 / 159& 221 / 159& 221 / 159& 221 / 159 \\ 
  HERA I+II NC $e^{+}p$ 460 \cite{Abramowicz:2015mha} & 208 / 204& 209 / 204& 210 / 204& 210 / 204 \\ 
  HERA I+II NC $e^{+}p$ 575 \cite{Abramowicz:2015mha} & 213 / 254& 213 / 254& 212 / 254& 212 / 254 \\
  HERA I+II NC $e^{+}p$ 820 \cite{Abramowicz:2015mha} & 66 / 70& 66 / 70& 65 / 70& 66 / 70 \\ 
  HERA I+II NC $e^{+}p$ 920 \cite{Abramowicz:2015mha} & 422 / 377& 424 / 377& 418 / 377& 419 / 377 \\ \hline
 {Charm H1-ZEUS} \cite{Abramowicz:1900rp} & - & 40 / 47& - & 38 / 47 \\ \hline
 { Correlated ${\bf \chi^2}$} & 111& 122& 111& 119 \\\hline
{\bf {Total $\bf{\chi^2}$ / dof}}  & ${\bf \frac{1335}{1131}}$ &  ${\bf \frac{1389}{1178}}$  & ${\bf \frac{1331}{1131}}$ &  ${\bf \frac{1378}{1178}}$  \\ \hline
\hline
    \end{tabular}
\vspace{-0.0cm}
\caption{\label{tab:data}{Data sets used in our NLO QCD analysis, with corresponding partial $\chi^2$ per data point for each data set, including  $\chi^2$ per degrees of freedom (dof) for the RT and RT OPT schemes.}}
\vspace{-0.4cm}
\end{center}
\end{table}

\section{\label{impapdfalphs}Impact of Charm Production Data on PDFs and $\alpha_s(M^2_Z)$ }
Now, we present the impact of charm cross section H1-ZEUS combined measurements data on simultaneous determination of PDFs and $\alpha_s(M^2_Z)$ in the RT and RTOPT schemes and for two separate scenarios. In the first scenario we fix $\alpha_s(M^2_Z)$ to 0.117 and develop our QCD fit analysis based on only  14 unknown free parameters, according to Eqs.~(14-18). Although in this scenario the value of $\chi^2_{TOTAL}$ / dof is reduced, according to Eqs.~(19-22), from 1.180 to 1.169,  we find nothing to show the impact of charm cross section H1-ZEUS combined measurements data on the PDFs. In the second scenario we consider the strong coupling $\alpha_s(M^2_Z)$ as an extra free parameter and refit our analysis, but this time with 15 unknown free parameters. Based on the second scenario, not only do we obtain $\sim 0.4$~\% and $\sim 0.9$~\%~improvement in the fit quality, without and with the charm flavour contribution, respectively, the same as the first scenario, but we also clearly find the impact of charm on the PDFs, especially on the gluon distribution. Some more details about the central role of the strong coupling in pQCD have been reported in Ref.~\cite{Vafaee:2017nze}.

 In Tables \ref{tab:pa1} and  \ref{tab:pa2}, we present next-to-leading order numerical values of parameters and their uncertainties for  the 
$xu_v$, $xd_v$, sea and gluon PDFs at the input scale of $Q^2_0 = 1.9$~GeV$^2$ for the two different scenarios.

\begin{table}[h]
\begin{center}
\begin{tabular}{|l|c|c|c|c|}
\hline
\hline
 \multicolumn{5}{|c|}{ {\bf First Scenario: The Strong Coupling, $\alpha_s(M^2_Z)$, is Fixed} }    \\ \hline
 { \bf Parameter} & {$~~~~$RT BASE$~~~~$} & { $~~~$RT TOTAL$~~~$} & {RTOPT BASE} & {RTOPT TOTAL} \\ \hline
  ${B_{u_v}}$ & $0.723 \pm 0.046$& $0.723 \pm 0.044$& $0.730 \pm 0.042$& $0.726 \pm 0.039$ \\ 
  ${C_{u_v}}$ & $4.841 \pm 0.088$& $4.833 \pm 0.087$& $4.827 \pm 0.086$& $4.823 \pm 0.084$ \\ 
  $E_{u_v}$ & $13.6 \pm 2.6$& $13.5 \pm 2.5$& $13.1 \pm 2.3$& $13.2 \pm 2.1$ \\ \hline
  ${B_{d_v}}$ & $0.818 \pm 0.095$& $0.826 \pm 0.094$& $0.825 \pm 0.094$& $0.822 \pm 0.095$ \\ 
  $C_{d_v}$ & $4.16 \pm 0.40$& $4.18 \pm 0.39$& $4.21 \pm 0.39$& $4.19 \pm 0.38$ \\ \hline
  $C_{\bar{U}}$ & $8.91 \pm 0.81$& $8.72 \pm 0.78$& $8.90 \pm 0.81$& $8.70 \pm 0.80$ \\ 
  $D_{\bar{U}}$ & $17.7 \pm 3.3$& $16.4 \pm 3.0$& $17.6 \pm 3.3$& $16.5 \pm 3.2$ \\ 
  $A_{\bar{D}}$ & $0.158 \pm 0.011$& $0.160 \pm 0.011$& $0.1561 \pm 0.0098$& $0.1594 \pm 0.0099$ \\ 
  $B_{\bar{D}}$ & $-0.1682 \pm 0.0082$& $-0.1666 \pm 0.0080$& $-0.1760 \pm 0.0074$& $-0.1732 \pm 0.0073$ \\ 
  $C_{\bar{D}}$ & $4.2 \pm 1.3$& $4.5 \pm 1.3$& $4.0 \pm 1.2$& $4.3 \pm 1.3$ \\\hline
  $B_g$ & $-0.11 \pm 0.16$& $-0.12 \pm 0.15$& $-0.07 \pm 0.13$& $-0.09 \pm 0.12$ \\ 
  $C_g$ & $11.2 \pm 1.7$& $10.7 \pm 1.4$& $12.3 \pm 1.7$& $11.8 \pm 1.5$  \\ 
  $A_g'$ &  $2.1 \pm 1.5$& $1.9 \pm 1.2$& $2.9 \pm 1.6$& $2.7 \pm 1.1$  \\ 
  ${B_g'}$ & $-0.206 \pm 0.079$& $-0.216 \pm 0.075$& $-0.142 \pm 0.083$& $-0.162 \pm 0.098$ \\  
  \hline 
  {${ \alpha_s(M^2_Z)}$} & $ 0.1176$ & $ 0.1176$ & $ 0.1176$ & $0.1176$ \\ \hline
\hline
    \end{tabular}
\vspace{-0.0cm}
\caption{\label{tab:pa1}{ {The NLO numerical values of parameters and their uncertainties for the $xu_v$, $xd_v$, $x\bar u$, $x\bar d$, $x\bar s$ and $xg$ PDFs at the initial scale of $Q^2_0 = 1.9$~GeV$^2$ in the first scenario, where the strong coupling, $\alpha_s(M_Z^2)$, is fixed to $0.117$.}}}
\vspace{-0.4cm}
\end{center}
\end{table}

\begin{table}[h]
\begin{center}
\begin{tabular}{|l|c|c|c|c|}
\hline
\hline
 \multicolumn{5}{|c|}{ {\bf Second Scenario: The Strong Coupling, $\alpha_s(M^2_Z)$, is Free} }    \\ \hline
 { \bf Parameter} & {$~~~~$RT BASE$~~~~$} & { $~~~$RT TOTAL$~~~$} & {RTOPT BASE} & {RTOPT TOTAL} \\ \hline
  ${B_{u_v}}$ & $0.712 \pm 0.047$& $0.725 \pm 0.048$& $0.710 \pm 0.045$& $0.709 \pm 0.043$ \\ 
  ${C_{u_v}}$ & $4.88 \pm 0.12$& $4.83 \pm 0.12$& $4.89 \pm 0.11$& $4.87 \pm 0.10$ \\ 
  $E_{u_v}$ & $13.9 \pm 2.3$& $13.4 \pm 2.2$& $13.7 \pm 2.2$& $13.7 \pm 2.3$ \\ \hline
  ${B_{d_v}}$ & $0.811 \pm 0.094$& $0.826 \pm 0.096$& $0.812 \pm 0.093$& $0.813 \pm 0.091$ \\ 
  $C_{d_v}$ &  $4.18 \pm 0.38$& $4.17 \pm 0.39$& $4.24 \pm 0.38$& $4.22 \pm 0.39$ \\ \hline
  $C_{\bar{U}}$ & $9.09 \pm 0.89$& $8.70 \pm 0.90$& $9.21 \pm 0.87$& $8.97 \pm 0.84$ \\ 
  $D_{\bar{U}}$ & $18.5 \pm 3.9$& $16.3 \pm 3.7$& $19.2 \pm 3.9$& $17.8 \pm 3.5$ \\ 
  $A_{\bar{D}}$ & $0.160 \pm 0.012$& $0.160 \pm 0.011$& $0.158 \pm 0.010$& $0.1610 \pm 0.0100$ \\ 
  $B_{\bar{D}}$ & $-0.1657 \pm 0.0099$& $-0.167 \pm 0.010$& $-0.1728 \pm 0.0083$& $-0.1704 \pm 0.0079$  \\ 
  $C_{\bar{D}}$ & $4.4 \pm 1.4$& $4.5 \pm 1.4$& $4.4 \pm 1.3$& $4.6 \pm 1.3$ \\\hline
  $B_g$ & $-0.13 \pm 0.11$& $-0.12 \pm 0.12$& $-0.10 \pm 0.12$& $-0.12 \pm 0.10$ \\ 
  $C_g$ & $11.8 \pm 2.2$& $10.6 \pm 2.0$& $13.5 \pm 2.3$& $12.7 \pm 2.3$  \\ 
  $A_g'$ & $2.3 \pm 1.1$& $1.87 \pm 0.84$& $3.4 \pm 1.6$& $3.0 \pm 1.6$ \\ 
  ${B_g'}$ & $-0.217 \pm 0.074$& $-0.215 \pm 0.073$& $-0.164 \pm 0.096$& $-0.182 \pm 0.074$ \\  
  \hline 
  {${ \alpha_s(M^2_Z)}$} & $0.1161 \pm 0.0037$& $0.1178 \pm 0.0038$& $0.1151 \pm 0.0032$& $0.1154 \pm 0.0028$ \\ \hline
\hline
    \end{tabular}
\vspace{-0.0cm}
\caption{\label{tab:pa2}{ {The NLO numerical values of parameters and their uncertainties for  the 
$xu_v$, $xd_v$, $x\bar u$, $x\bar d$, $x\bar s$ and $xg$ PDFs at the initial scale of $Q^2_0 = 1.9$~GeV$^2$ in the second scenario, where the strong coupling, $\alpha_s(M_Z^2)$, is taken as an extra free parameter.}}}
\vspace{-0.4cm}
\end{center}
\end{table}

 According to the numerical results in Table~\ref{tab:pa2}, when we add the charm cross section H1-ZEUS combined measurements data to the  HERA run I and II combined data, the numerical value of $\alpha_s(M^2_Z)$ changes from $0.1161 \pm 0.0037$ to $0.1178 \pm 0.0038$  and from $0.1151 \pm 0.0032$ to $0.1154 \pm 0.0028$, for the RT and RTOPT schemes, without and with charm flavour data included, respectively.  If we compare our results for  $\alpha_s(M^2_Z)$ for RT TOTAL and RTOPT TOTAL with the world average value, $\alpha_s(M^2_Z)=0.1185 \pm 0.0006$, which was recently reported by the PDG \cite{Agashe:2014kda}, we  find a good agreement with the world average value. Of course, it should be noted, since the PDG value of $\alpha_s(M^2_Z)$ is extracted by global fits to a variety of experimental data, it has a much smaller uncertainty. In other words, although our QCD analysis has been performed based on only two data sets, our numerical results for the strong coupling are in good agreement with the world average value. Also, these values of strong coupling show the impact of the RT and RTOPT schemes on the determination of $\alpha_s(M^2_Z)$, when considered as an extra free parameter.

\clearpage
\section{\label{Results}Results}
According to Table~\ref{tab:f1}, in going from the RT scheme to the RTOPT scheme, we get $\sim 0.4$~\% and $\sim 0.9$~\%~improvement in the fit quality, without and with the charm flavour contributions included, respectively. Also, according to Table~\ref{tab:f2}, in going from the RT scheme to the RTOPT scheme, we get $\sim 0.9$~\% and $\sim 2.0$~\%~improvement in the  $\alpha_s(M^2_Z)$ value, without and with the charm flavour contributions respectively. 
\begin{table}[h]
\begin{center}
\begin{tabular}{|l|c|c|c|c|}
\hline
\hline
 \multicolumn{3}{|c|}{ {\bf First Scenario: The Strong Coupling, $\alpha_s(M^2_Z)$ is Fixed} }    \\ \hline
 {Scheme} & $\chi^2_{\rm TOTAL}/dof$ & ${\alpha_s(M^2_Z)}$ \\ \hline 
  {RT BASE} & {${ 1335/1131}$}& $0.1176$  \\
  {RT TOTAL} & {${1389/1178}$}& $0.1176$  \\
  {RTOPT BASE} & {${1331/1131}$}& $0.1176$  \\ 
  {RTOPT TOTAL} & {${1378/1178}$}& $0.1176$  \\ \hline
\hline
    \end{tabular}
\vspace{-0.0cm}
\caption{\label{tab:f1}{ Comparison of the numerical values of $\frac{\chi^2_{\rm TOTAL}}{dof}$ for the RT and RTOPT schemes in the first scenario, where the strong coupling, $\alpha_s(M_Z^2)$, is fixed to $0.117$. RTOPT TOTAL has the best fit quality, as an impact of adding charm cross section H1-ZEUS combined measurements data to HERA I and II combined data. }}
\vspace{-0.4cm}
\end{center}
\end{table} 

\begin{table}[h]
\begin{center}
\begin{tabular}{|l|c|c|c|c|}
\hline
\hline
 \multicolumn{3}{|c|}{ {\bf Second Scenario: The Strong Coupling, $\alpha_s(M^2_Z)$ is Free} }    \\ \hline
 {Scheme} & $\chi^2_{\rm TOTAL}/dof$ & ${\alpha_s(M^2_Z)}$ \\ \hline 
  {RT BASE}  & {${1335/1130}$}& $0.1161 \pm 0.0037$  \\
  {RT TOTAL} & {${1389/1177}$}& $0.1178 \pm 0.0038$  \\
  {RTOPT BASE} & {${1330/1130}$}& $0.1151 \pm 0.0032$  \\  
  {RTOPT TOTAL} & {${1377/1177}$}& $0.1154 \pm 0.0028$ \\ \hline
\hline
    \end{tabular}
\vspace{-0.0cm}
\caption{\label{tab:f2}{Comparison of the numerical values of $\frac{\chi^2_{\rm TOTAL}}{dof}$ and ${\alpha_s(M^2_Z)}$ for the RT and RTOPT schemes in the second scenario, where the strong coupling, $\alpha_s(M_Z^2)$, is taken as an extra free parameter. RTOPT TOTAL has the best fit quality and improvement in oupling, $\alpha_s(M^2_Z)$, as an impact of adding charm cross section H1-ZEUS combined measurements data to HERA I and II combined data.}}
\vspace{-0.4cm}
\end{center}
\end{table}

\begin{figure*}
\includegraphics[width=0.19\textwidth]{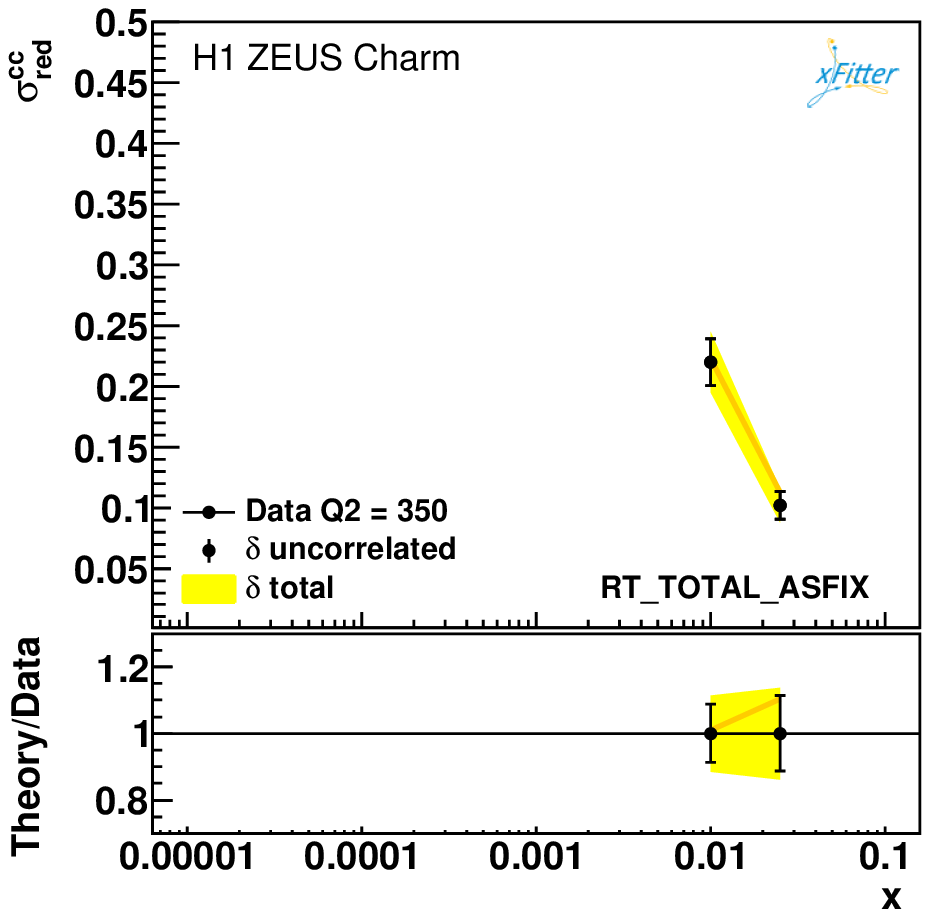}
\includegraphics[width=0.19\textwidth]{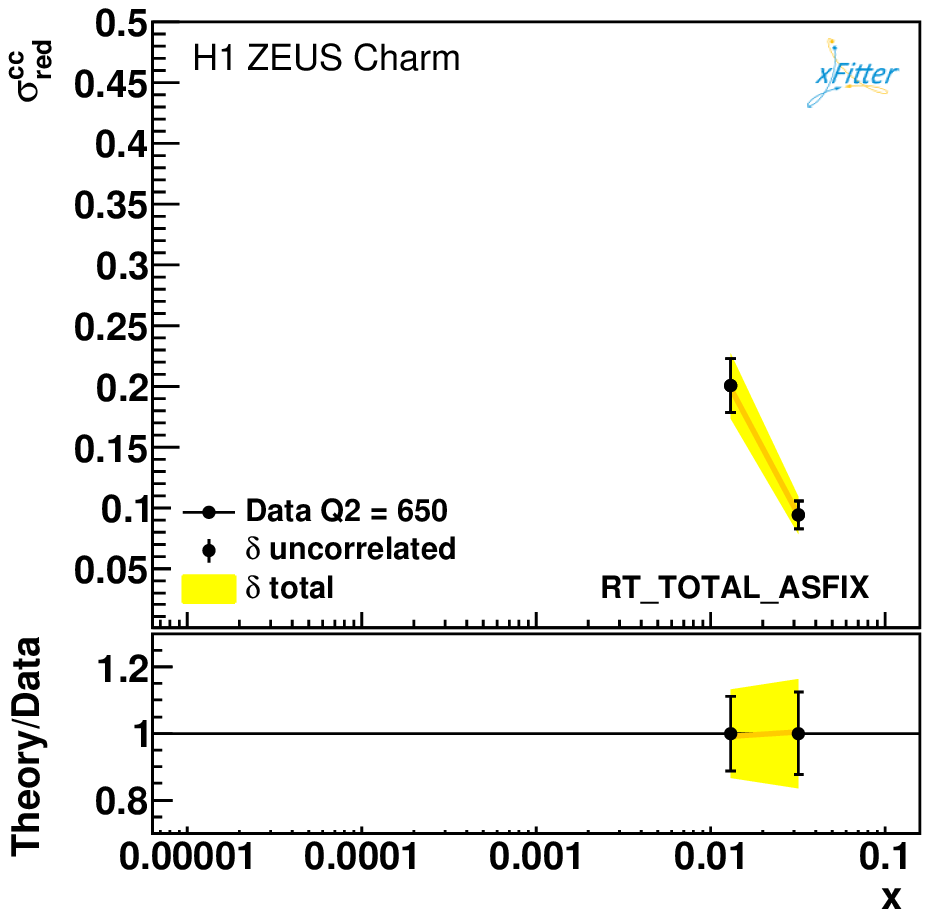}
\includegraphics[width=0.19\textwidth]{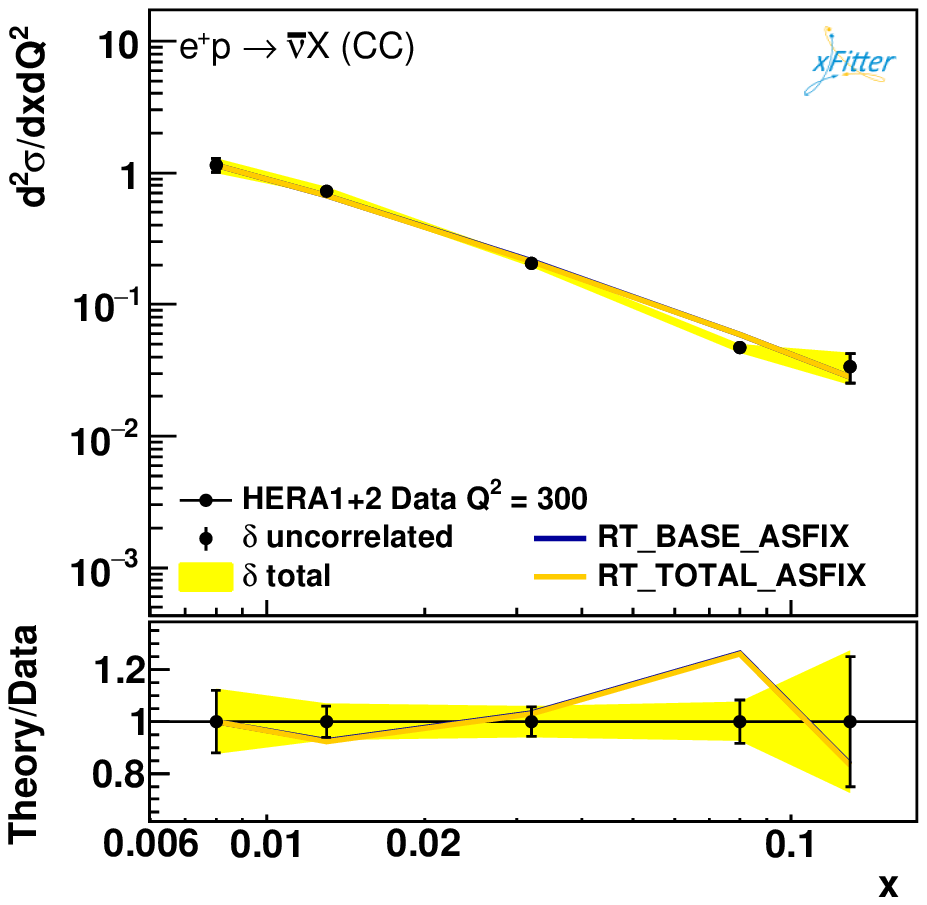}
\includegraphics[width=0.19\textwidth]{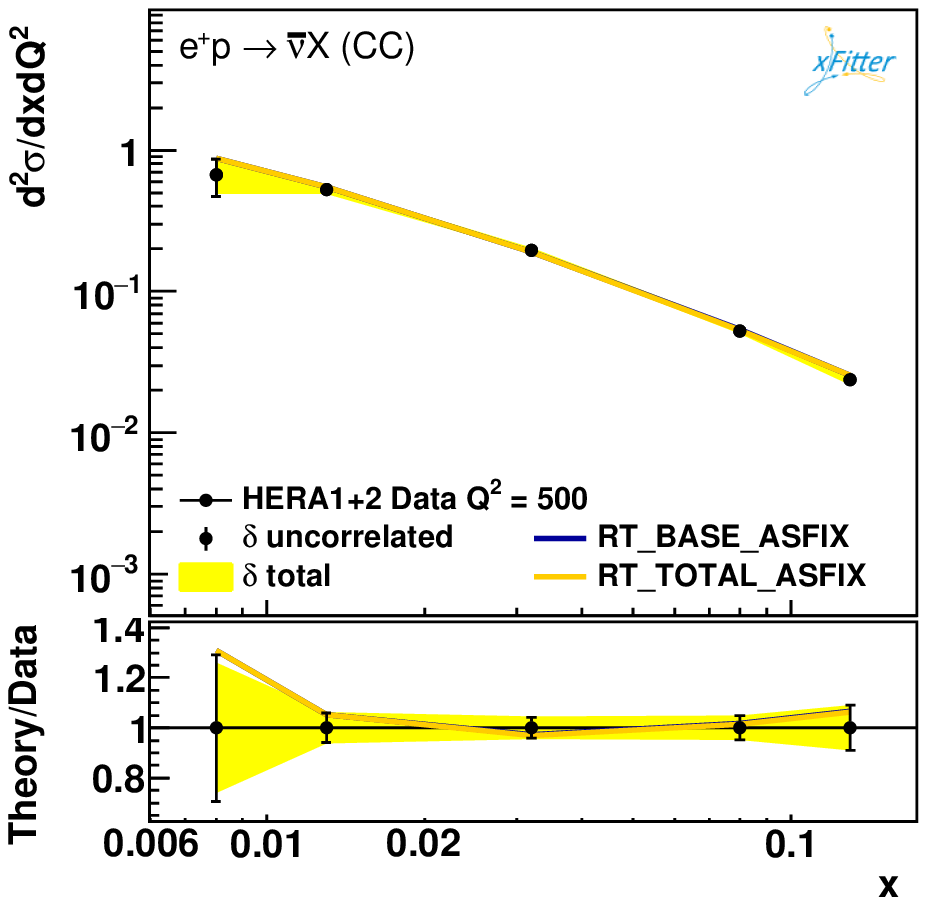}
\includegraphics[width=0.19\textwidth]{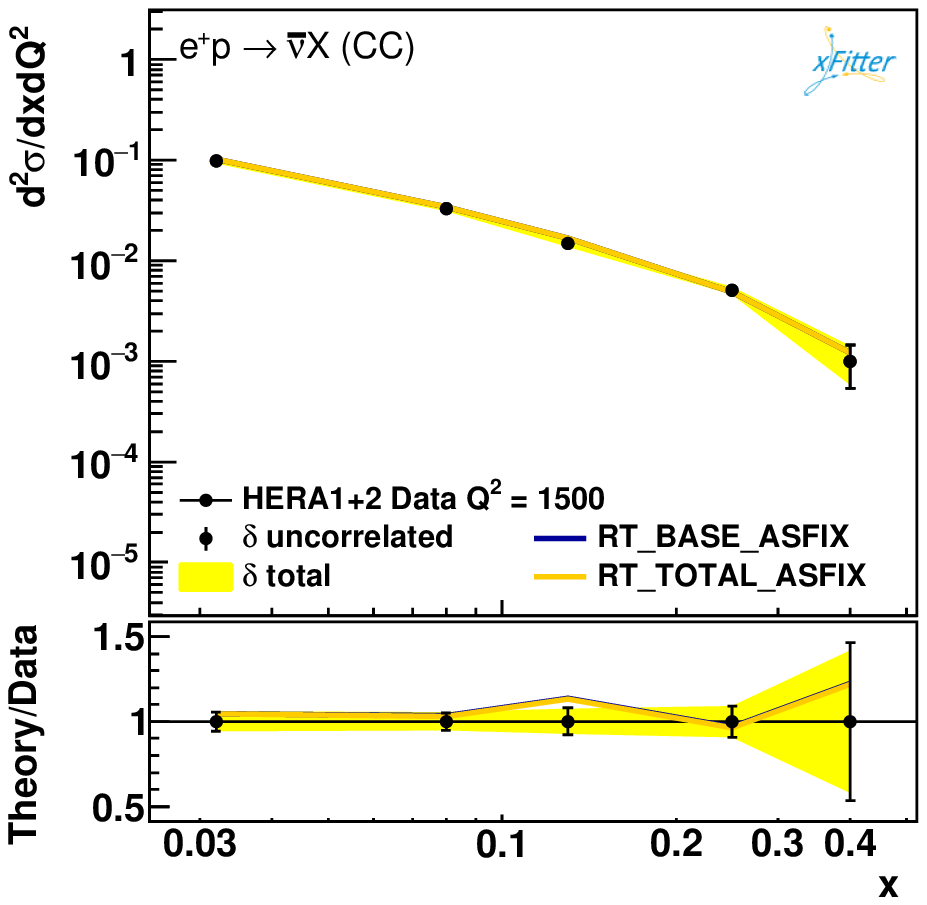}

\includegraphics[width=0.19\textwidth]{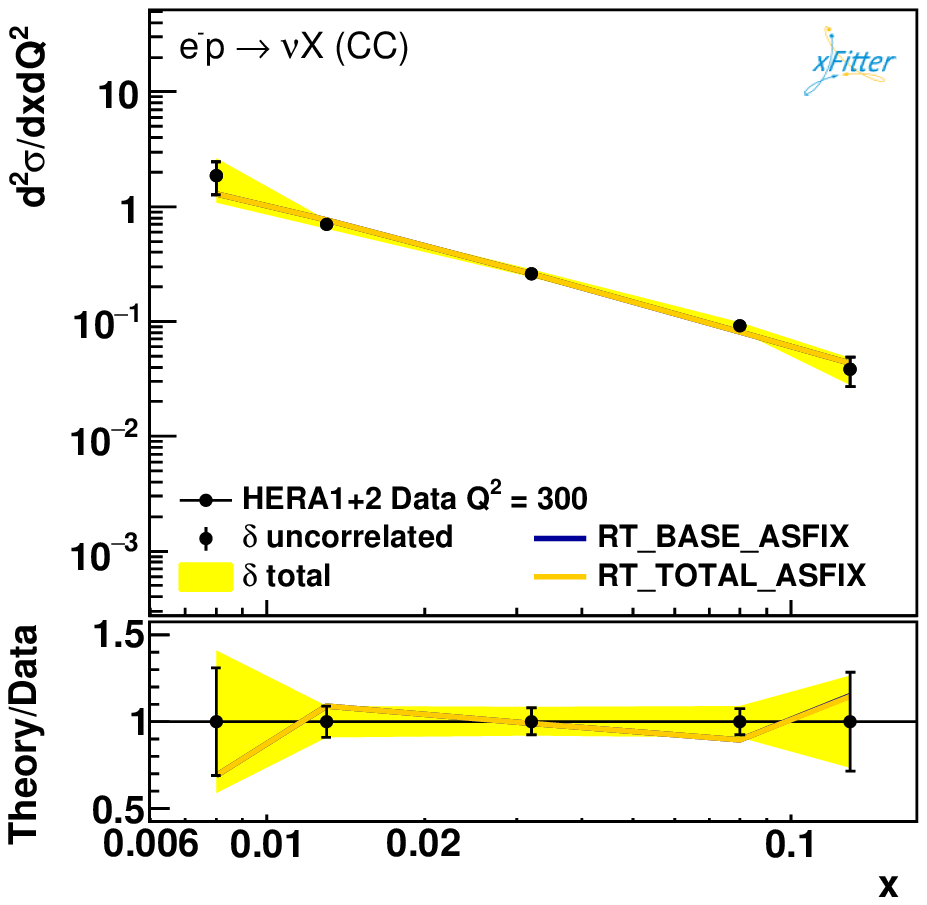}
\includegraphics[width=0.19\textwidth]{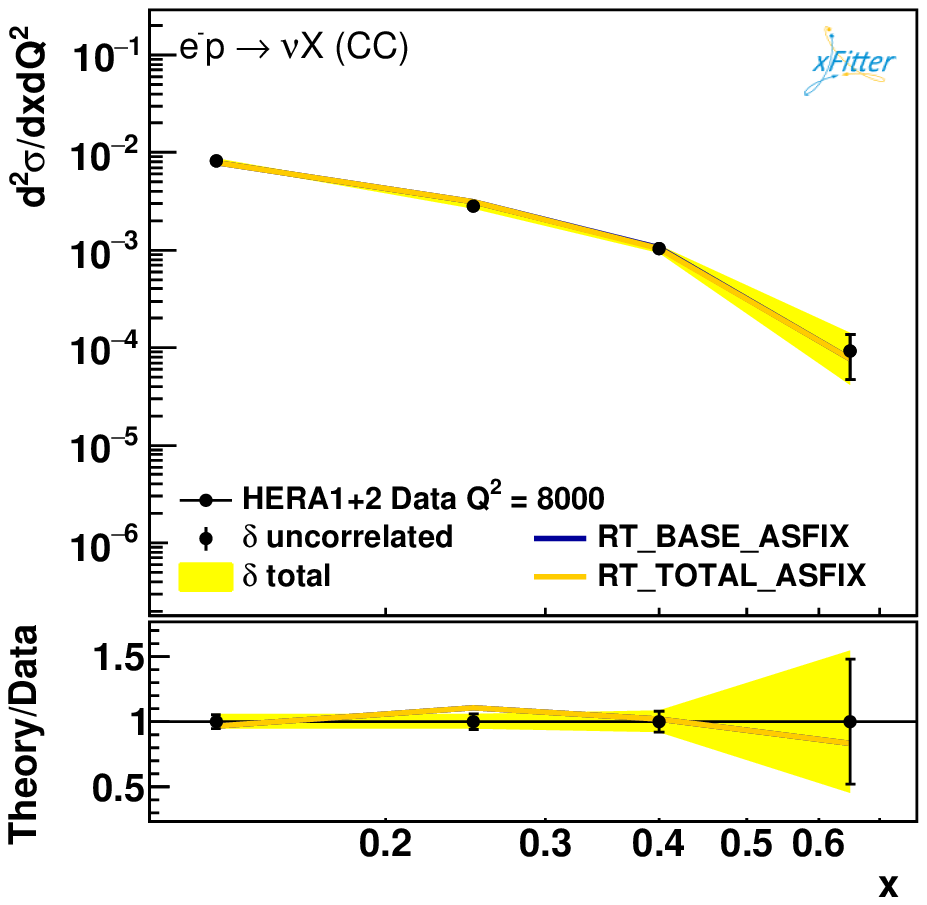}
\includegraphics[width=0.19\textwidth]{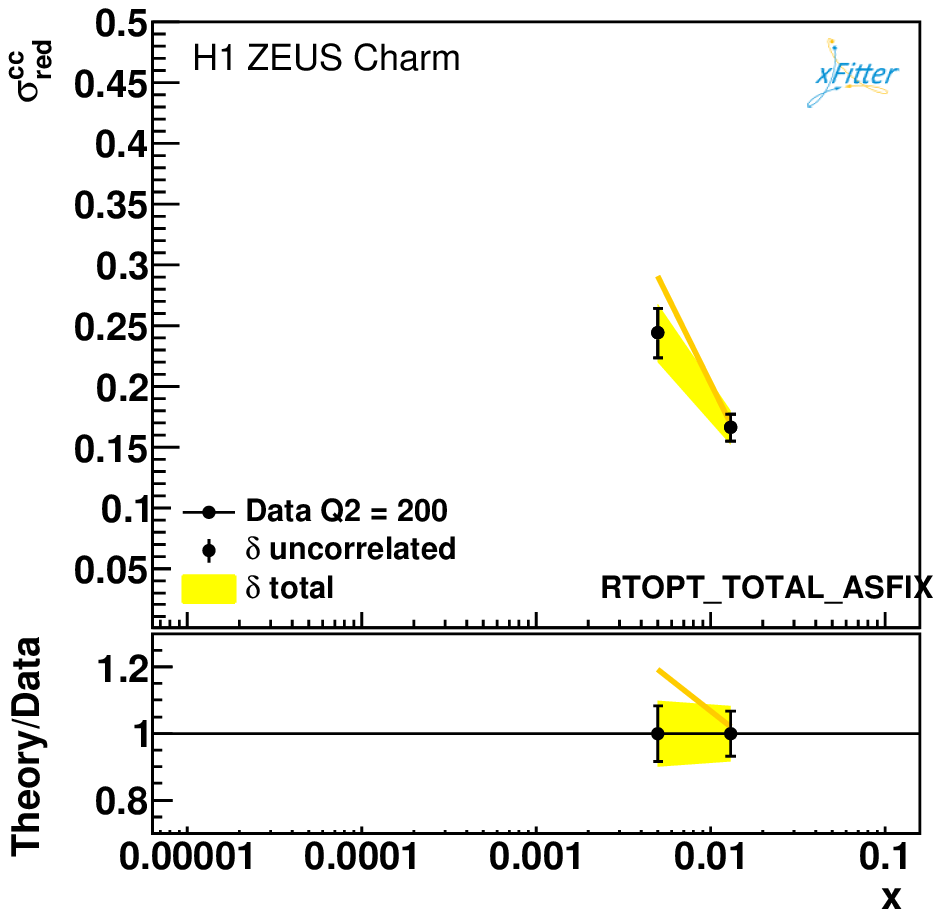}
\includegraphics[width=0.19\textwidth]{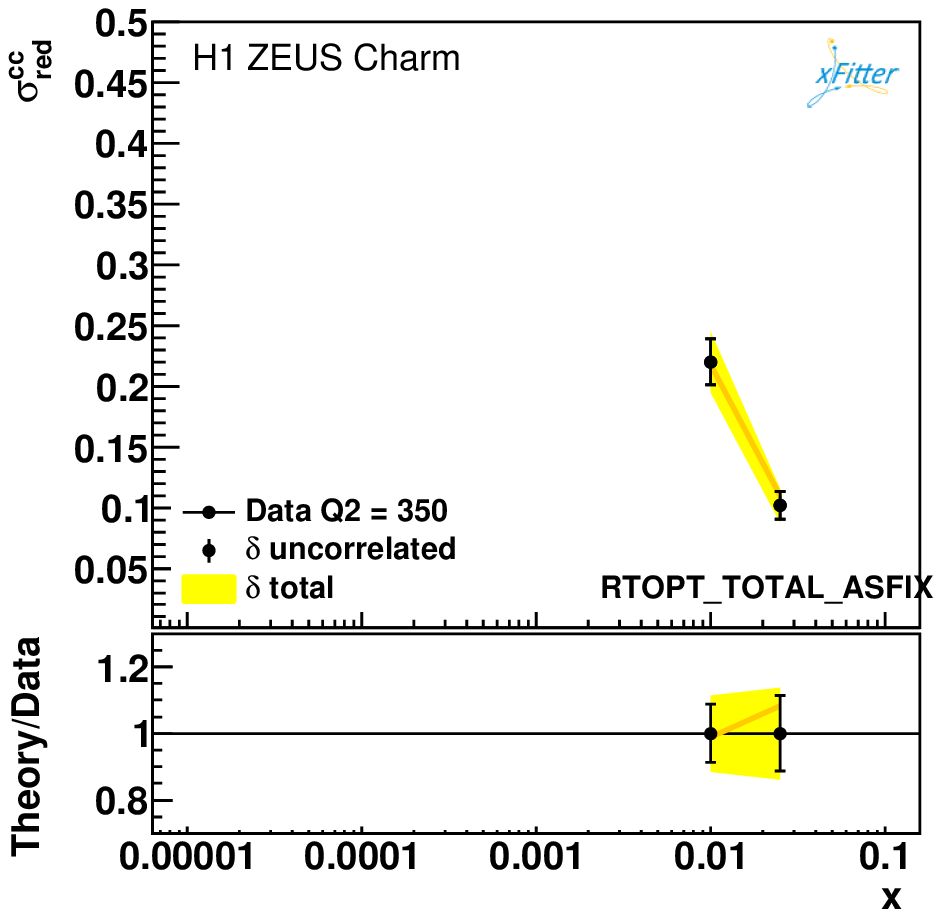}
\includegraphics[width=0.19\textwidth]{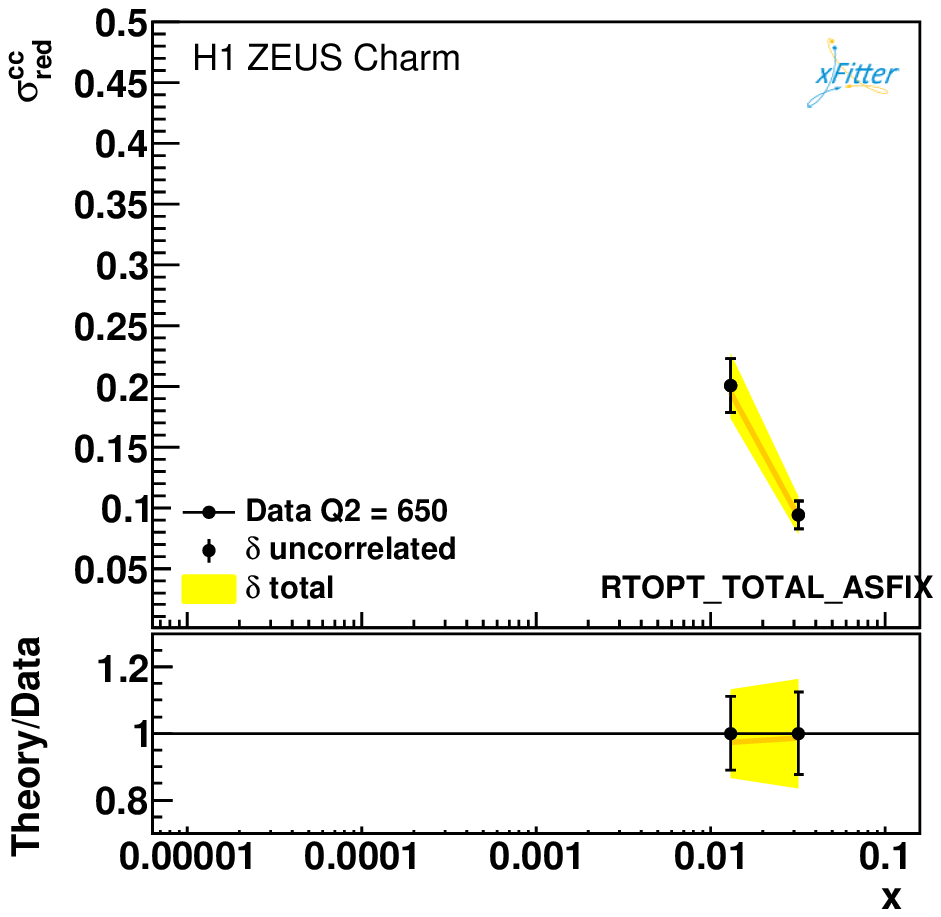}

\includegraphics[width=0.19\textwidth]{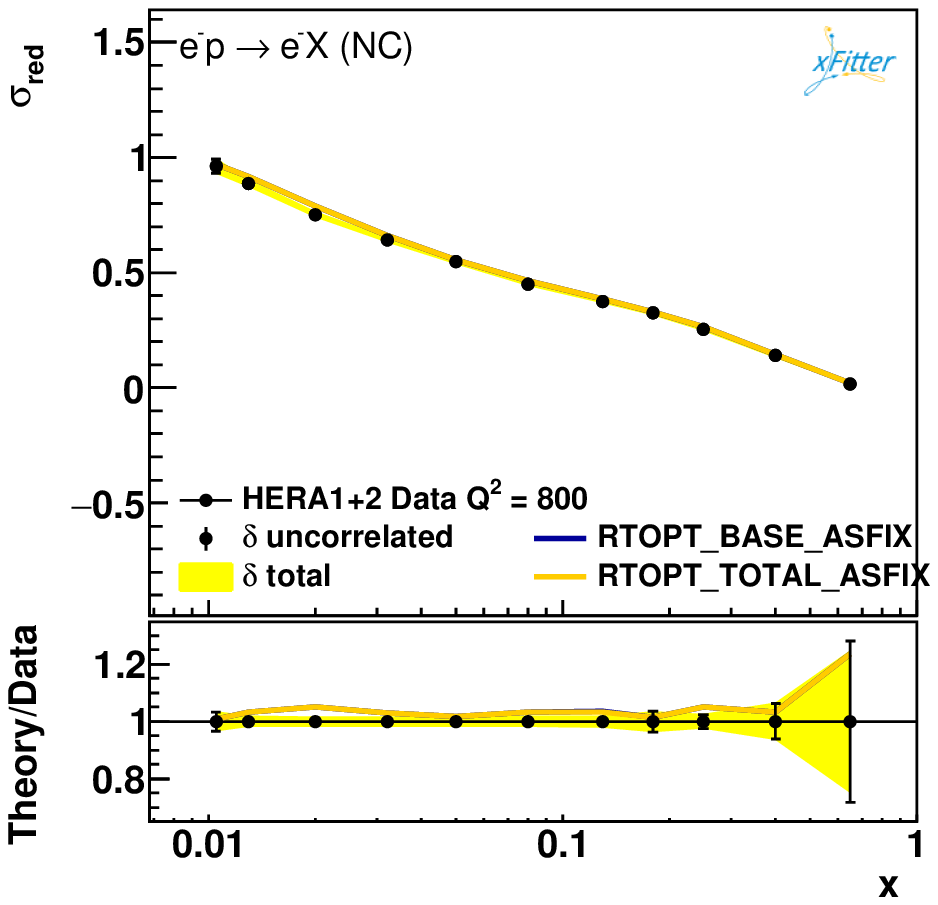}
\includegraphics[width=0.19\textwidth]{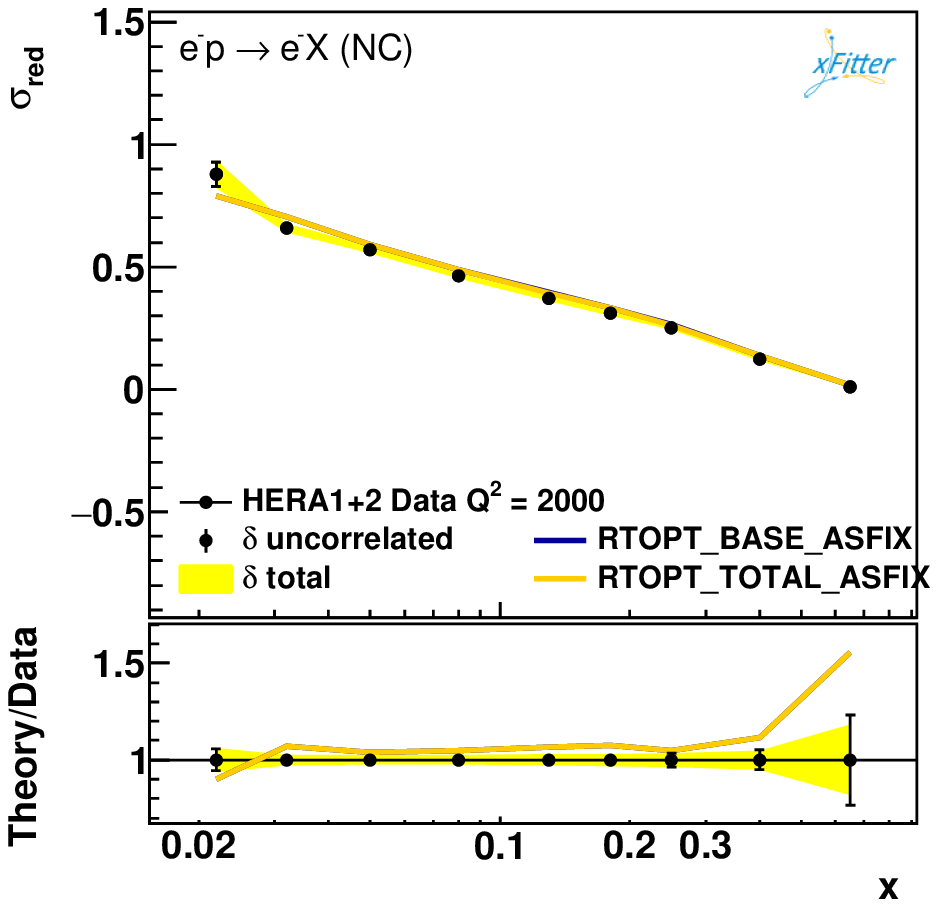}
\includegraphics[width=0.19\textwidth]{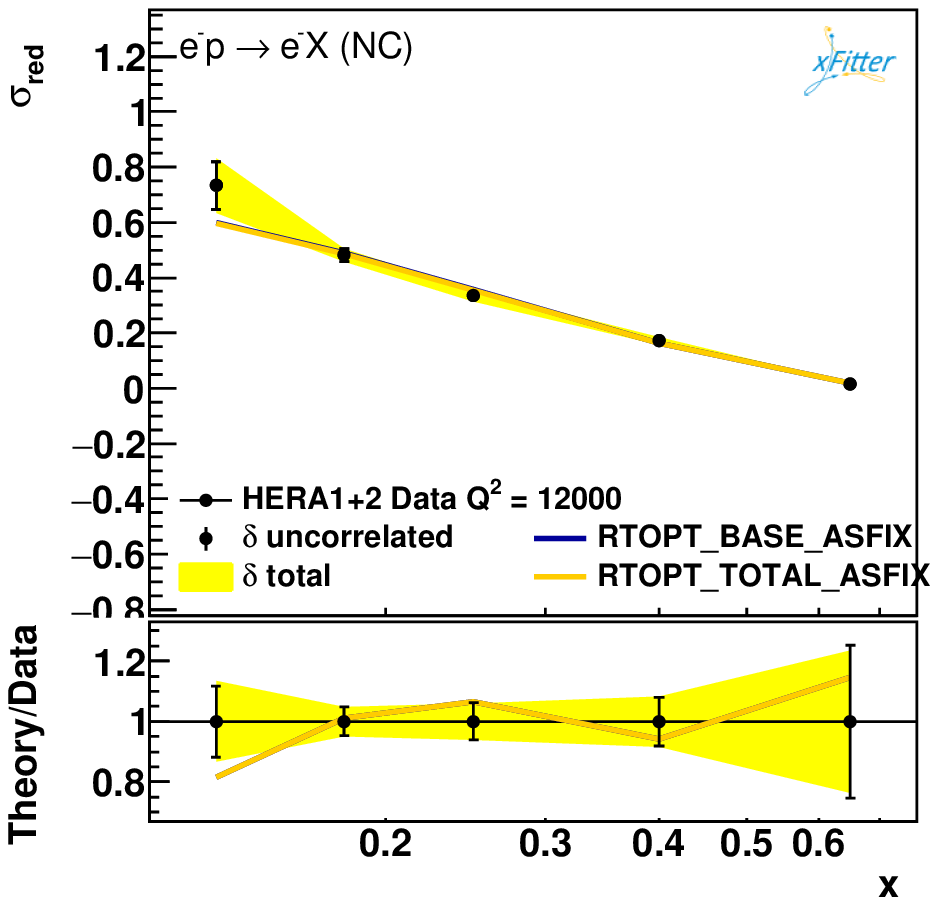}
\includegraphics[width=0.19\textwidth]{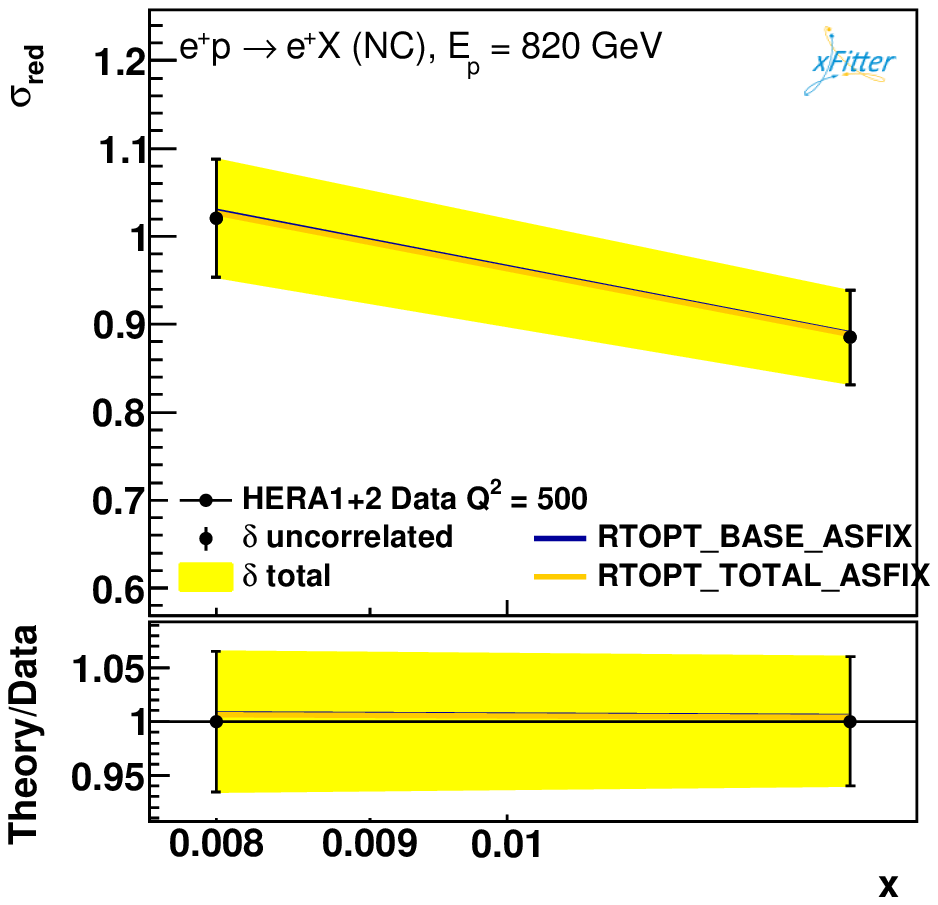}
\includegraphics[width=0.19\textwidth]{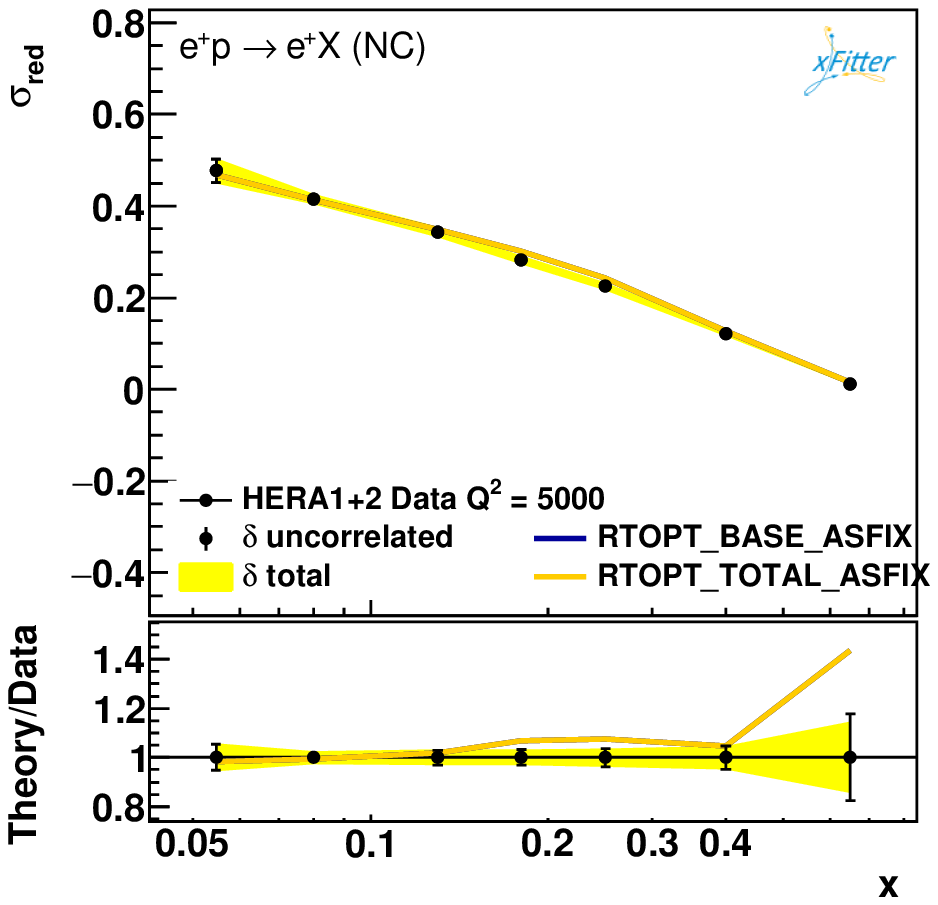}

\includegraphics[width=0.19\textwidth]{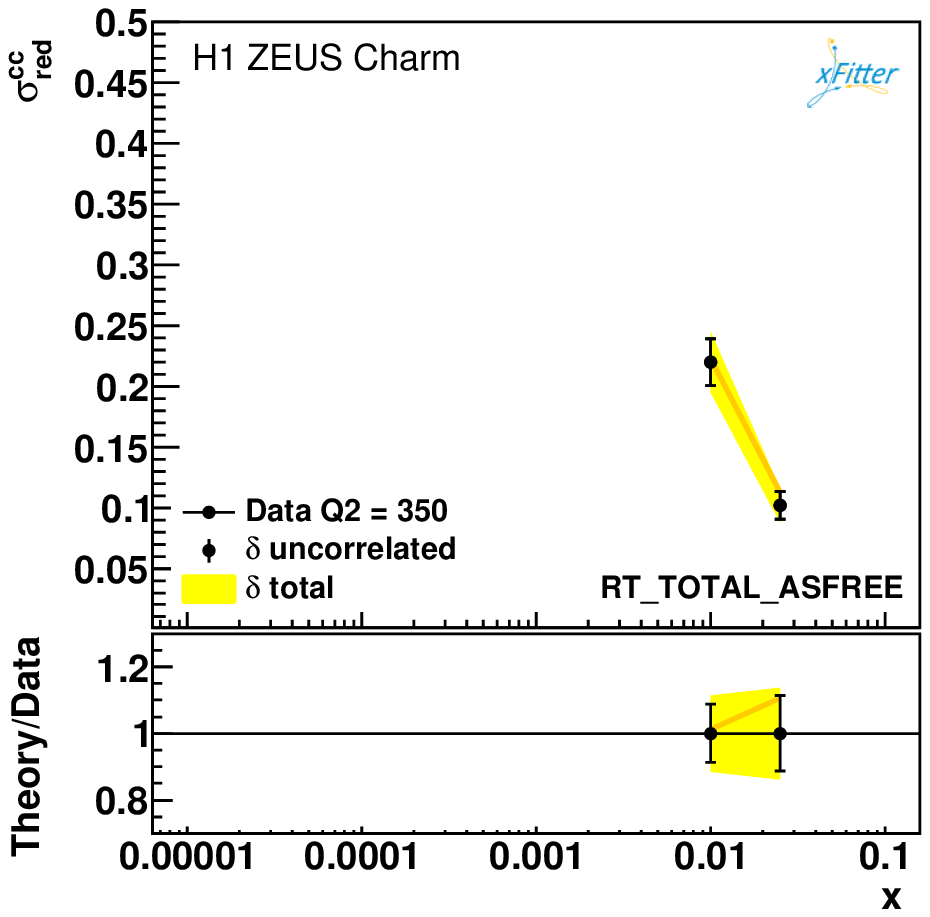}
\includegraphics[width=0.19\textwidth]{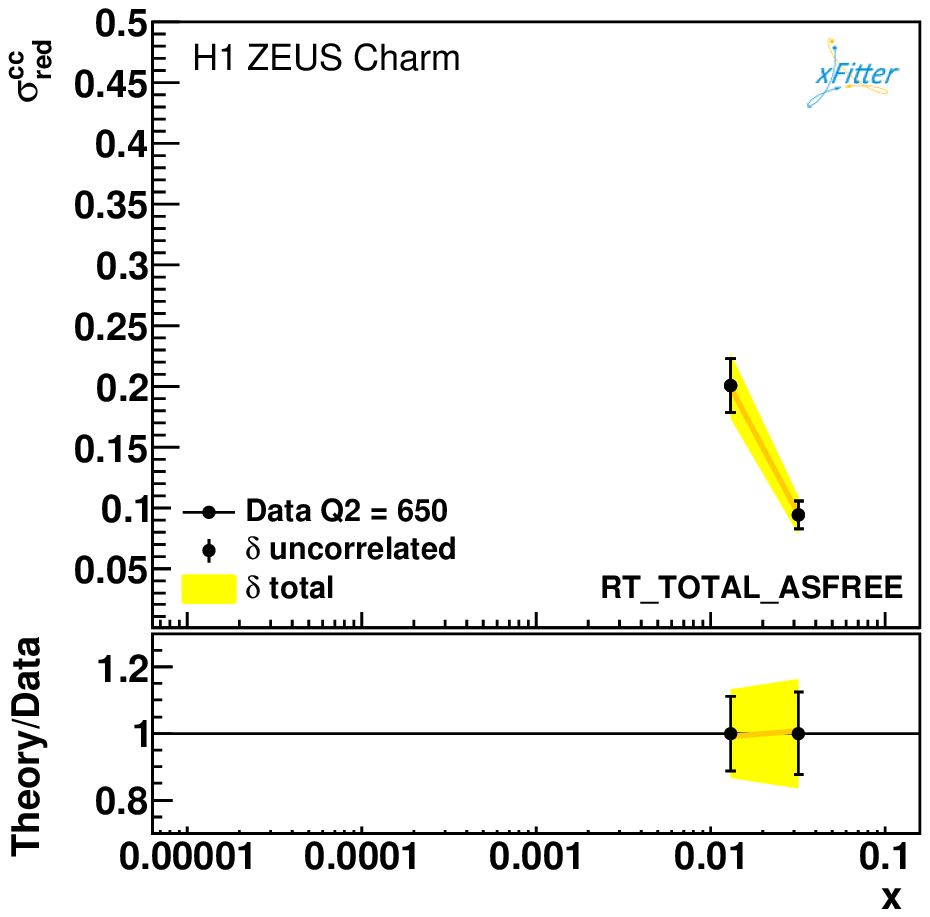}
\includegraphics[width=0.19\textwidth]{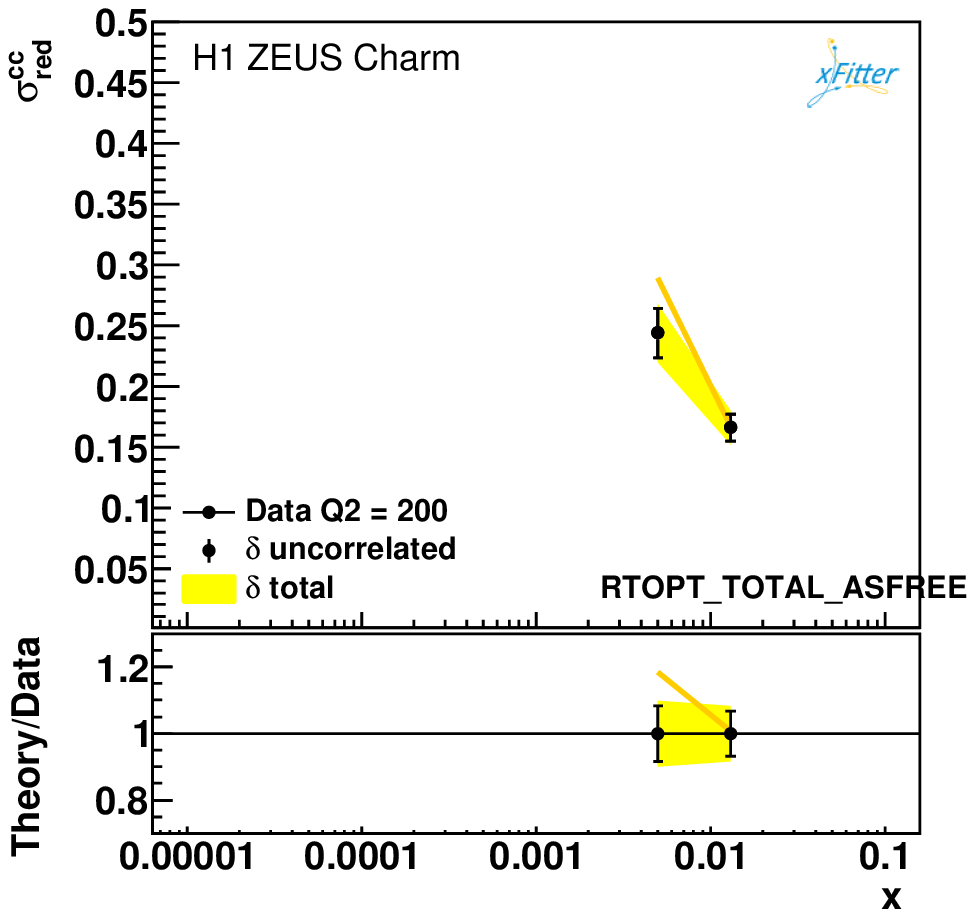}
\includegraphics[width=0.19\textwidth]{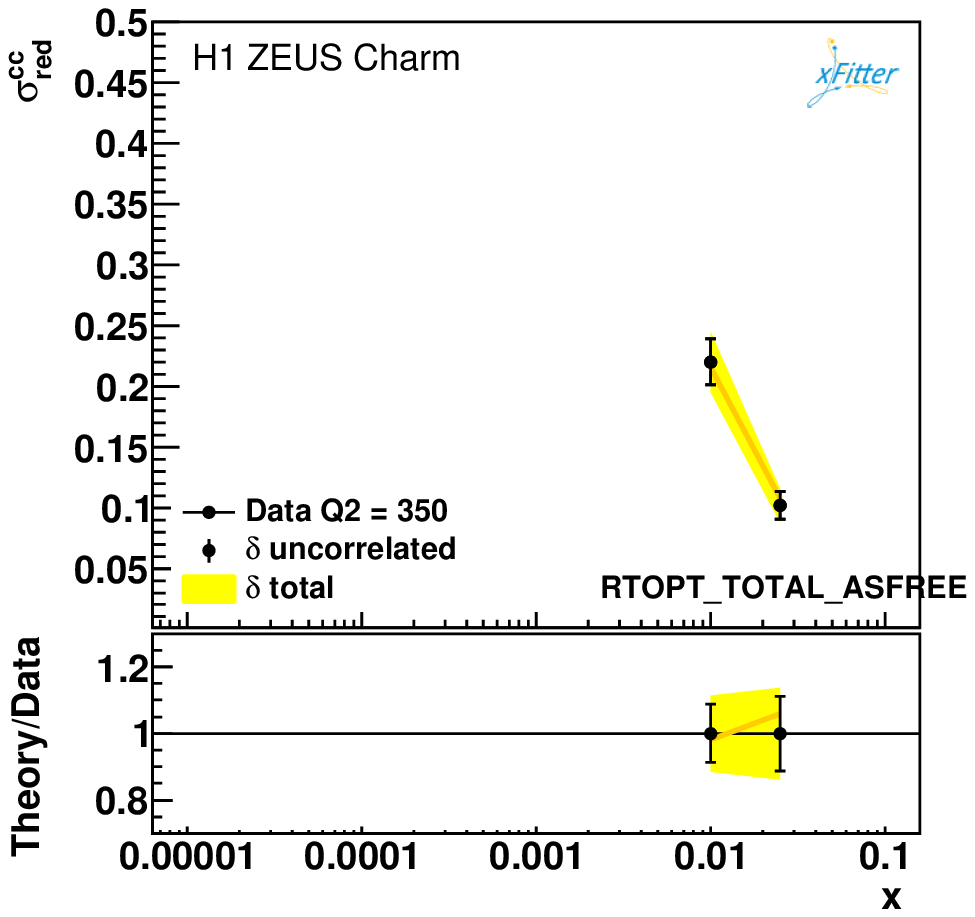}
\includegraphics[width=0.19\textwidth]{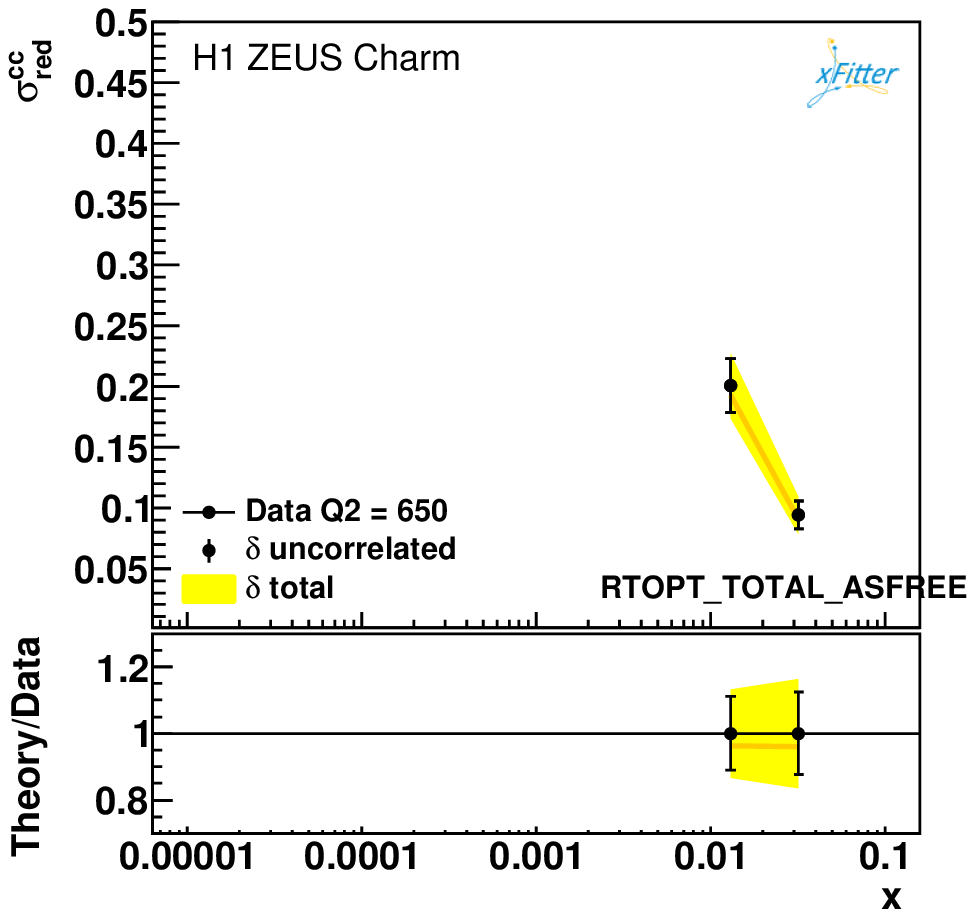}

\includegraphics[width=0.19\textwidth]{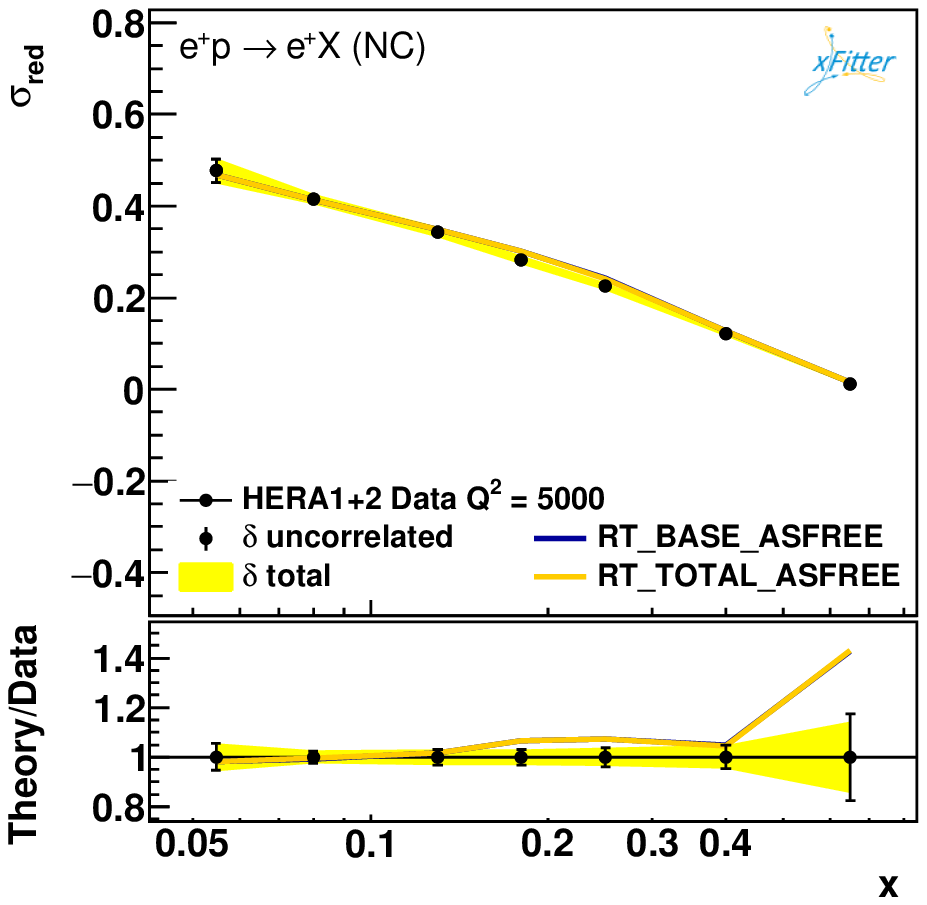}
\includegraphics[width=0.19\textwidth]{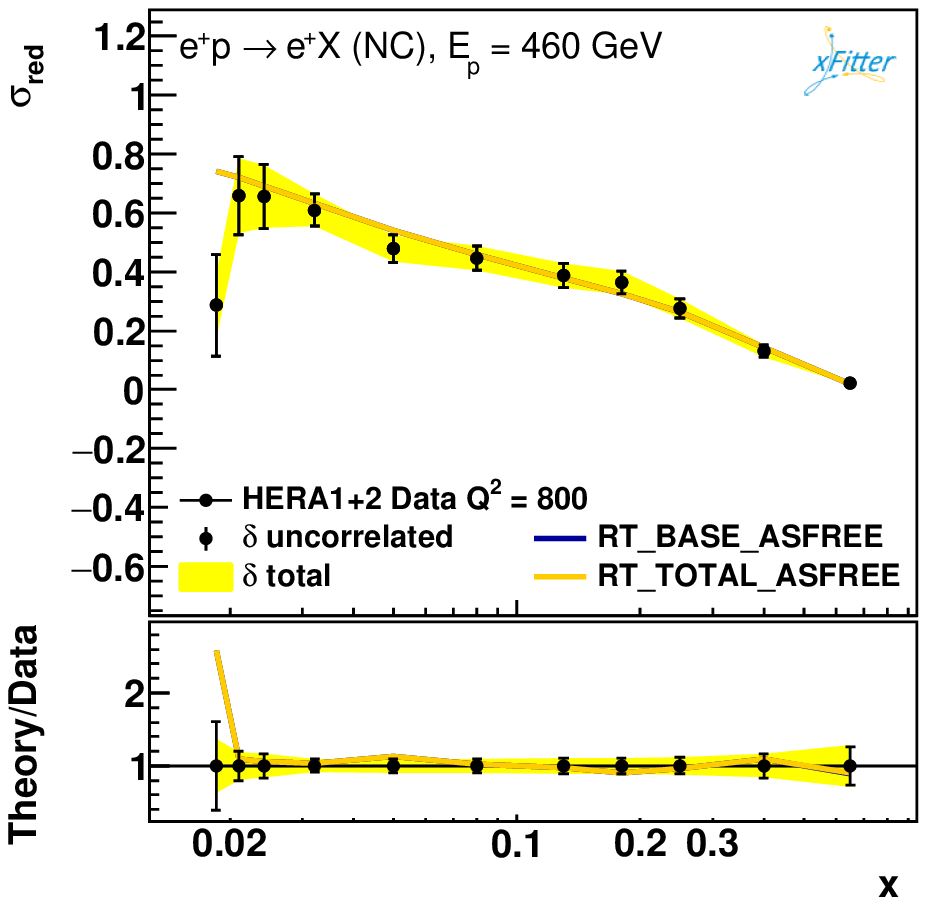}
\includegraphics[width=0.19\textwidth]{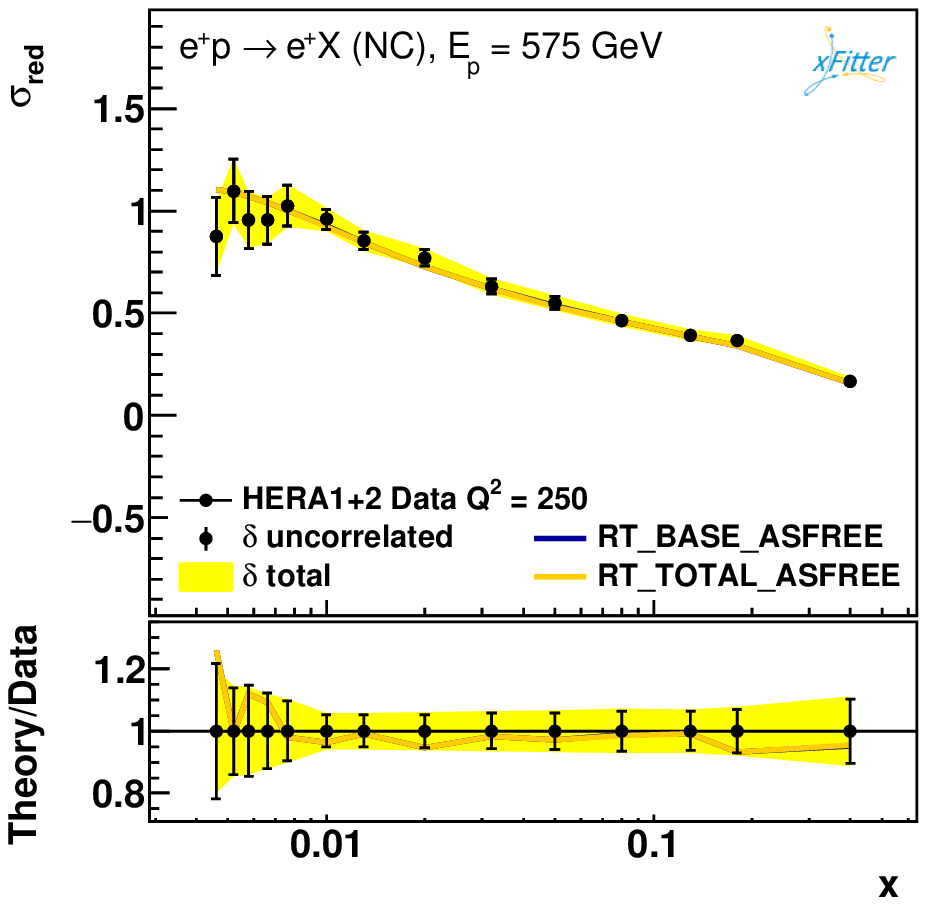}
\includegraphics[width=0.19\textwidth]{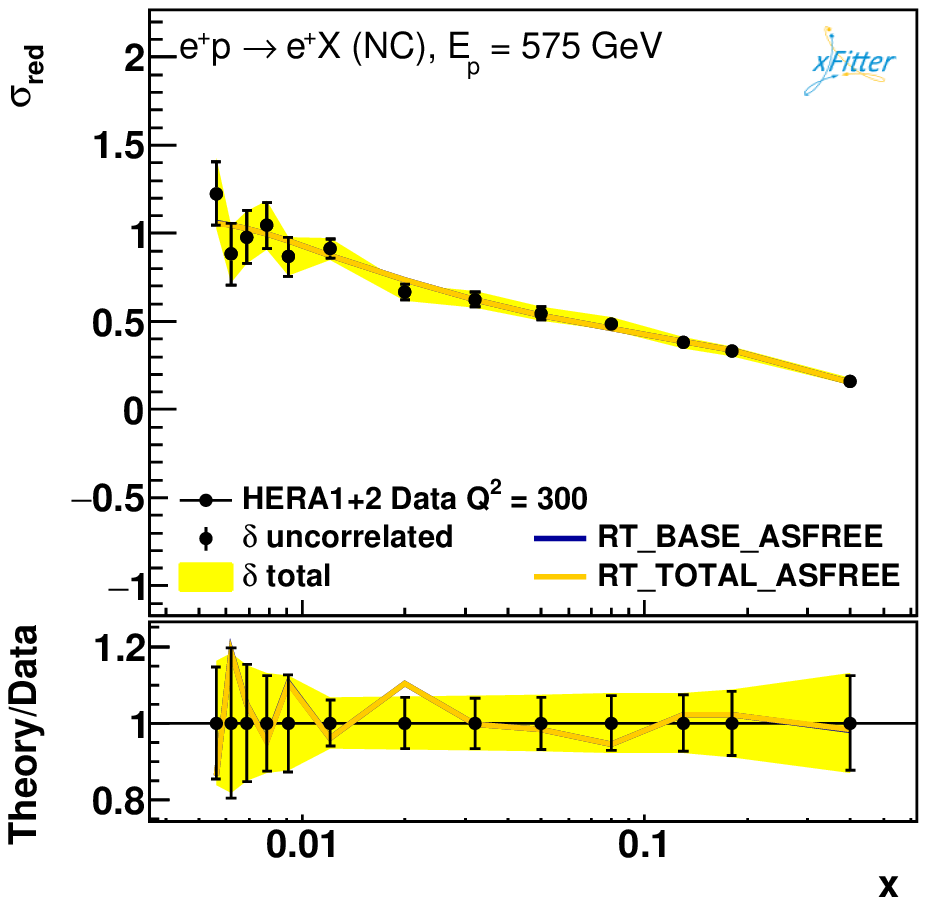}
\includegraphics[width=0.19\textwidth]{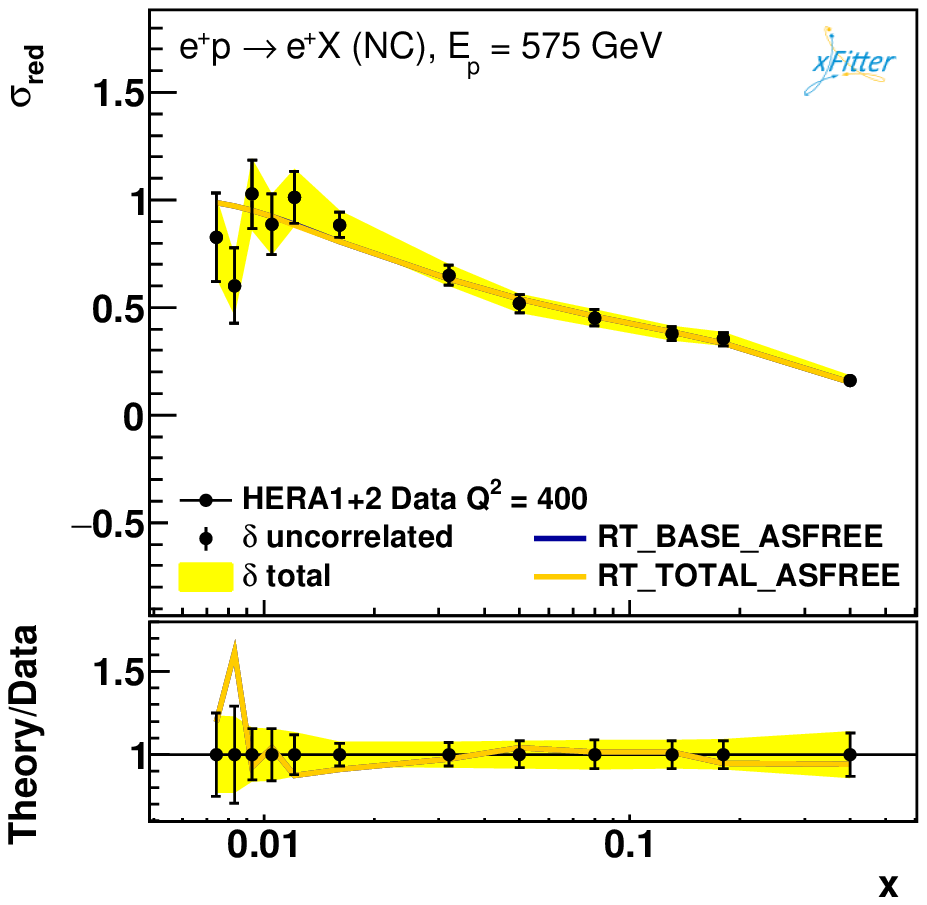}

\includegraphics[width=0.19\textwidth]{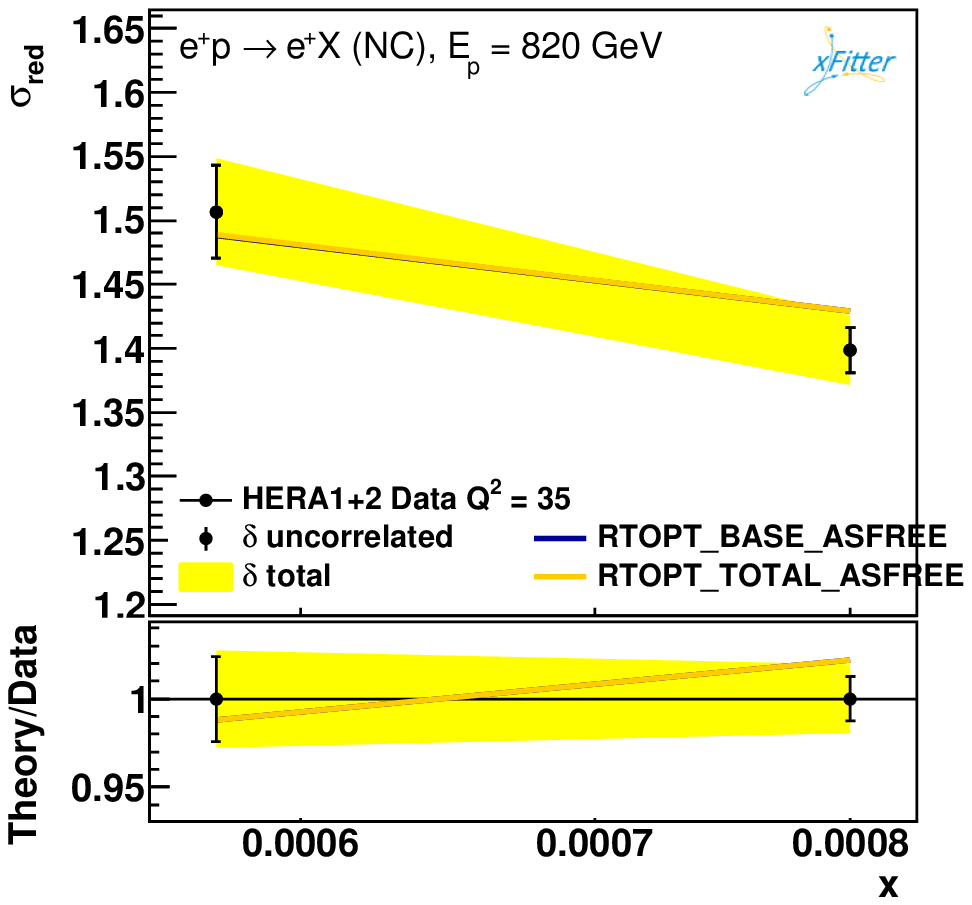}
\includegraphics[width=0.19\textwidth]{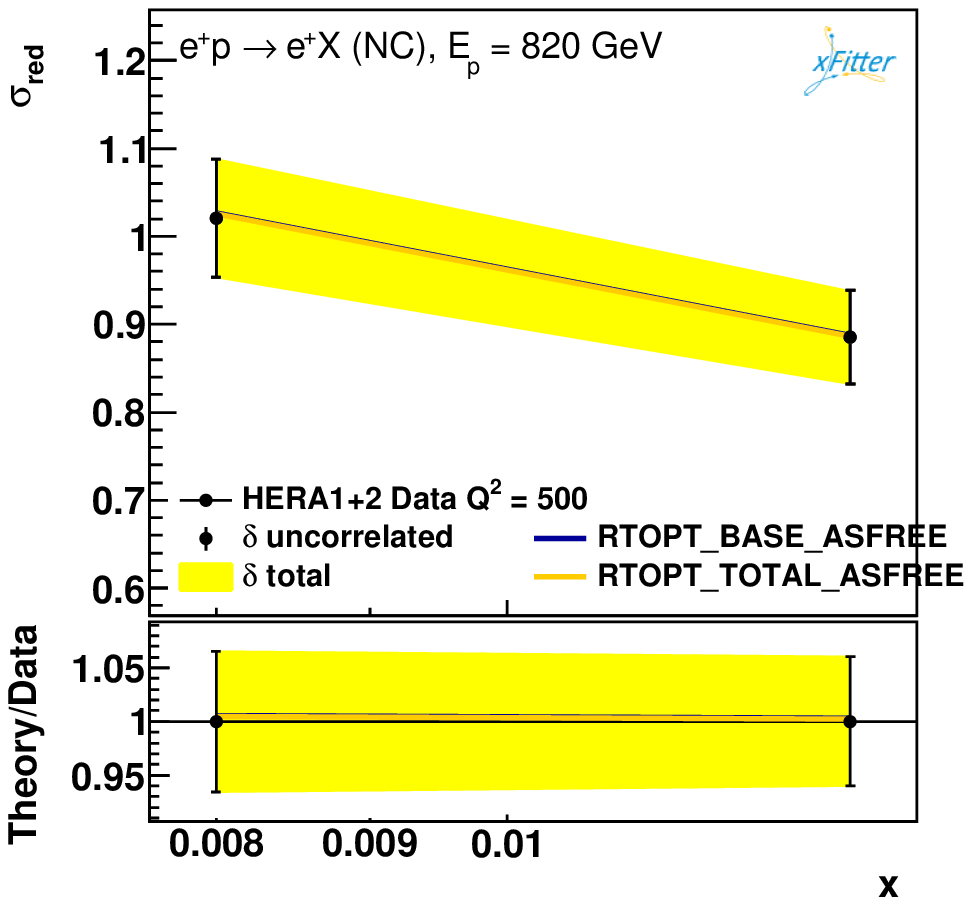}
\includegraphics[width=0.19\textwidth]{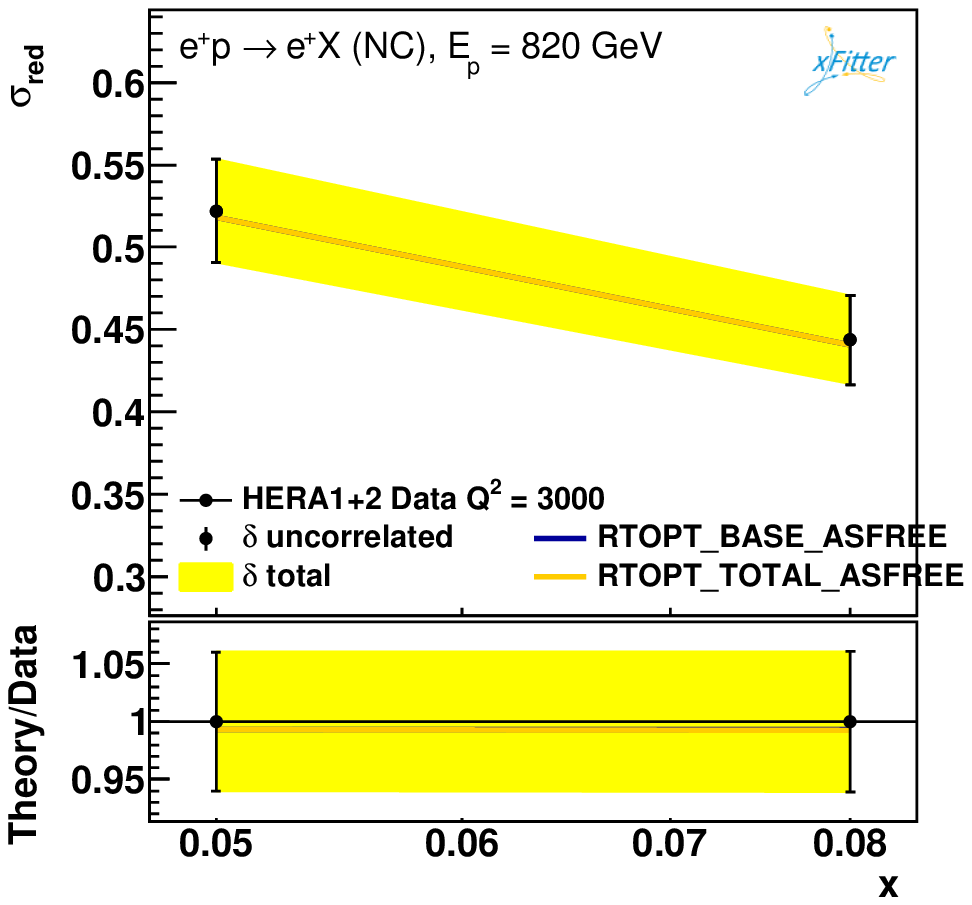}
\includegraphics[width=0.19\textwidth]{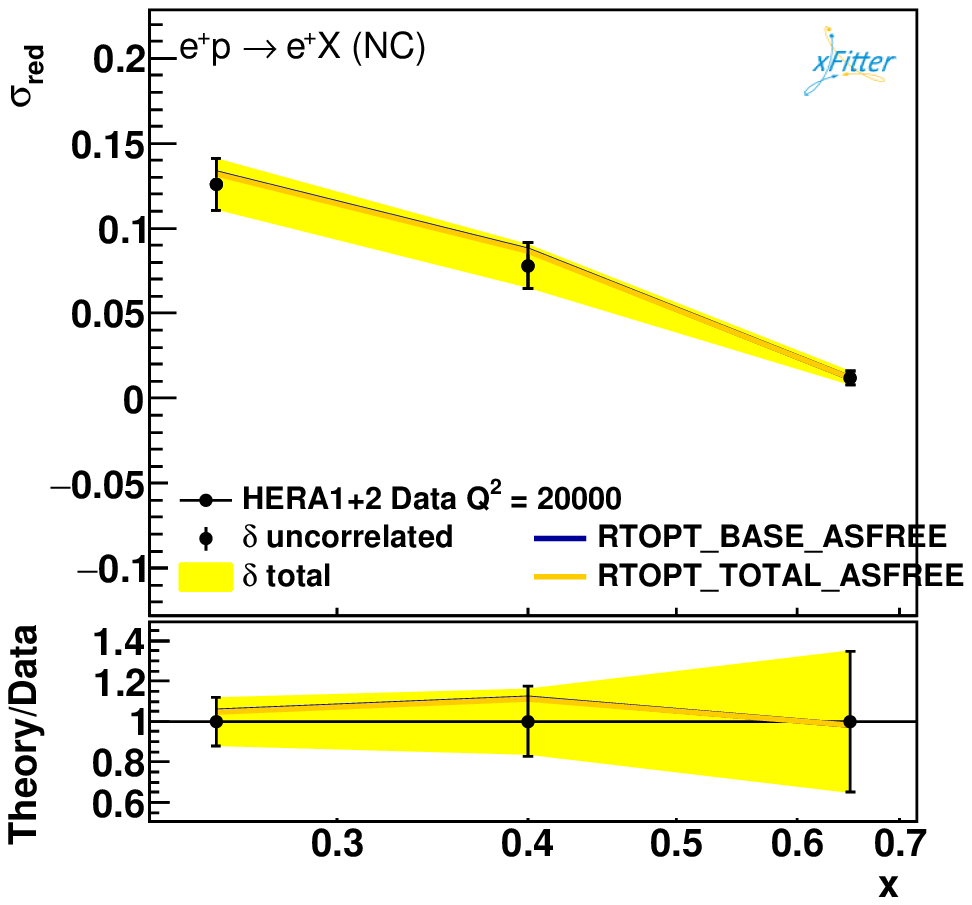}
\includegraphics[width=0.19\textwidth]{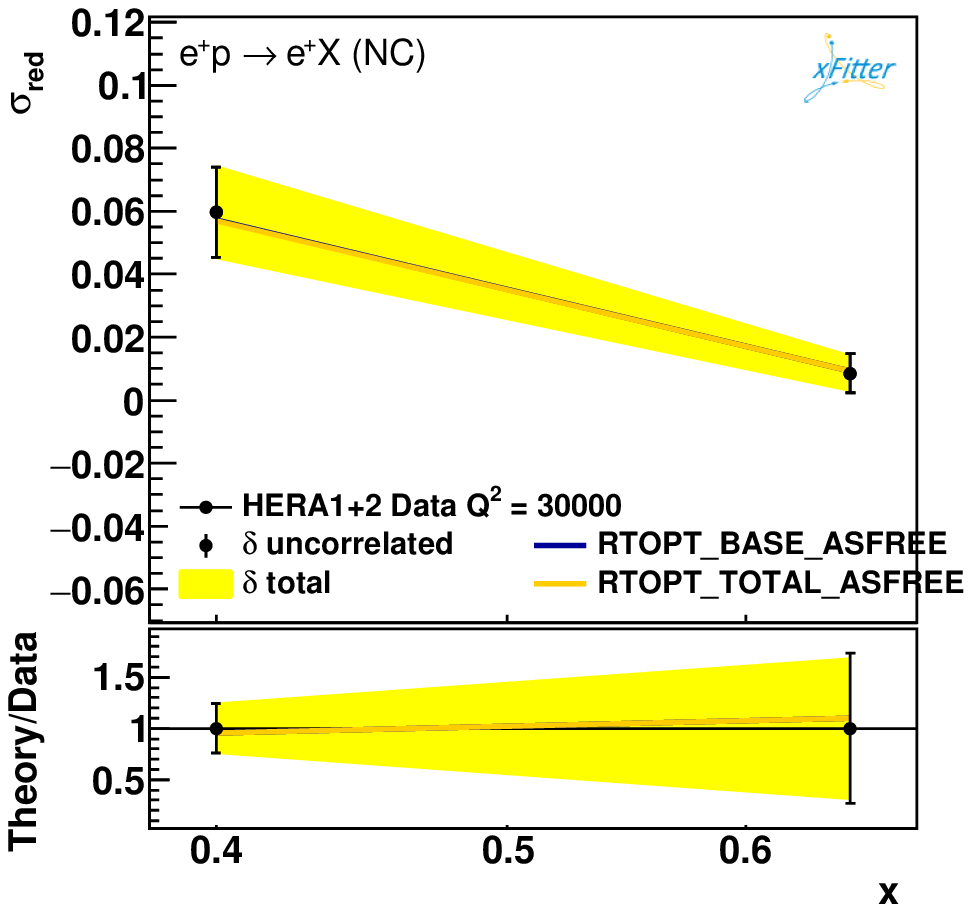}

\caption{Illustrations of the consistency of HERA combined measurements of the reduced DIS  $e^{\pm}p$ data \cite{Abramowicz:2015mha} and the theory predictions as a function of $x$ and for different values of $Q^2$.}
\label{fig:1}
\end{figure*}

 In Fig.~\ref{fig:1},  we illustrate the consistency of HERA measurements of the reduced deep inelastic  $e^{\pm}p$  scattering  cross sections data \cite{Abramowicz:2015mha} and charm production reduced cross section measurements data \cite{Abramowicz:1900rp} with the theory predictions as a function of $x$ and for different values of $Q^2$. According to our QCD analysis, we have good agreement between the theoretical and experimental data. The uncertainties on the cross sections in Fig.~\ref{fig:1} are obtained using Hessian error propagation. The corresponding $\frac{\chi^2_{\rm TOTAL}}{dof}$ values for each of the data sets in Fig.~\ref{fig:1} are listed in Table~\ref{tab:data}. 

\begin{figure*}
\includegraphics[width=0.23\textwidth]{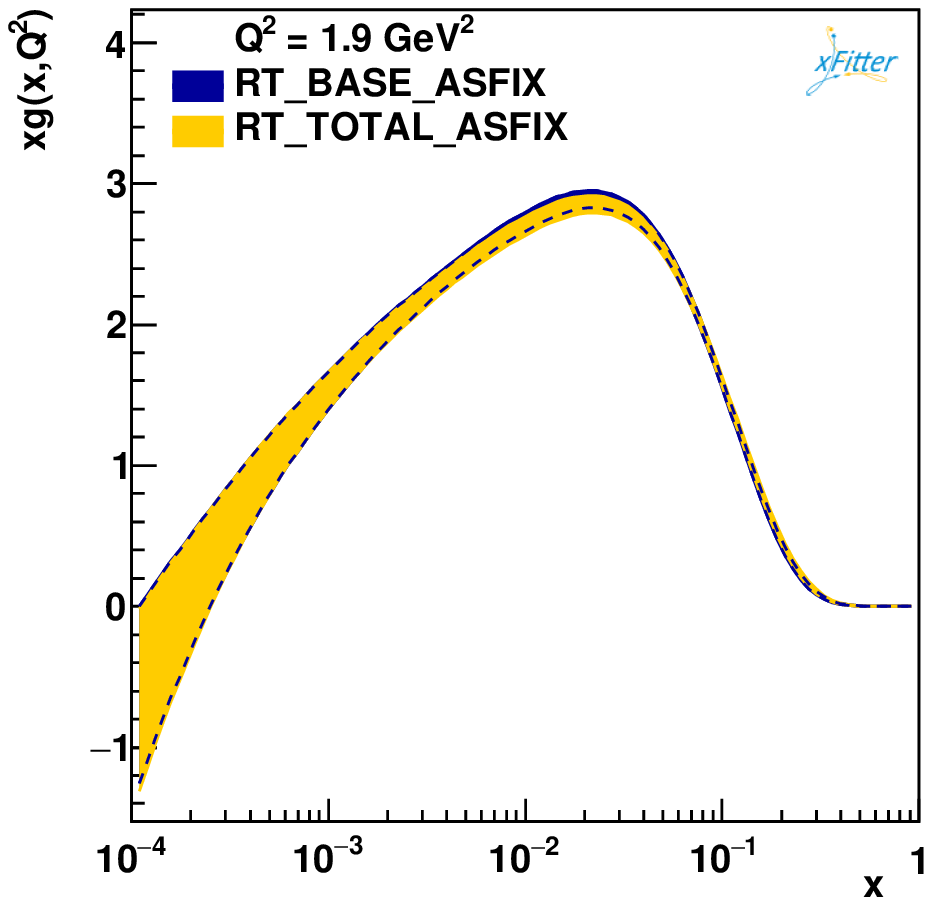}
\includegraphics[width=0.23\textwidth]{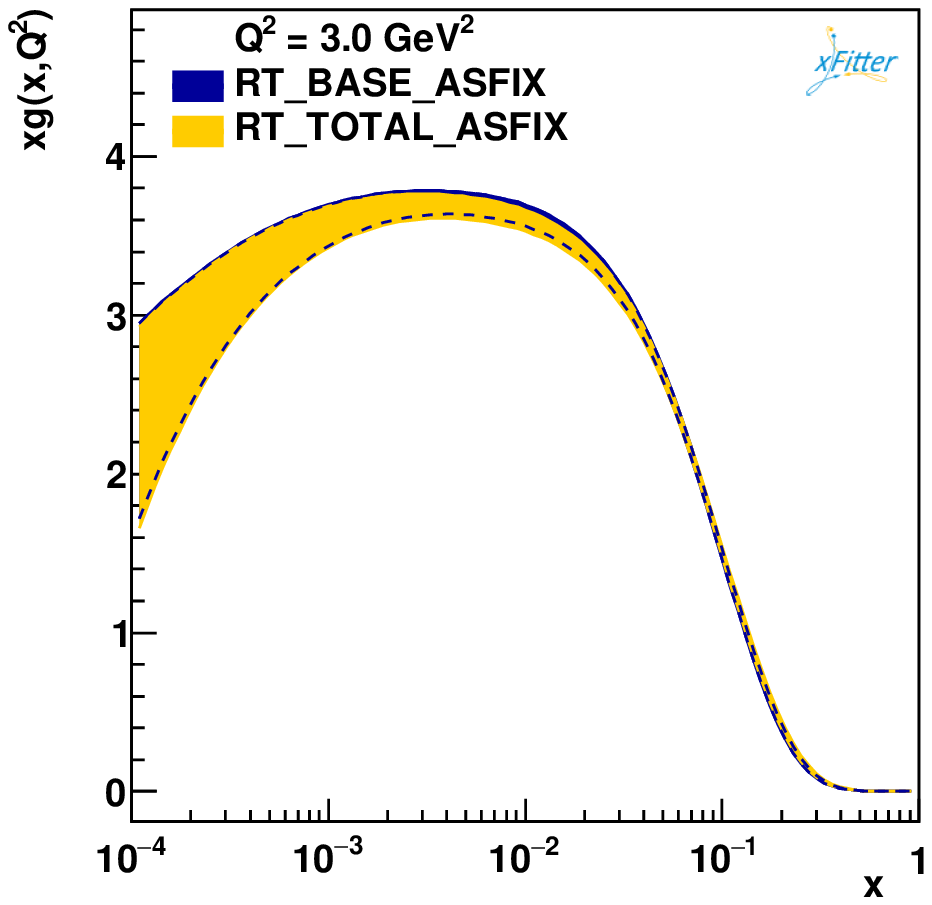}
\includegraphics[width=0.23\textwidth]{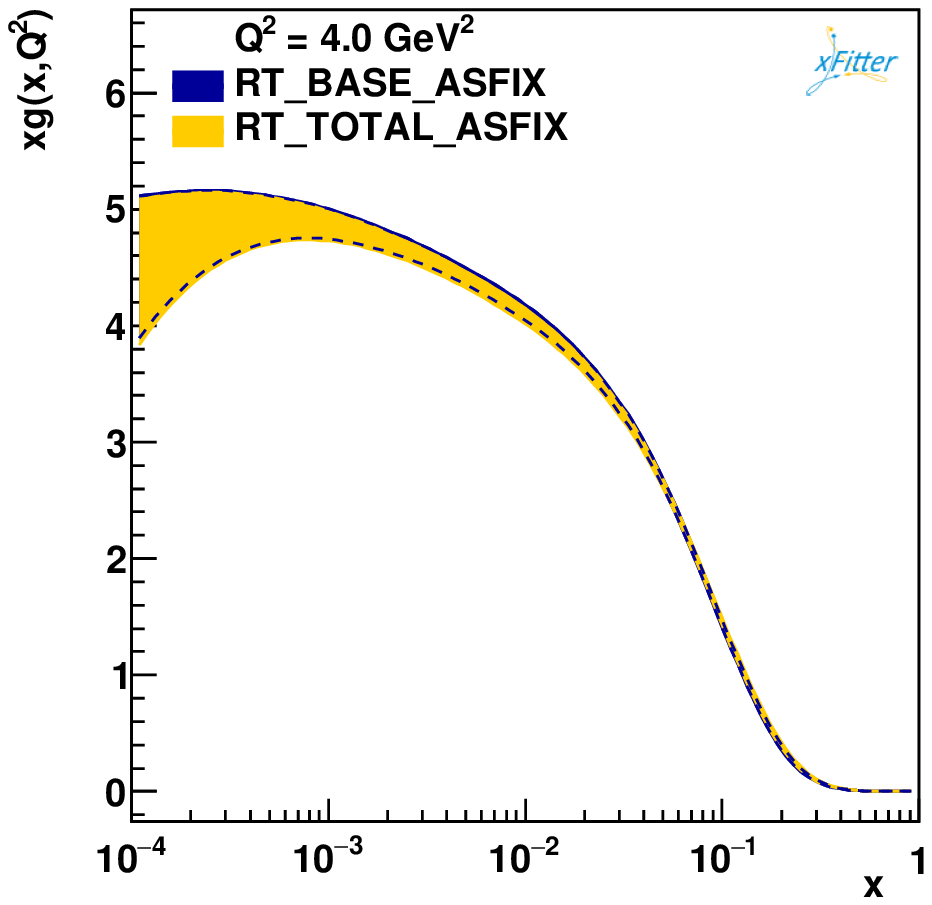}
\includegraphics[width=0.23\textwidth]{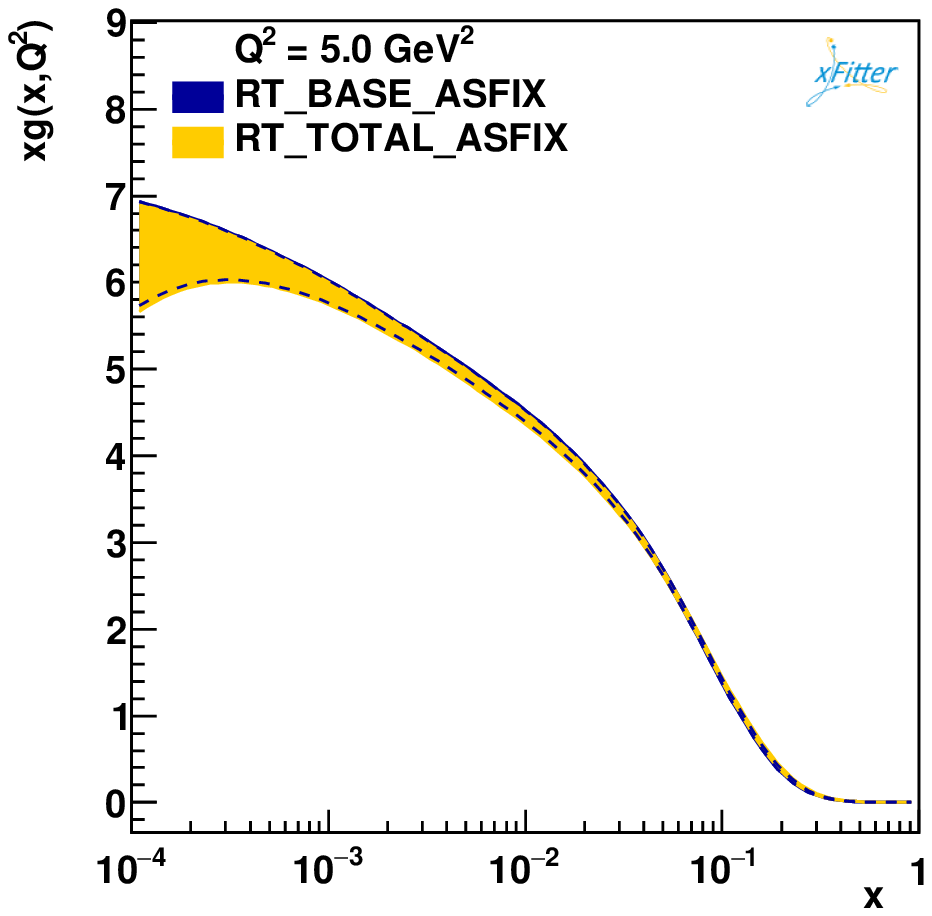}

\includegraphics[width=0.23\textwidth]{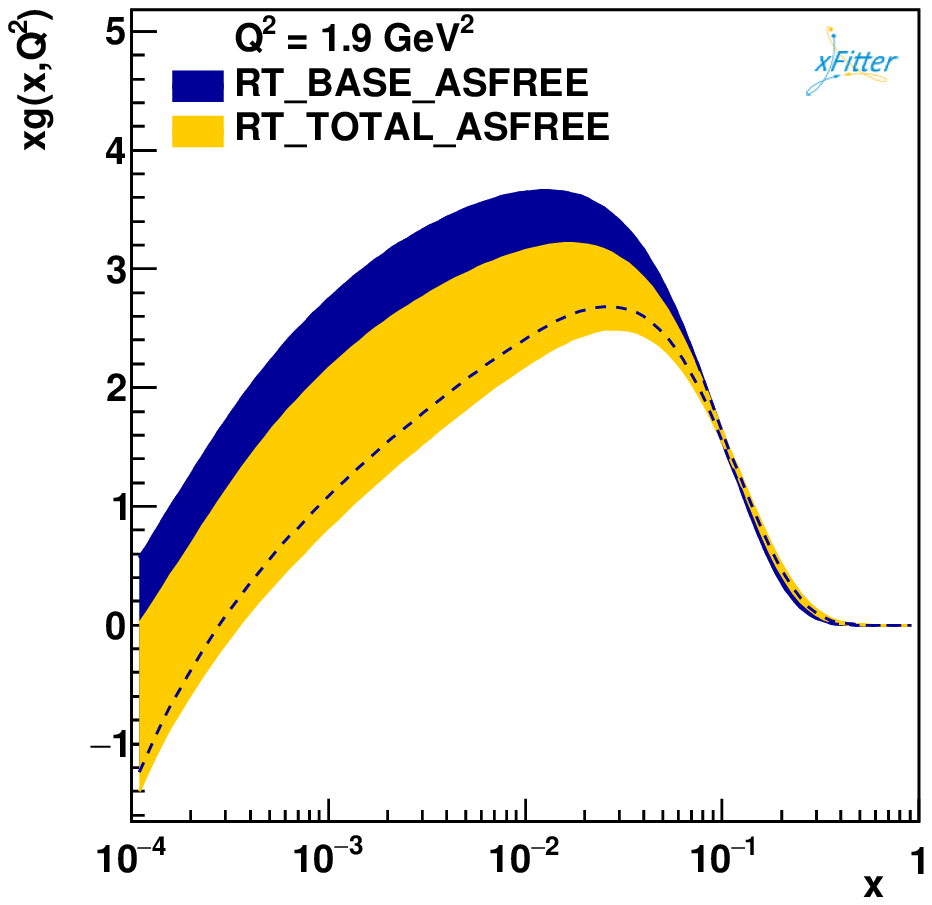}
\includegraphics[width=0.23\textwidth]{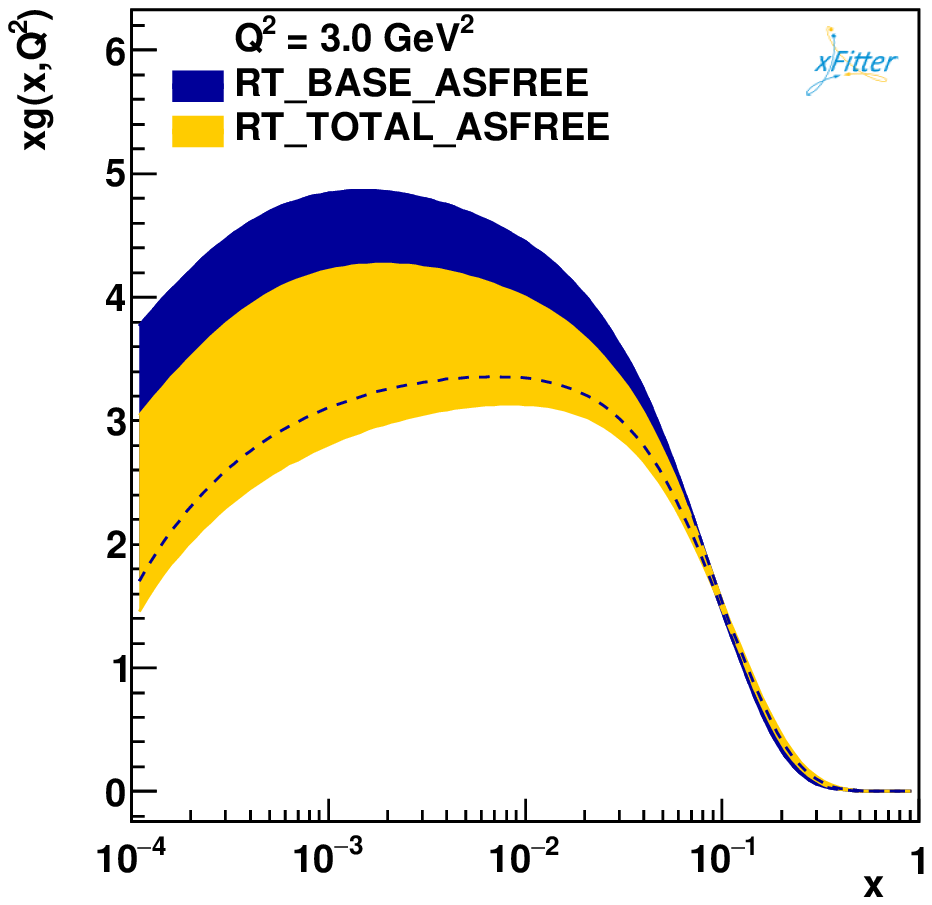}
\includegraphics[width=0.23\textwidth]{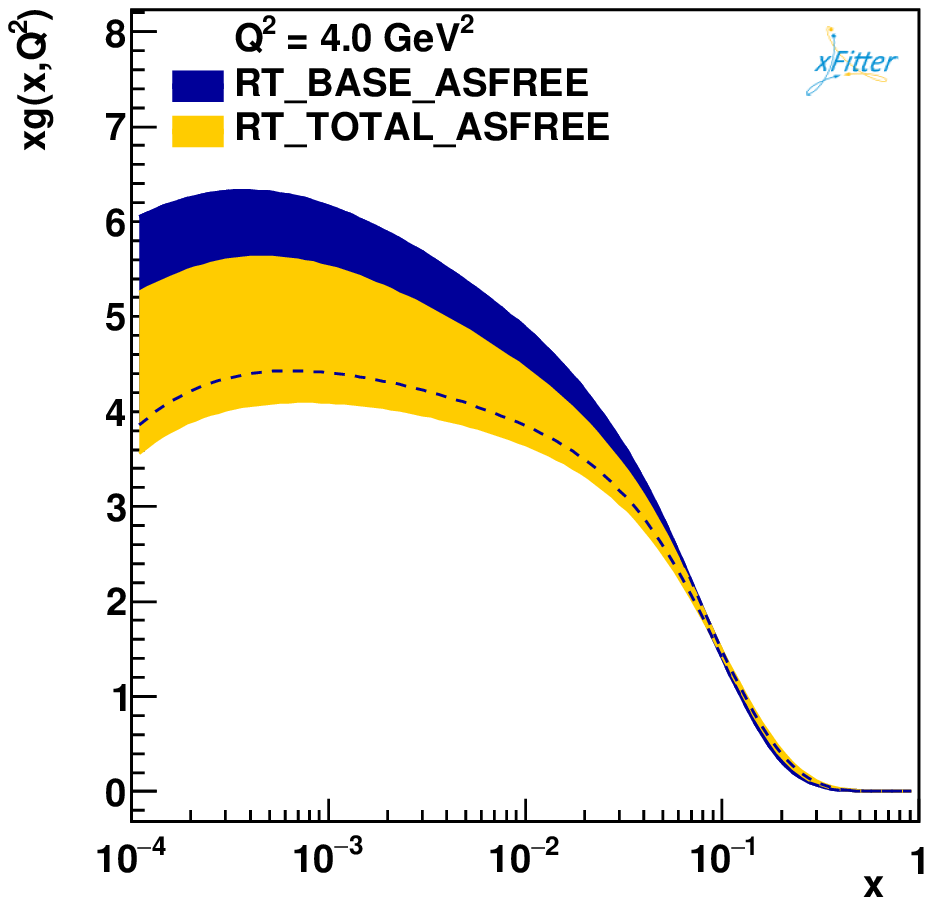}
\includegraphics[width=0.23\textwidth]{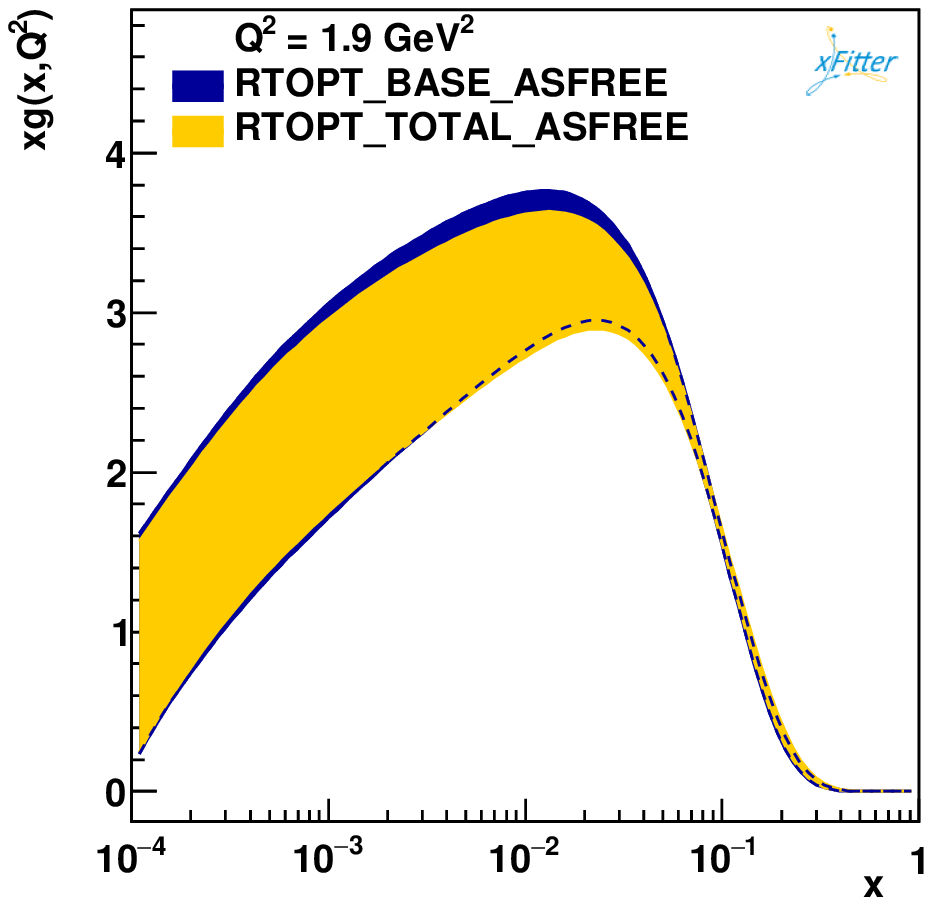}

\caption{The gluon PDFs as extracted for the RT scheme in two separate scenarios. These distributions are plotted at the starting value of $Q_0^2$ = 1.9~GeV$^2$ and $Q^2$ = 3, 4 and 5~GeV$^2$, as a function of $x$. The upper four  diagrams correspond to the first scenario, where the strong coupling, $\alpha_s(M_Z^2)$, is fixed and we find no impact of adding charm H1-ZEUS combined data to the HERA I and II combined data. The lower four diagrams  correspond to the second scenario, where we consider the strong coupling, $\alpha_s(M_Z^2)$, as an extra free parameter, clearly revealing the impact of adding charm H1-ZEUS combined data to the HERA I and II combined data.}
\label{fig:2}
\end{figure*}

\begin{figure*}
\includegraphics[width=0.23\textwidth]{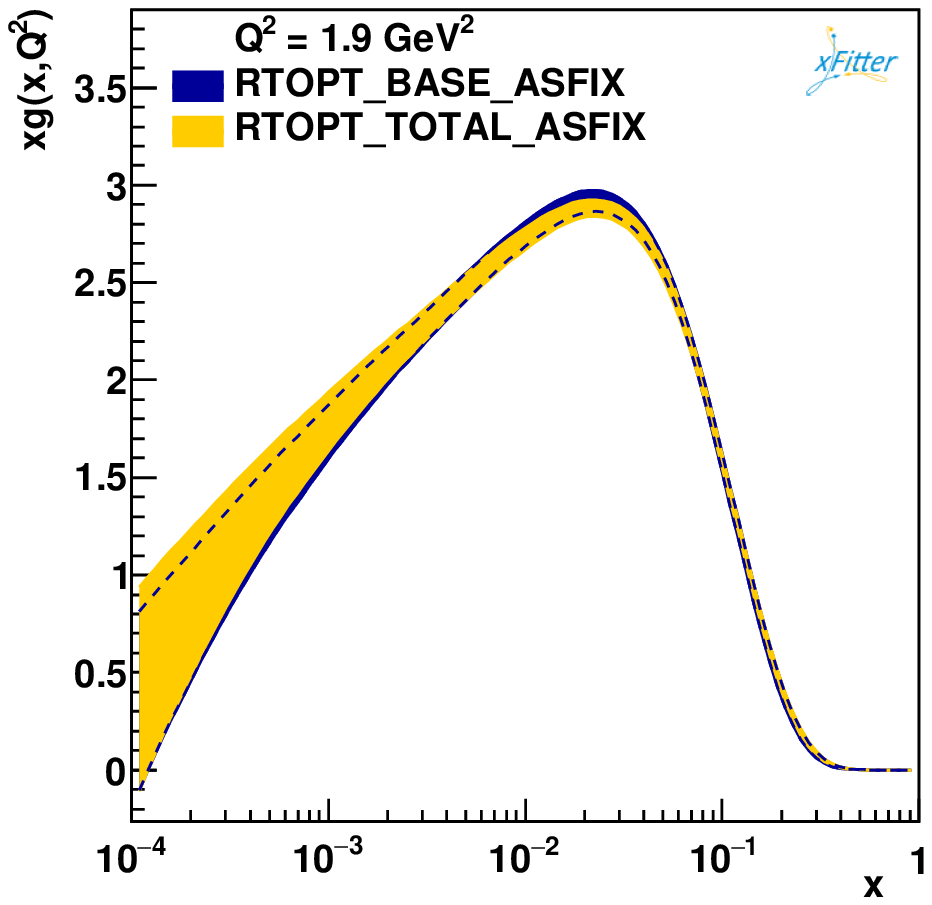}
\includegraphics[width=0.23\textwidth]{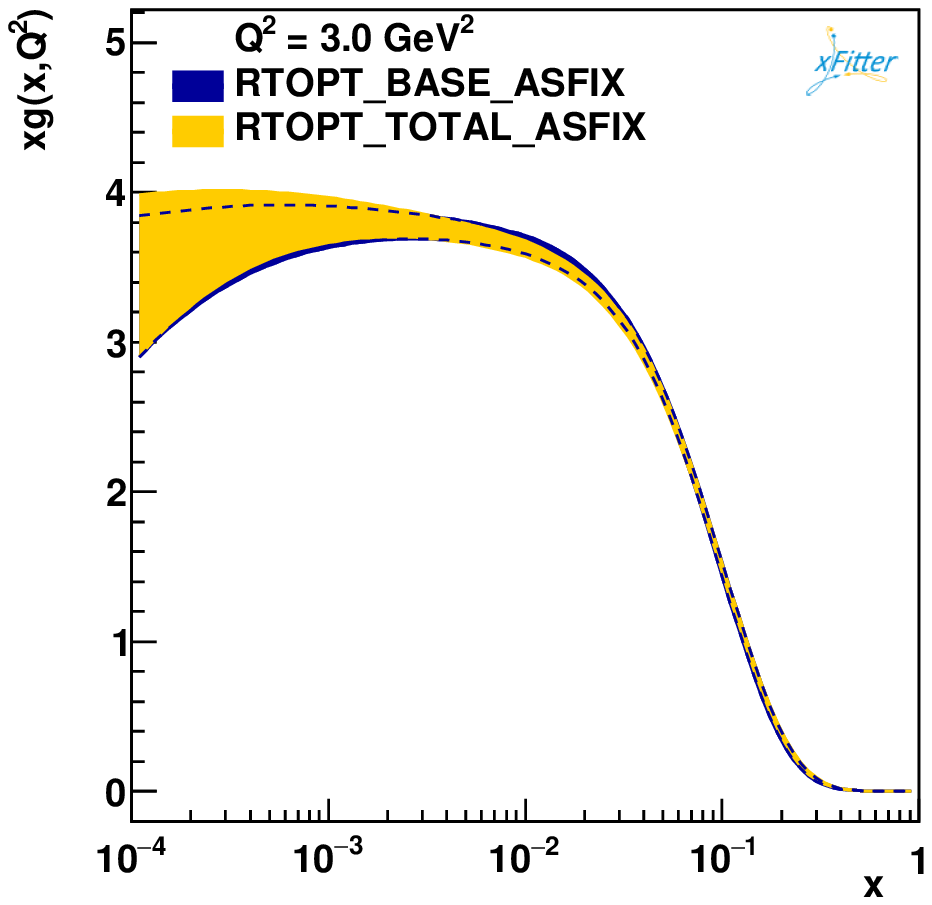}
\includegraphics[width=0.23\textwidth]{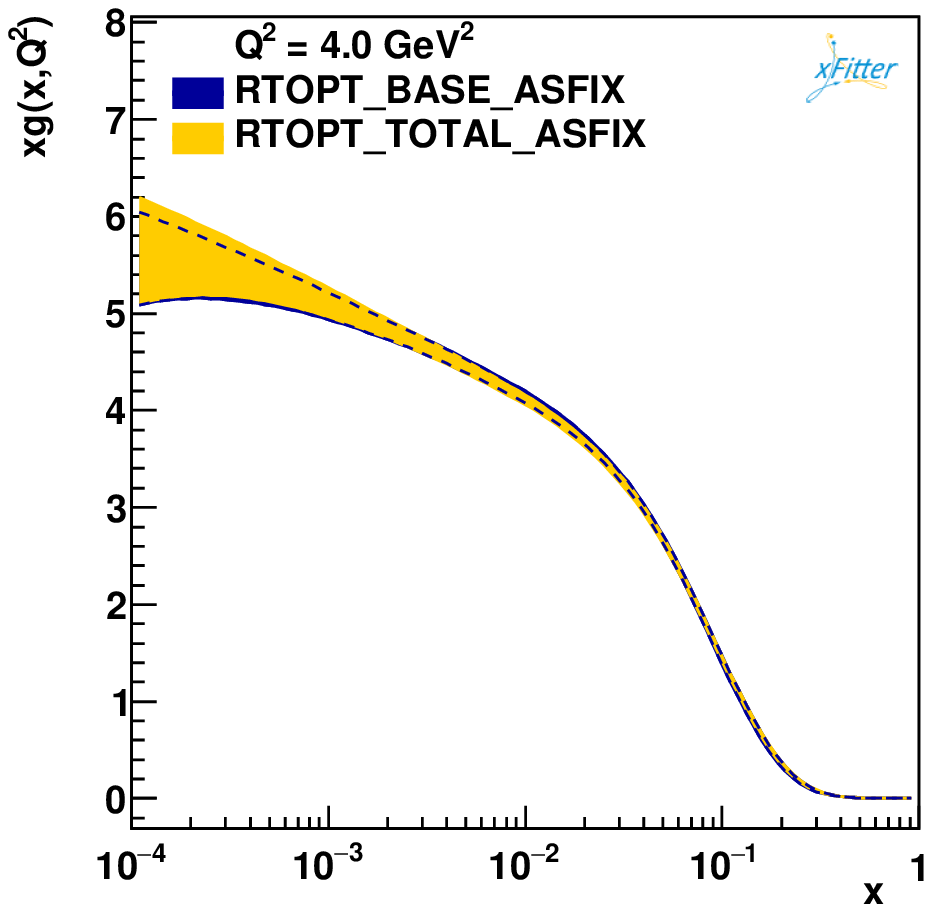}
\includegraphics[width=0.23\textwidth]{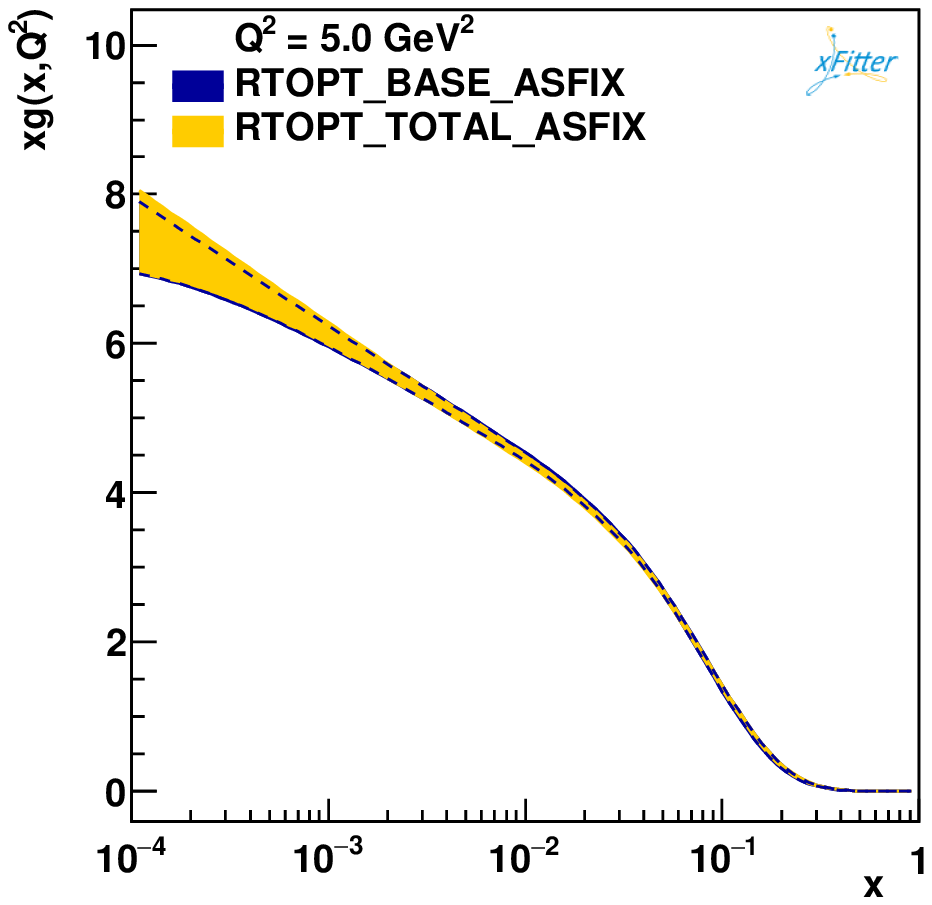}

\includegraphics[width=0.23\textwidth]{./fig/g5}
\includegraphics[width=0.23\textwidth]{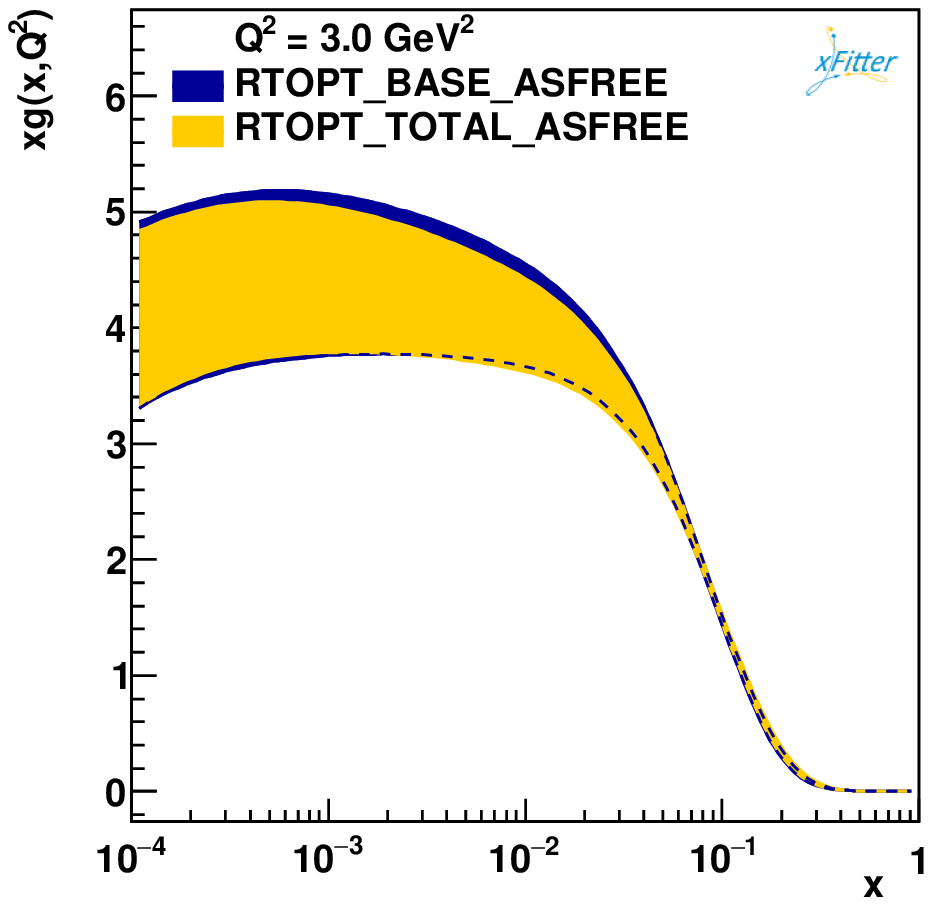}
\includegraphics[width=0.23\textwidth]{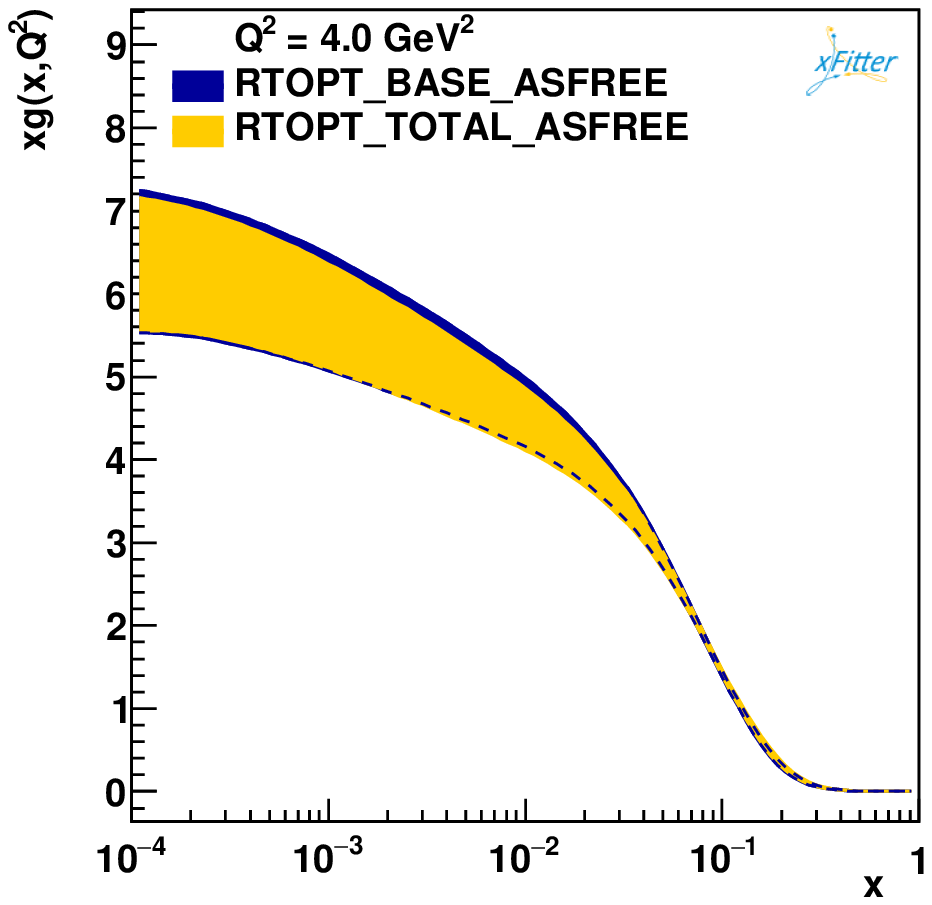}
\includegraphics[width=0.23\textwidth]{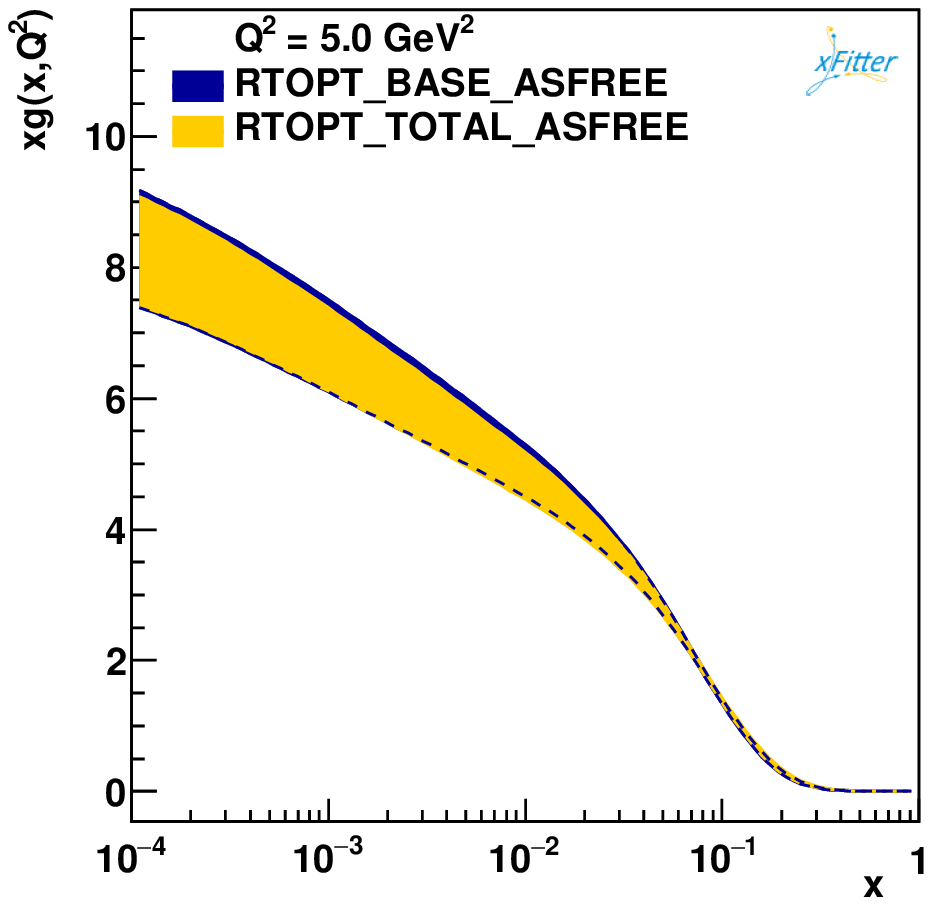}

\caption{The gluon PDFs as extracted for the RTOPT scheme in two separate scenarios, at the starting value of $Q_0^2$ = 1.9~GeV$^2$ and $Q^2$ = 3, 4 and 5~GeV$^2$, as a function of $x$. The impact of adding charm data can be seen only in the four lower diagrams, where the strong coupling, $\alpha_s(M_Z^2)$, is considered as an extra free parameter.}
\label{fig:3}
\end{figure*}

 The impact of charm cross section H1-ZEUS combined measurements data on HERA I and II combined data for gluon distribution functions are shown in Figs. \ref{fig:2} and \ref{fig:3}, at the starting value of $Q_0^2$ = 1.9~GeV$^2$ and $Q^2$ = 3, 4 and 5~GeV$^2$, in the RT and RTOPT schemes and for two separate scenarios. Clearly, in the first scenario, where the strong coupling $\alpha_s(M_Z^2)$ is fixed, we find no impact from adding charm H1-ZEUS combined data to the HERA I and II combined data. In the second scenario, however, where we consider the strong coupling $\alpha_s(M_Z^2)$ as an extra free parameter, we clearly find the impact of adding charm H1-ZEUS combined data to the HERA I and II combined data.
  
\begin{figure*}
\includegraphics[width=0.23\textwidth]{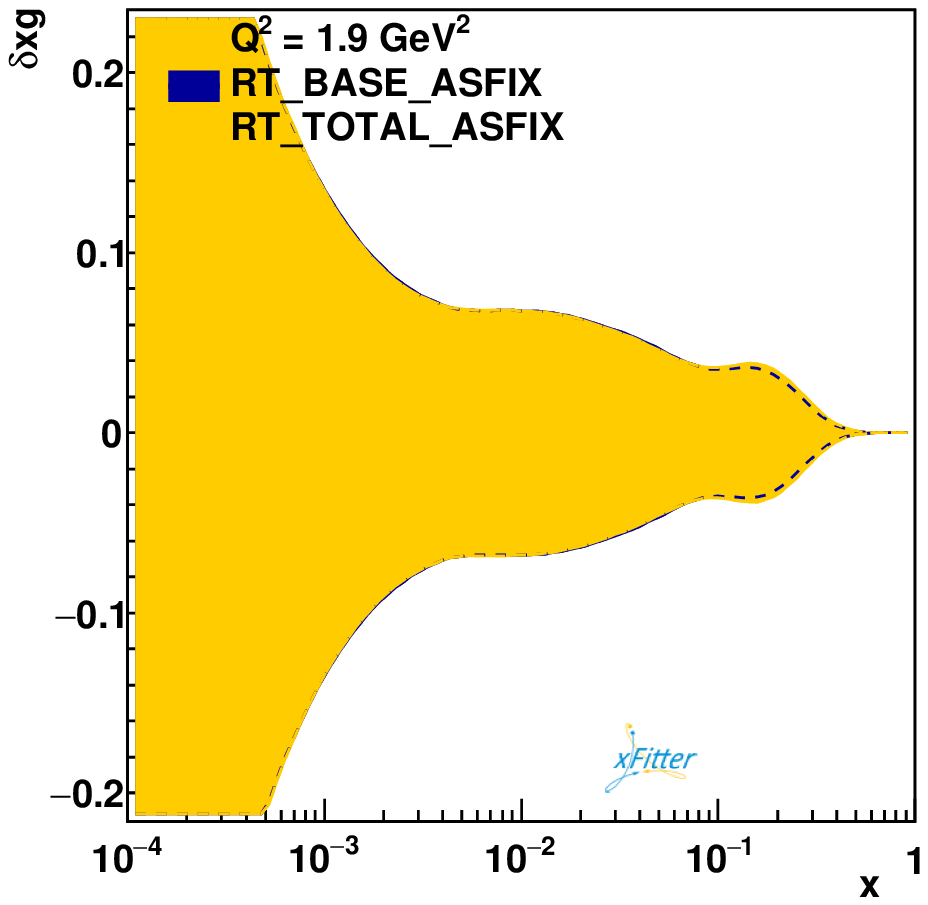}
\includegraphics[width=0.23\textwidth]{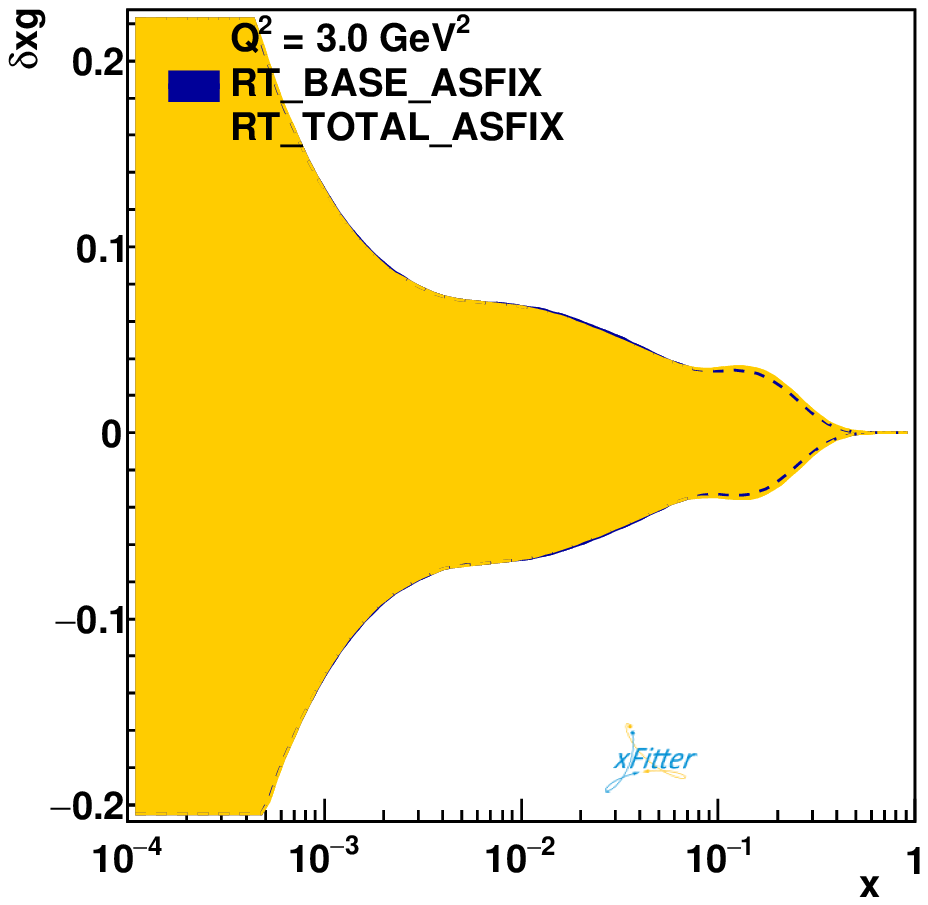}
\includegraphics[width=0.23\textwidth]{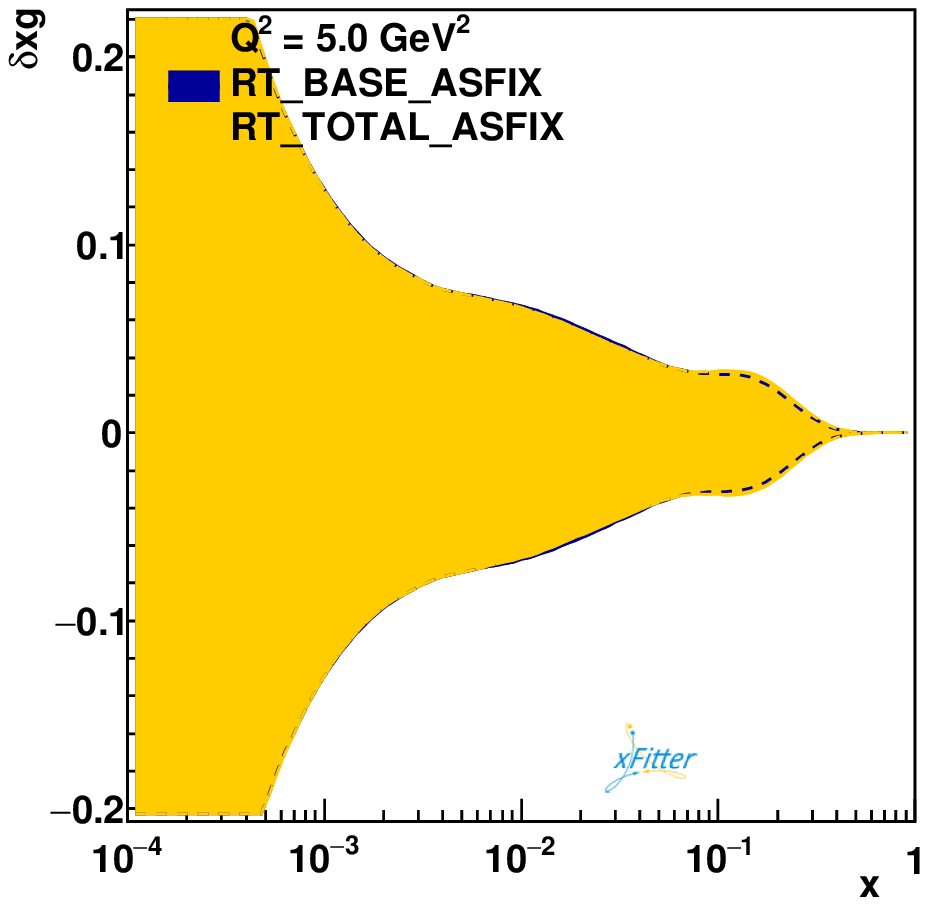}
\includegraphics[width=0.23\textwidth]{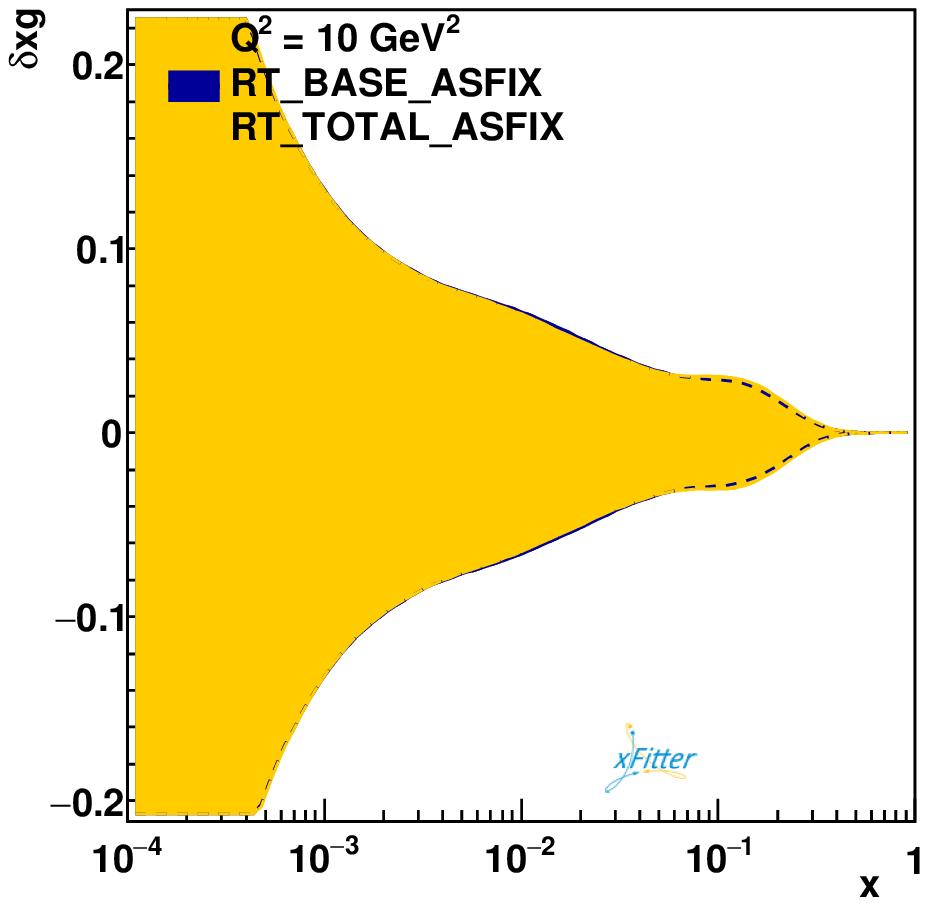}

\includegraphics[width=0.23\textwidth]{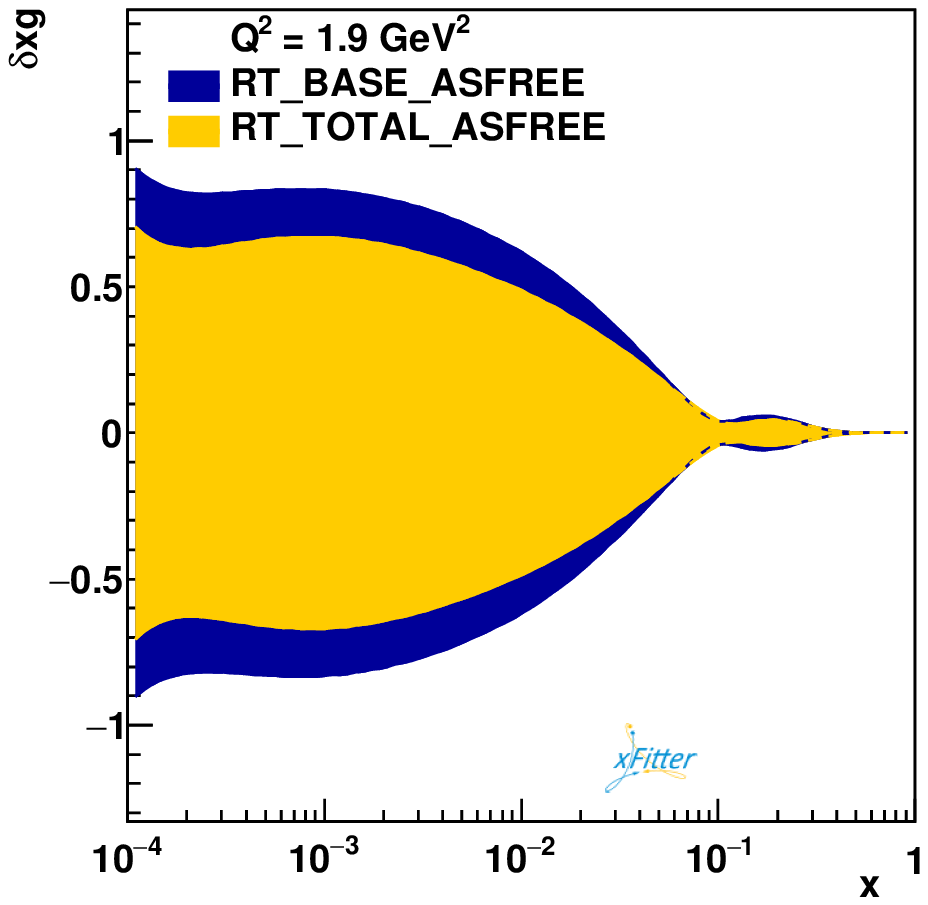}
\includegraphics[width=0.23\textwidth]{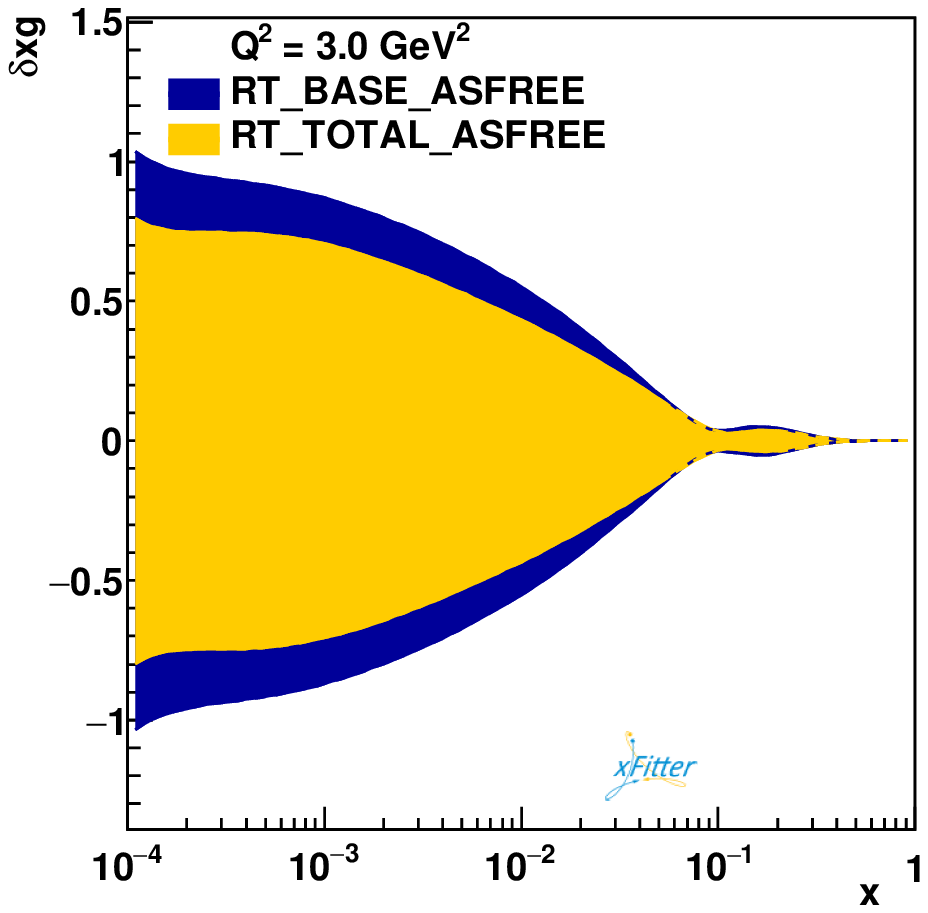}
\includegraphics[width=0.23\textwidth]{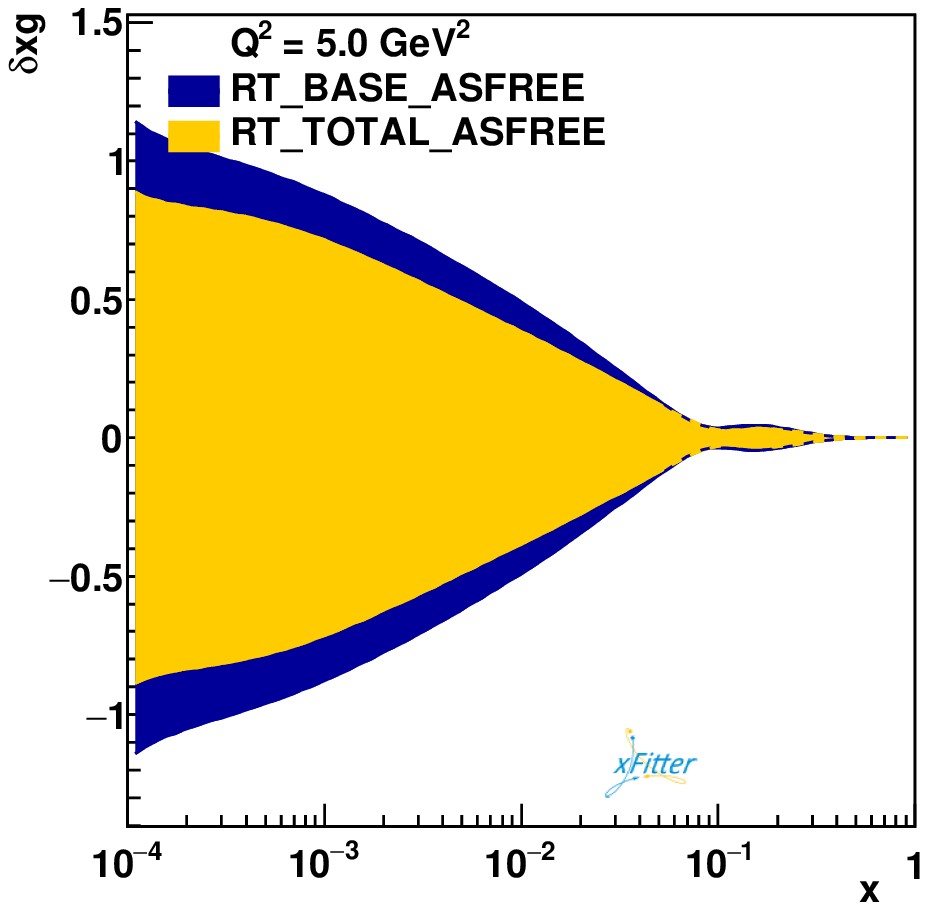}
\includegraphics[width=0.23\textwidth]{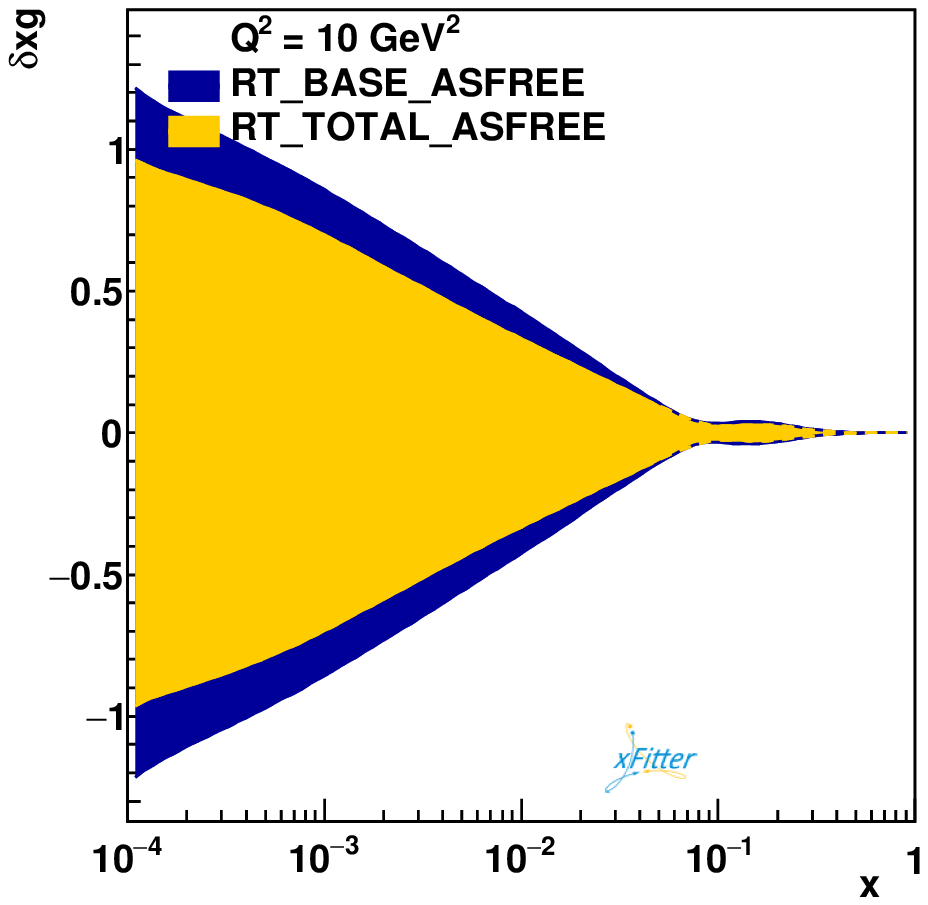}

\caption{The partial gluon PDFs as extracted for the RT scheme in two separate scenarios. These PDFs are plotted at the starting value of $Q_0^2$ = 1.9~GeV$^2$ and $Q^2$ = 3, 5 and 10~GeV$^2$, as a function of $x$. The upper four diagrams are based on a fixed strong coupling, and do not show the impact of adding charm flavour. The lower four diagrams, based on the second scenario, where $\alpha_s(M_Z^2)$ is considered as an extra free parameter, clearly show this impact. }
\label{fig:4}
\end{figure*}

\begin{figure*}
\includegraphics[width=0.23\textwidth]{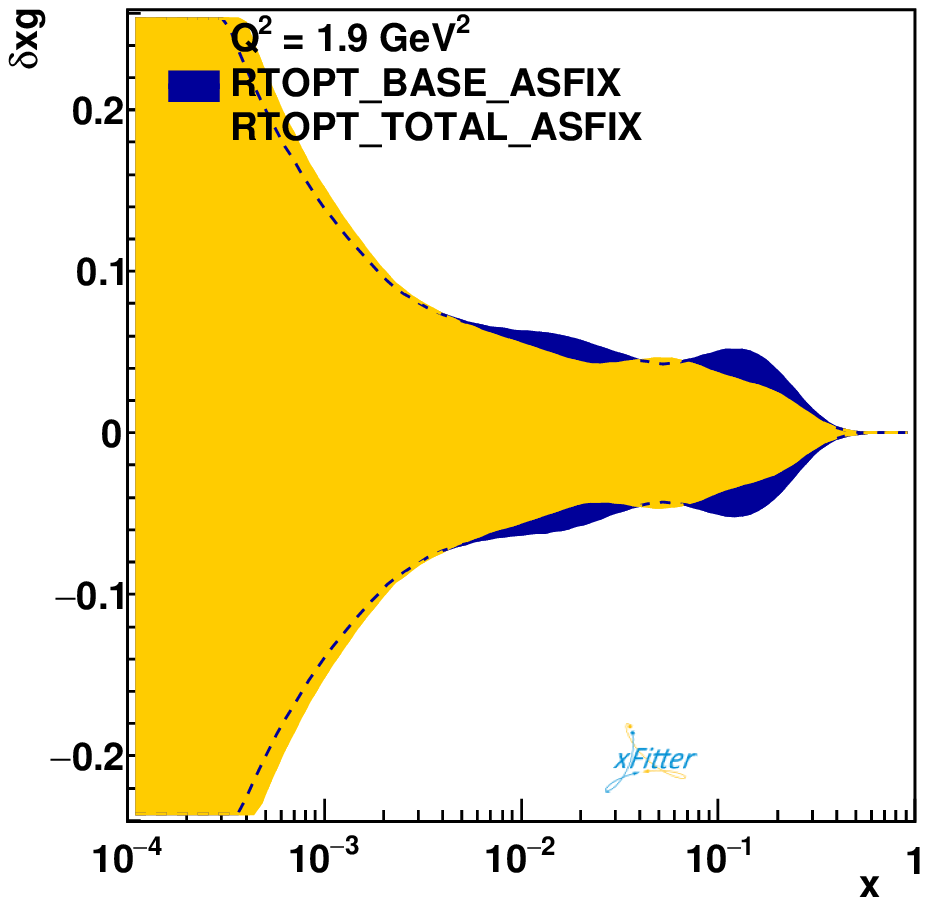}
\includegraphics[width=0.23\textwidth]{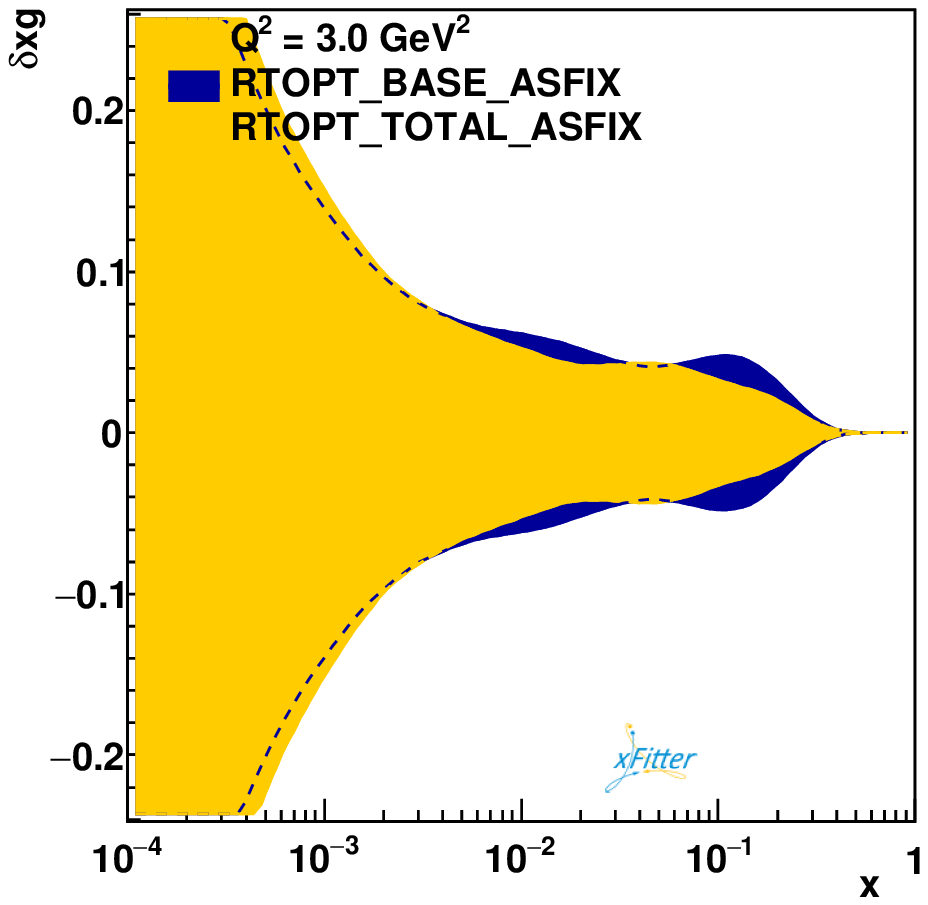}
\includegraphics[width=0.23\textwidth]{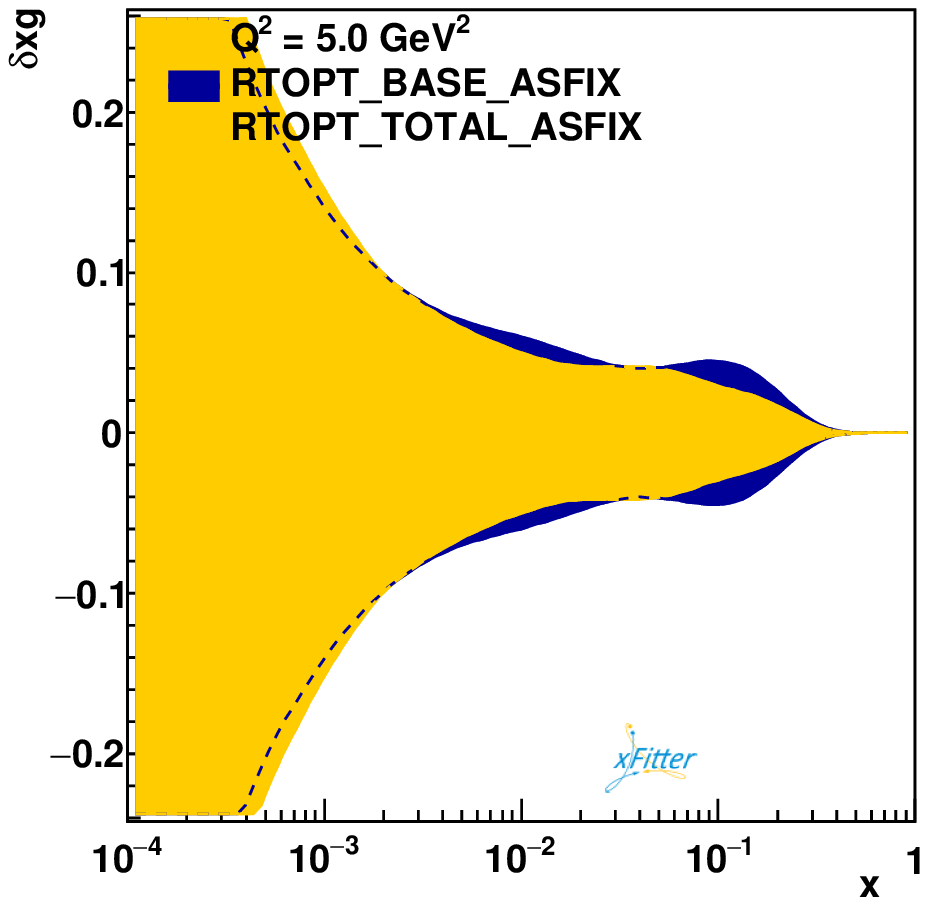}
\includegraphics[width=0.23\textwidth]{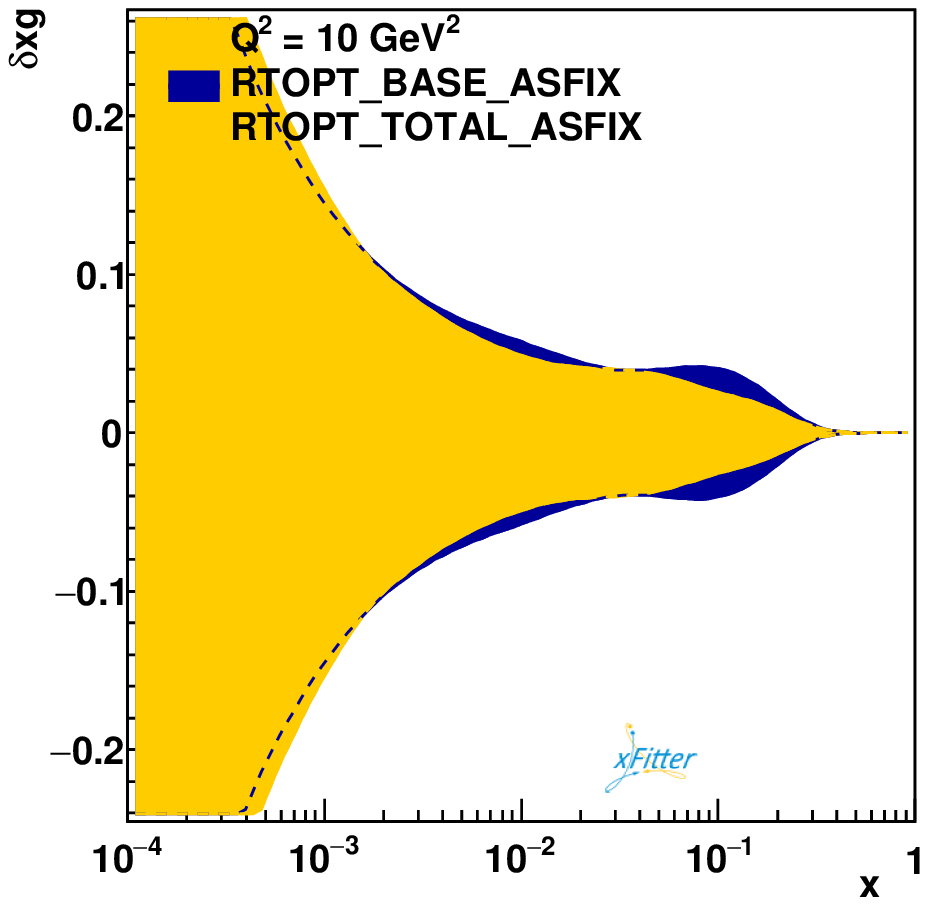}

\includegraphics[width=0.23\textwidth]{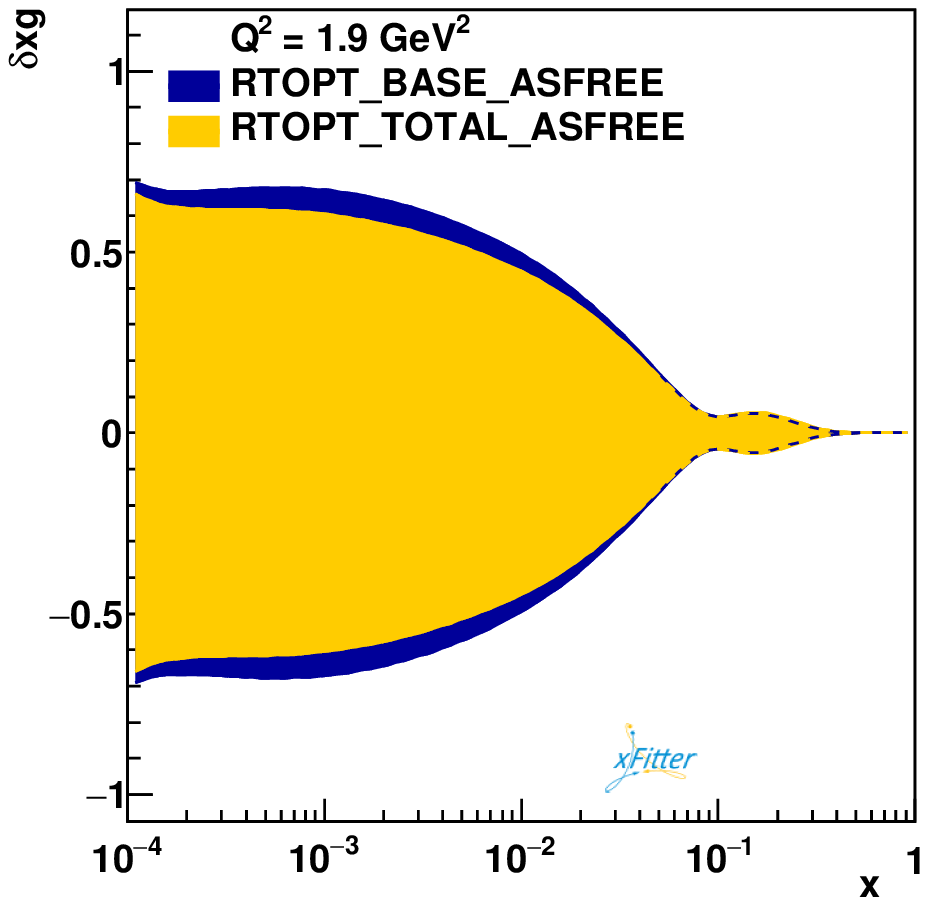}
\includegraphics[width=0.23\textwidth]{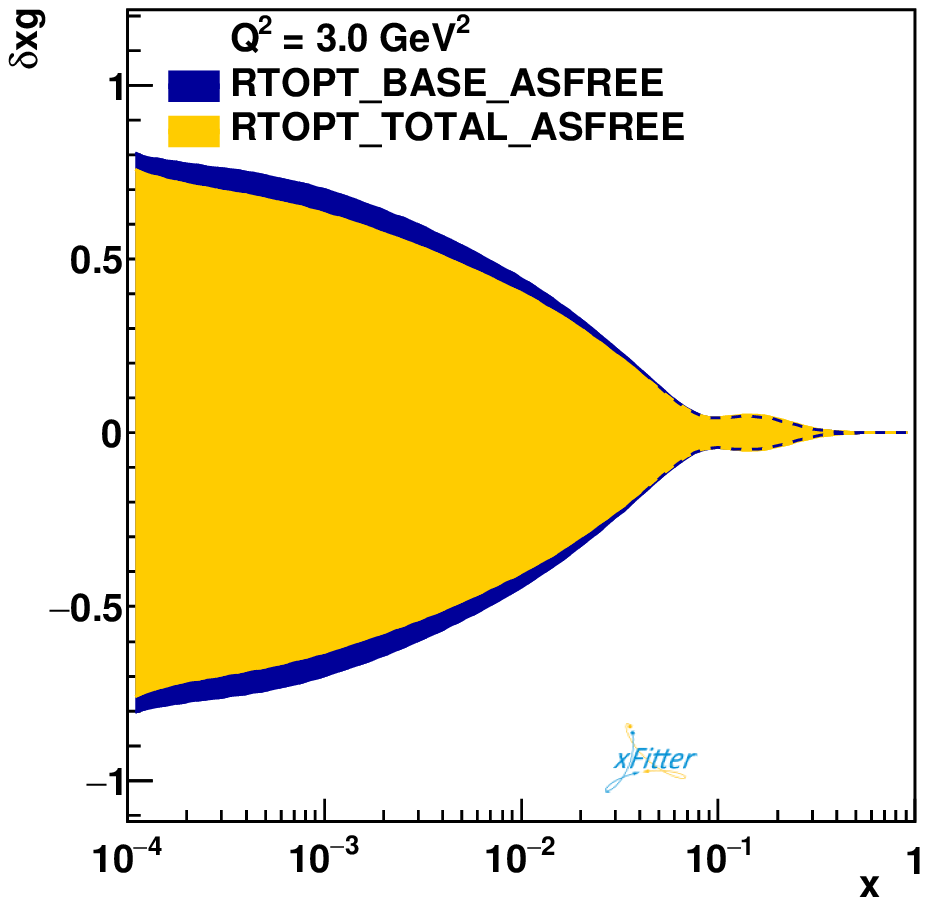}
\includegraphics[width=0.23\textwidth]{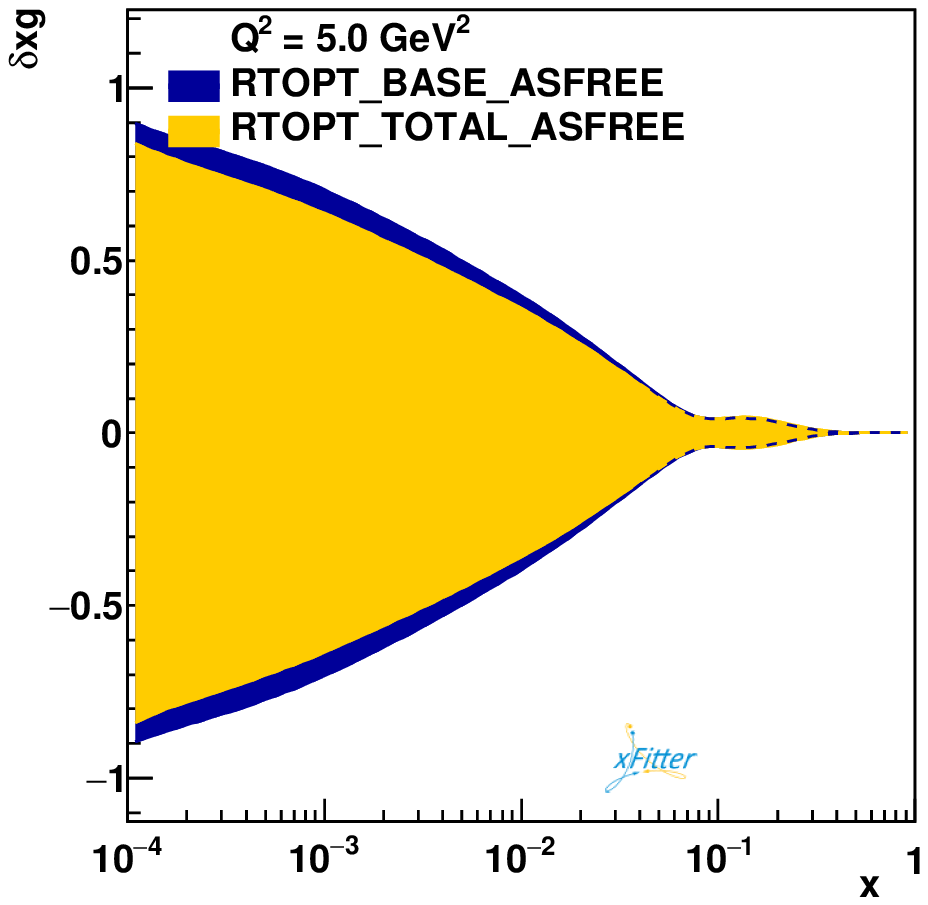}
\includegraphics[width=0.23\textwidth]{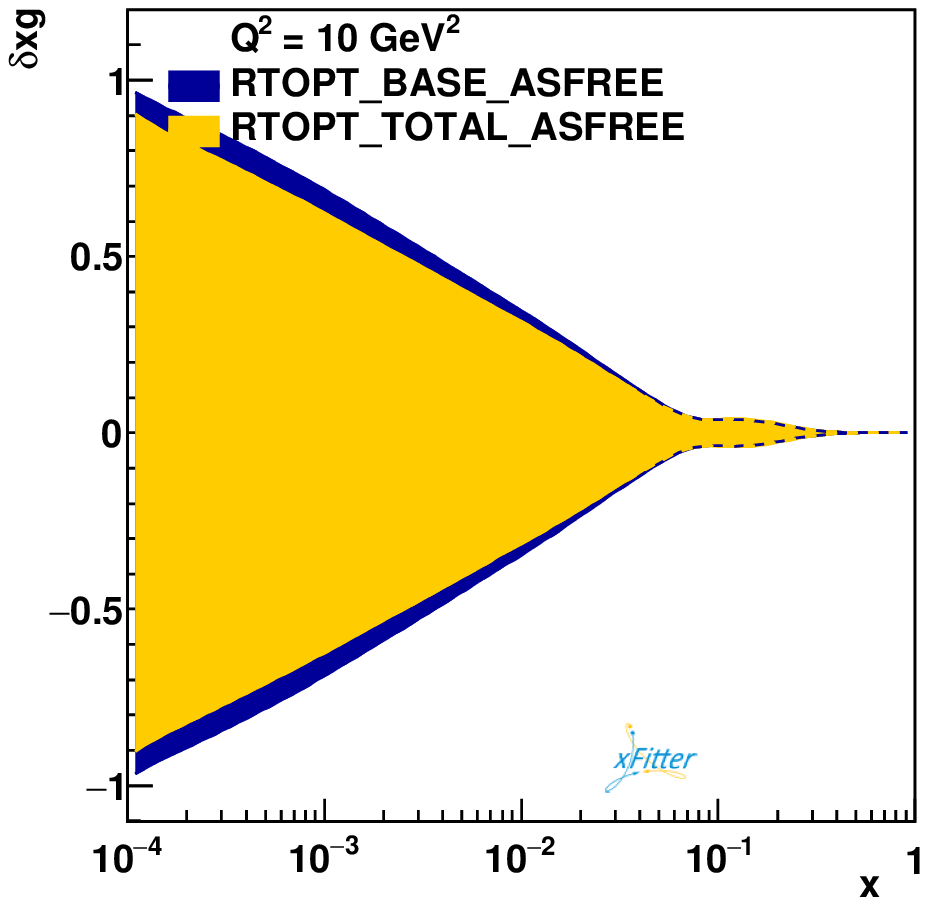}

\caption{The partial of gluon PDFs as extracted for the RTOPT scheme, at the starting value of $Q_0^2$ = 1.9~GeV$^2$ and $Q^2$ = 3, 5 and 10~GeV$^2$, as a function of $x$. The impact of adding charm cross section H1-ZEUS combined  data can bee seen in the four lower diagrams, which have been plotted based on the second scenario.}
\label{fig:5}
\end{figure*}

 The partial gluon distribution functions are shown in Figs. \ref{fig:4} and \ref{fig:5}, at $Q^2$ = 1.9, 3, 5 and 10~GeV$^2$ in the RT and RTOPT schemes and for two separate scenarios. The impact of adding charm H1-ZEUS combined data to the HERA I and II combined data can be seen only in the second scenario, where the strong coupling, $\alpha_s(M_Z^2)$, is considered as an extra free parameter.

\begin{figure*}
\includegraphics[width=0.23\textwidth]{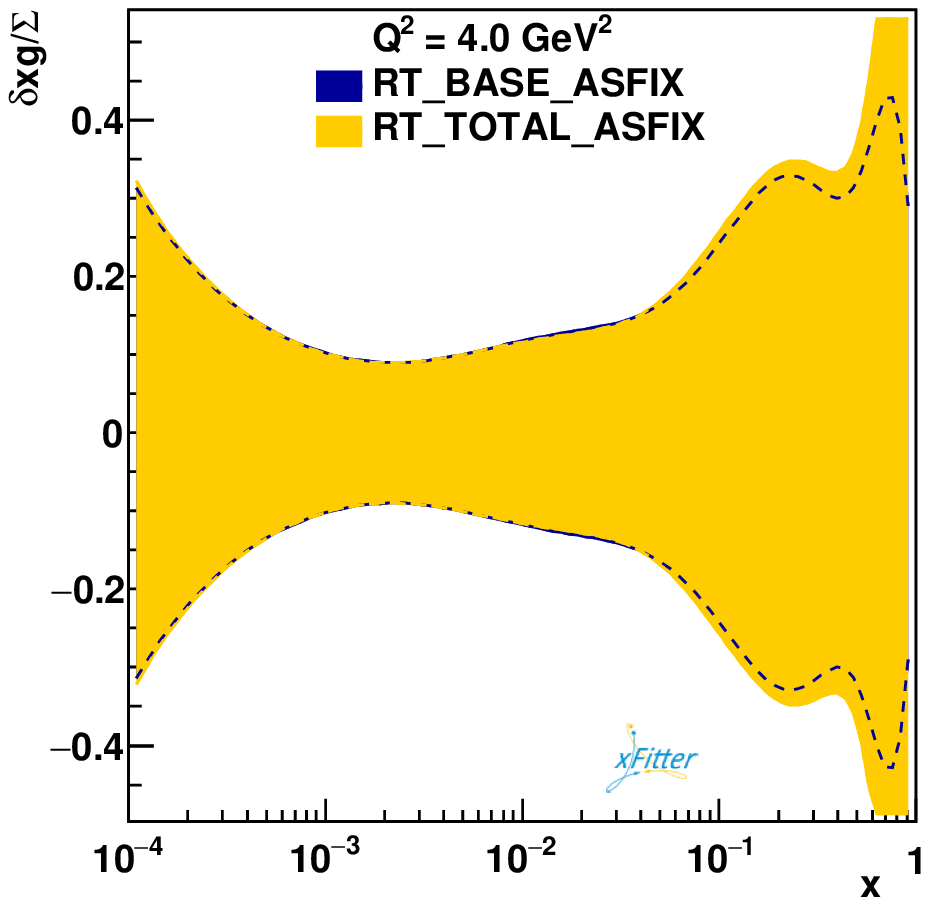}
\includegraphics[width=0.23\textwidth]{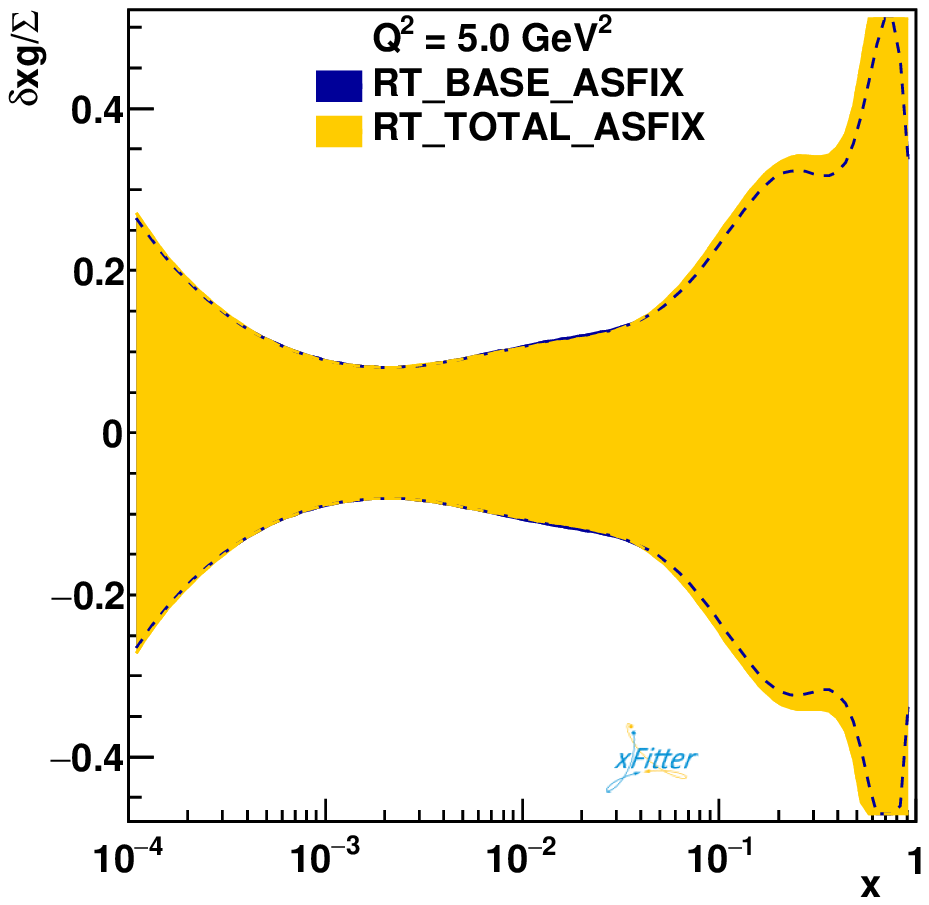}
\includegraphics[width=0.23\textwidth]{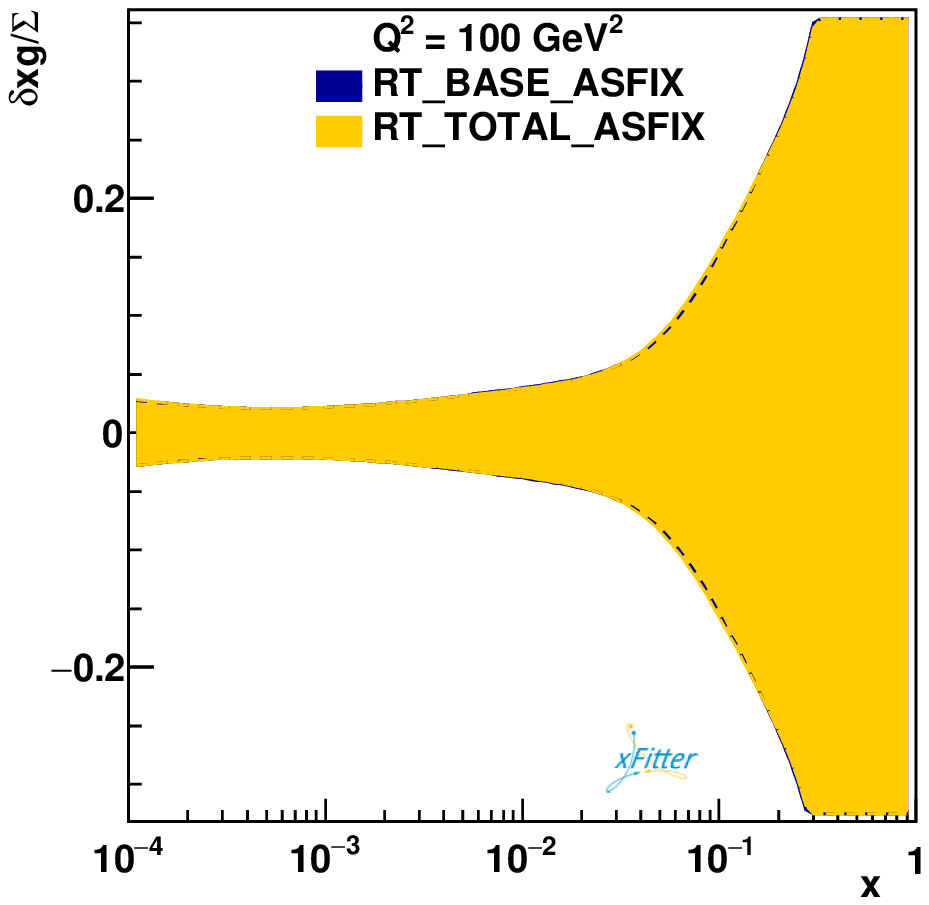}
\includegraphics[width=0.23\textwidth]{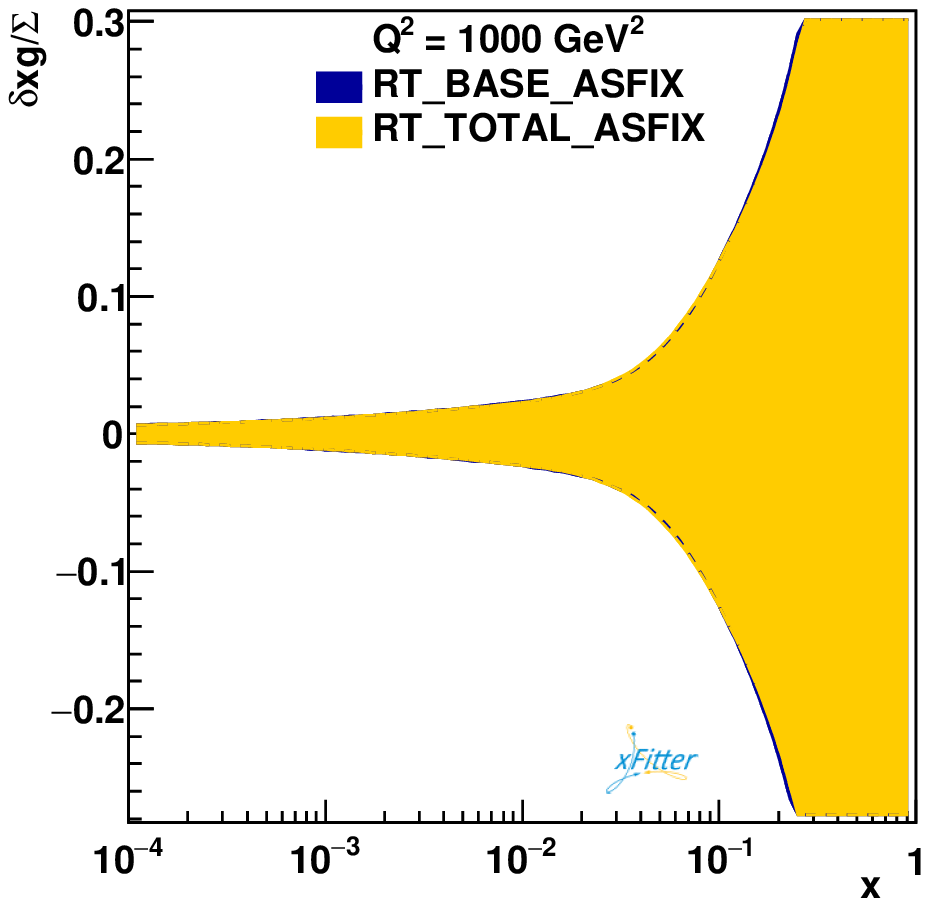}

\includegraphics[width=0.23\textwidth]{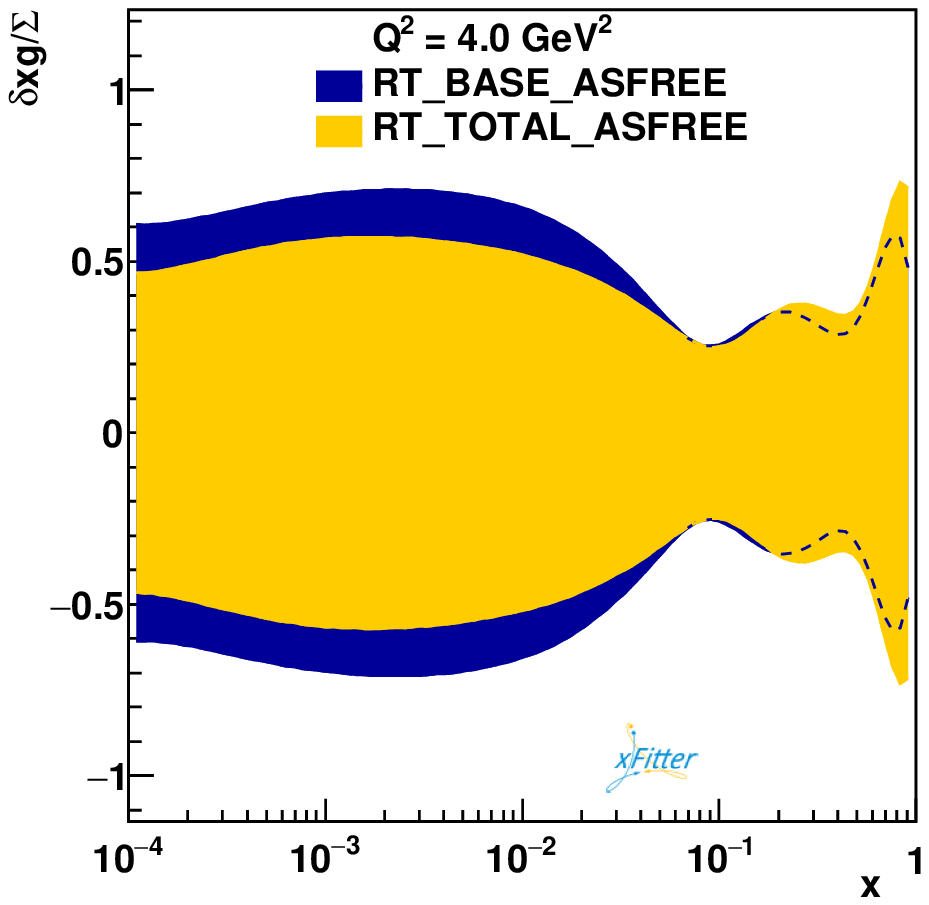}
\includegraphics[width=0.23\textwidth]{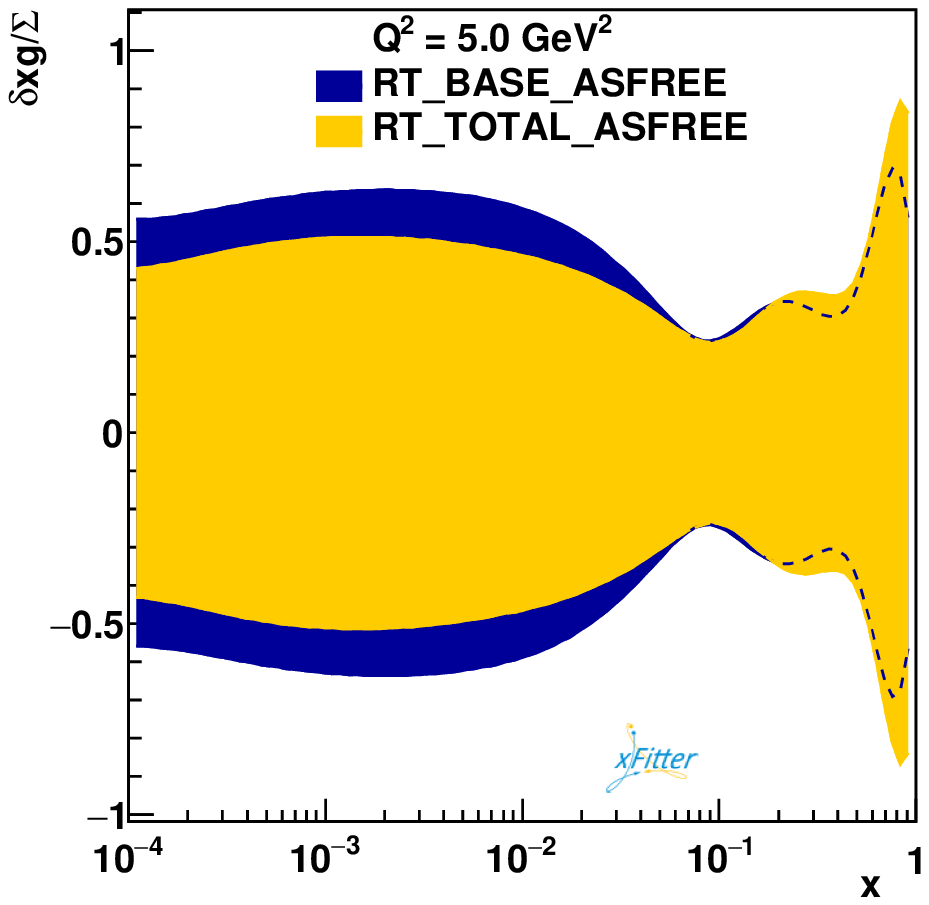}
\includegraphics[width=0.23\textwidth]{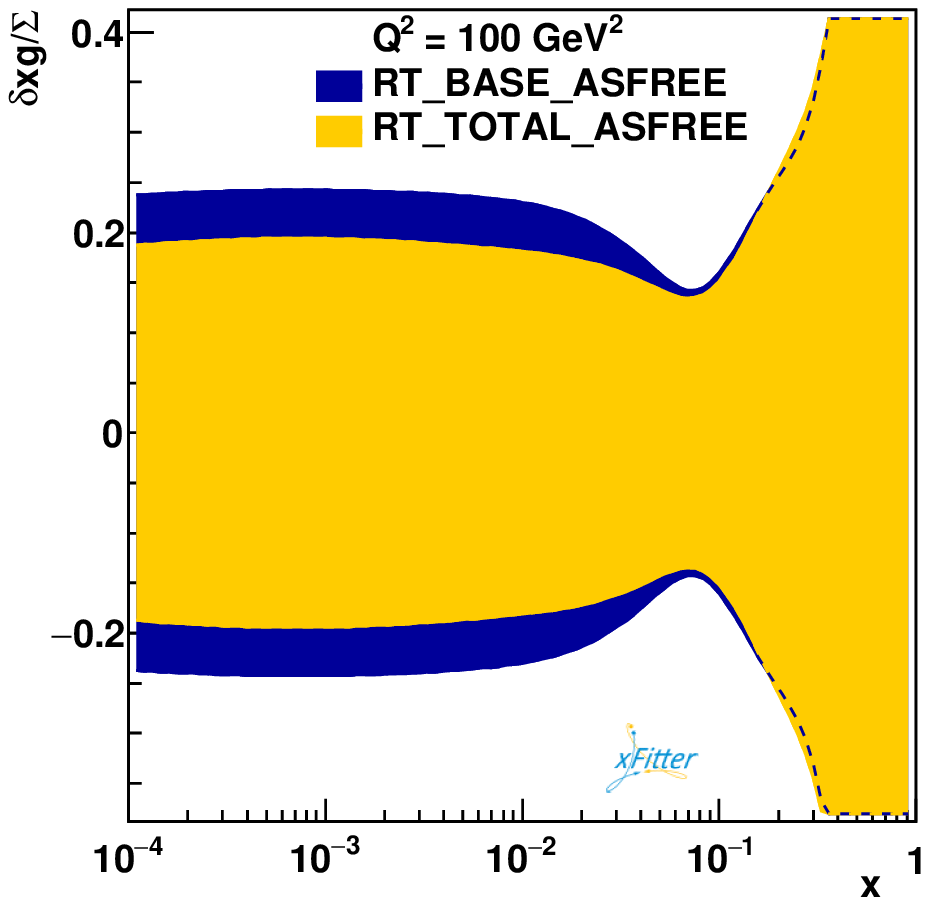}
\includegraphics[width=0.23\textwidth]{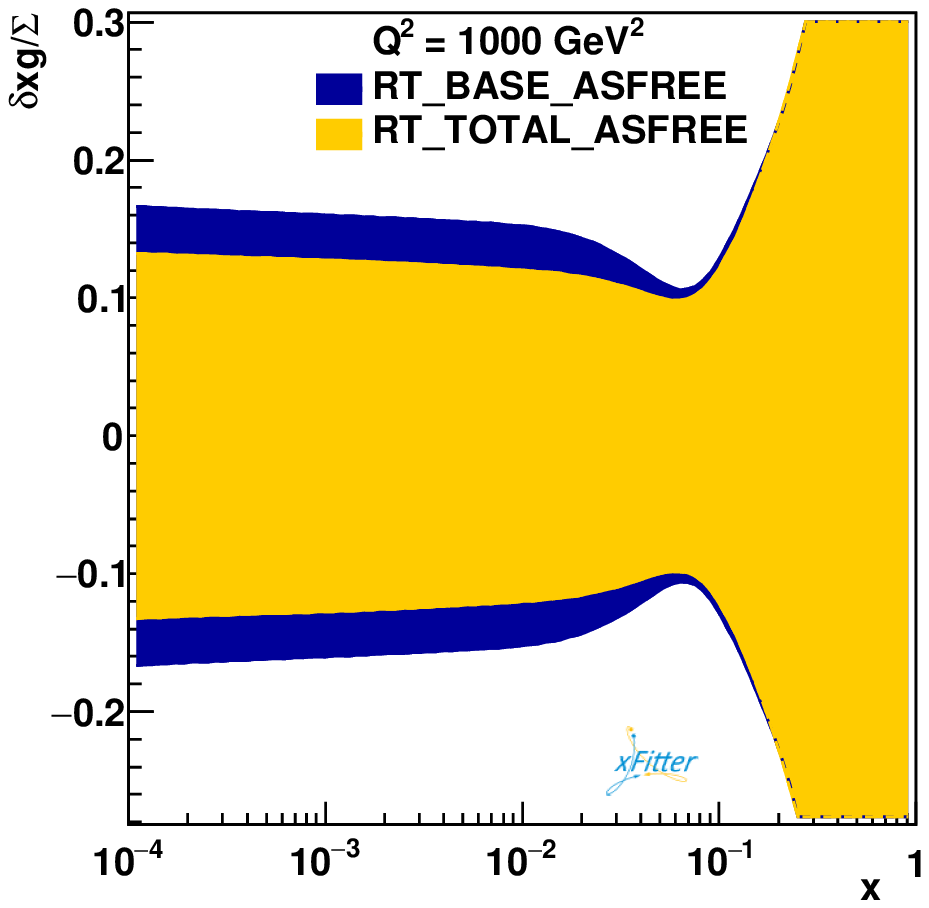}

\caption{The partial ratio of gluon distributions over $\Sigma$-PDFs, as extracted for the RT scheme in two separate scenarios, at $Q^2$ = 4, 5, 100 and 1000~GeV$^2$, as a function of $x$. Only the four lower diagrams, corresponding to second scenario, show the impact of charm flavour on the PDFs.}
\label{fig:6}
\end{figure*}

\begin{figure*}
\includegraphics[width=0.23\textwidth]{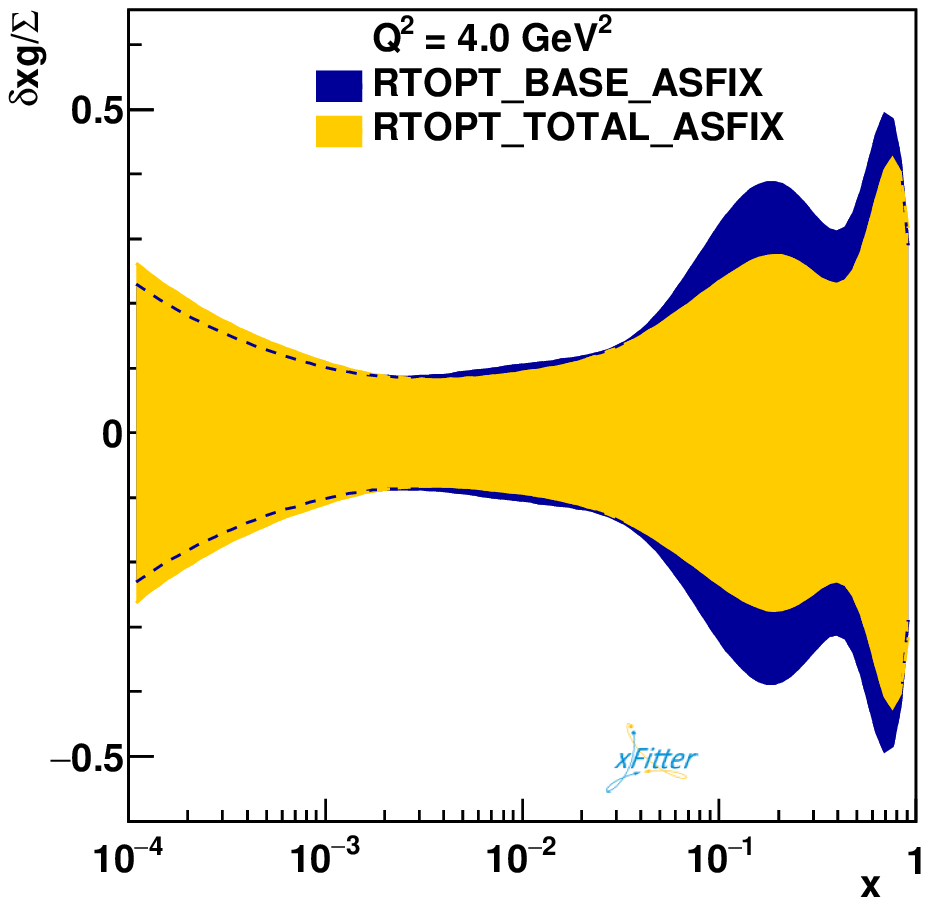}
\includegraphics[width=0.23\textwidth]{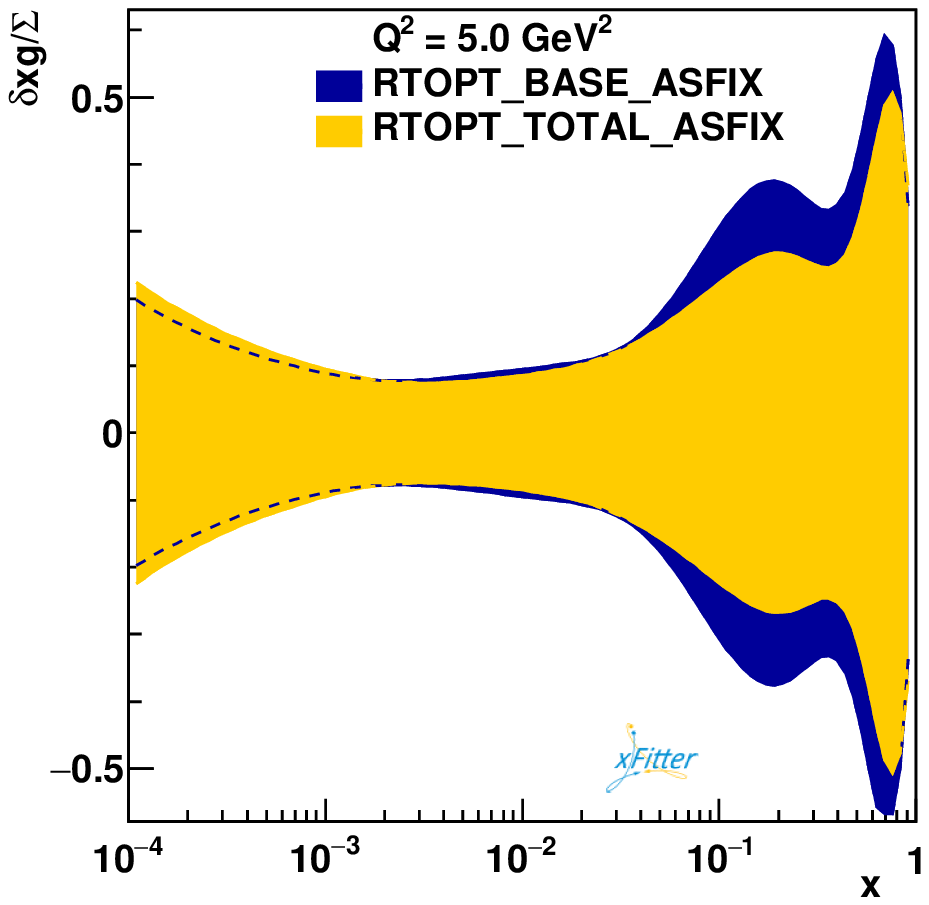}
\includegraphics[width=0.23\textwidth]{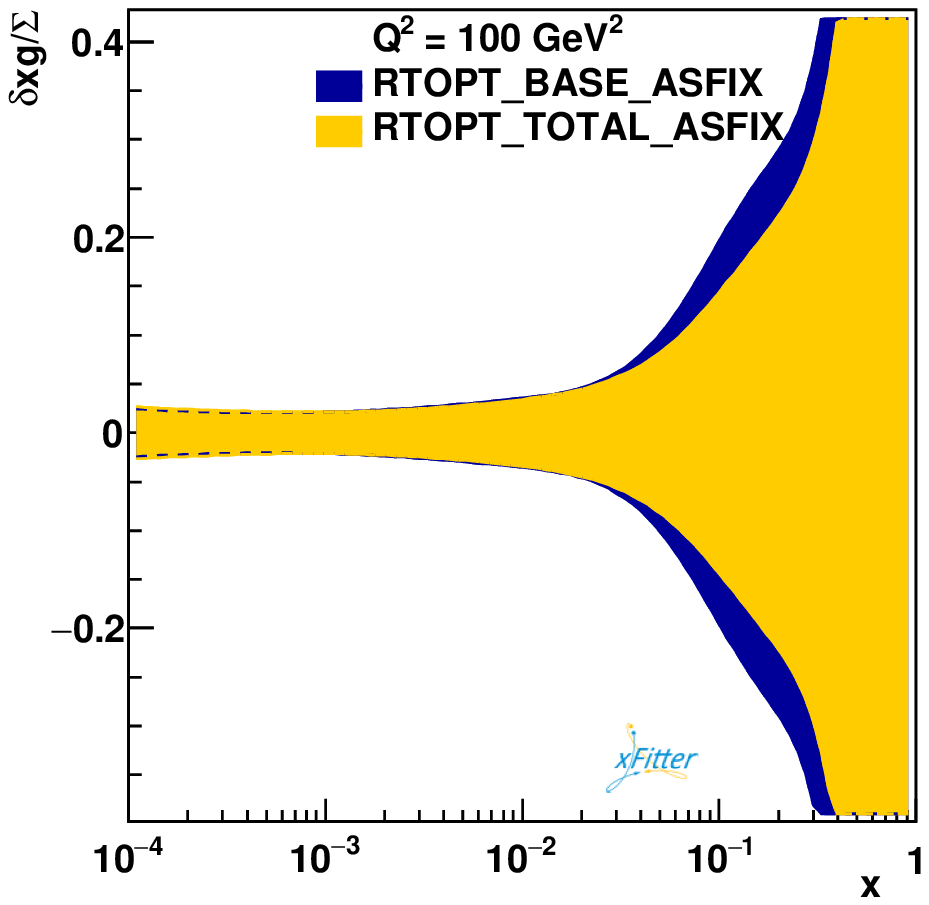}
\includegraphics[width=0.23\textwidth]{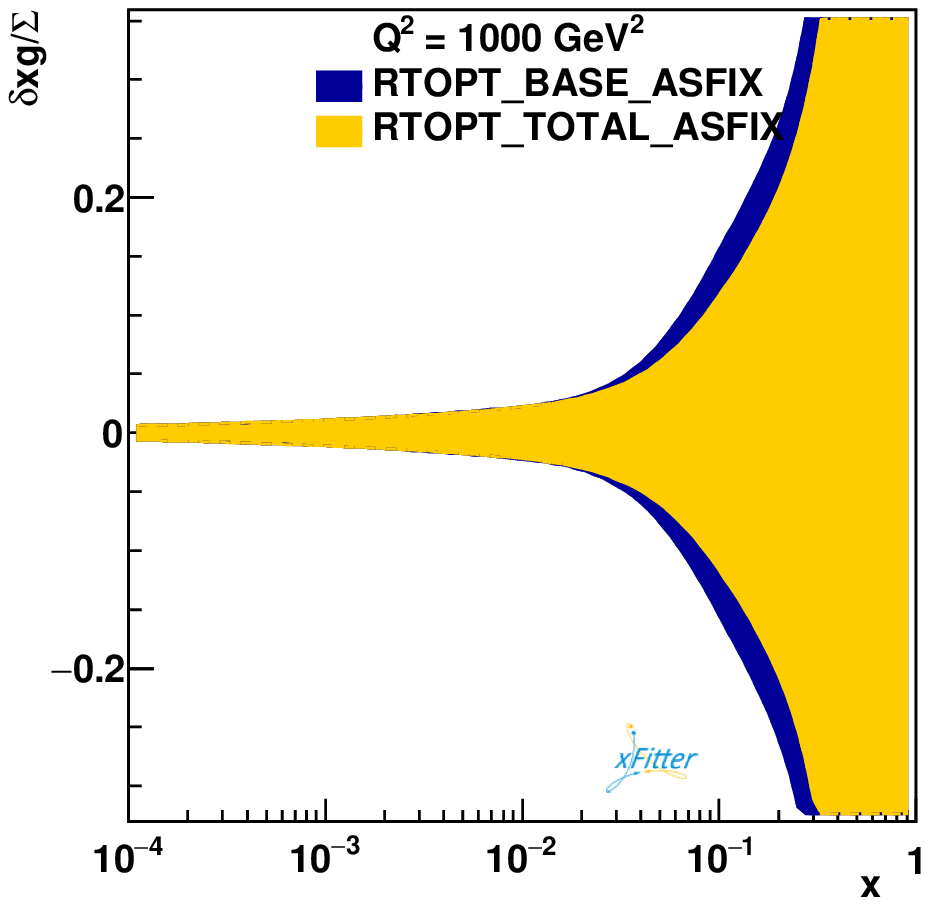}

\includegraphics[width=0.23\textwidth]{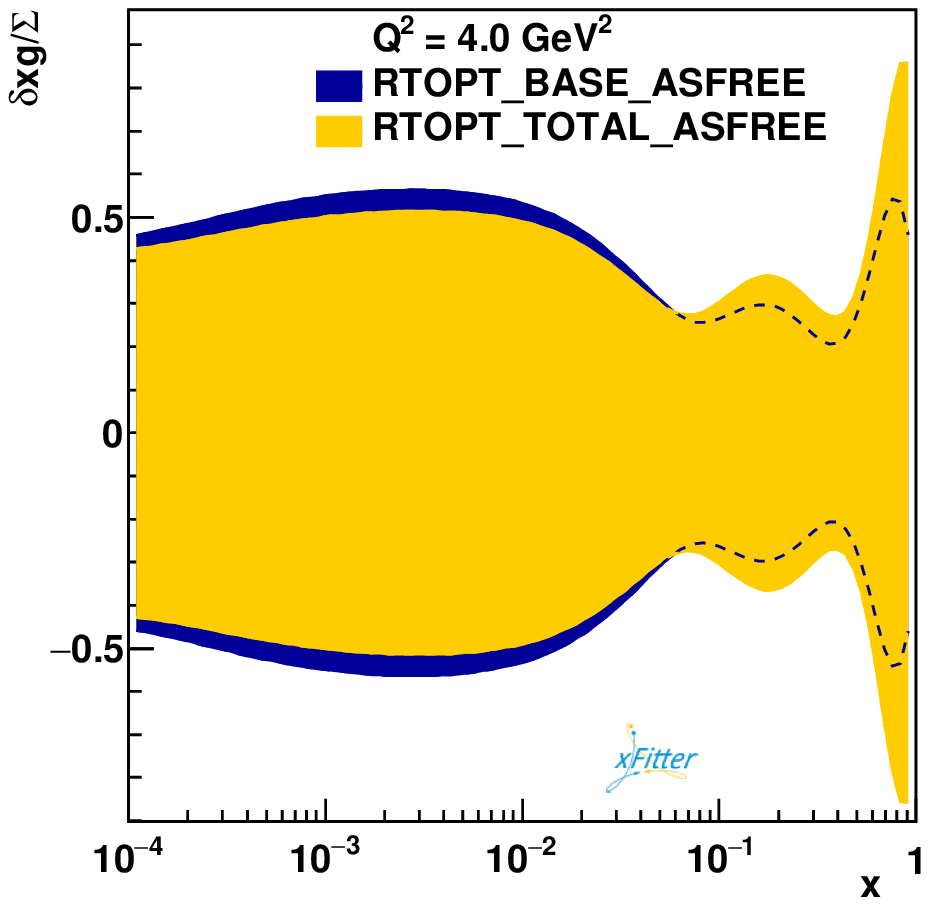}
\includegraphics[width=0.23\textwidth]{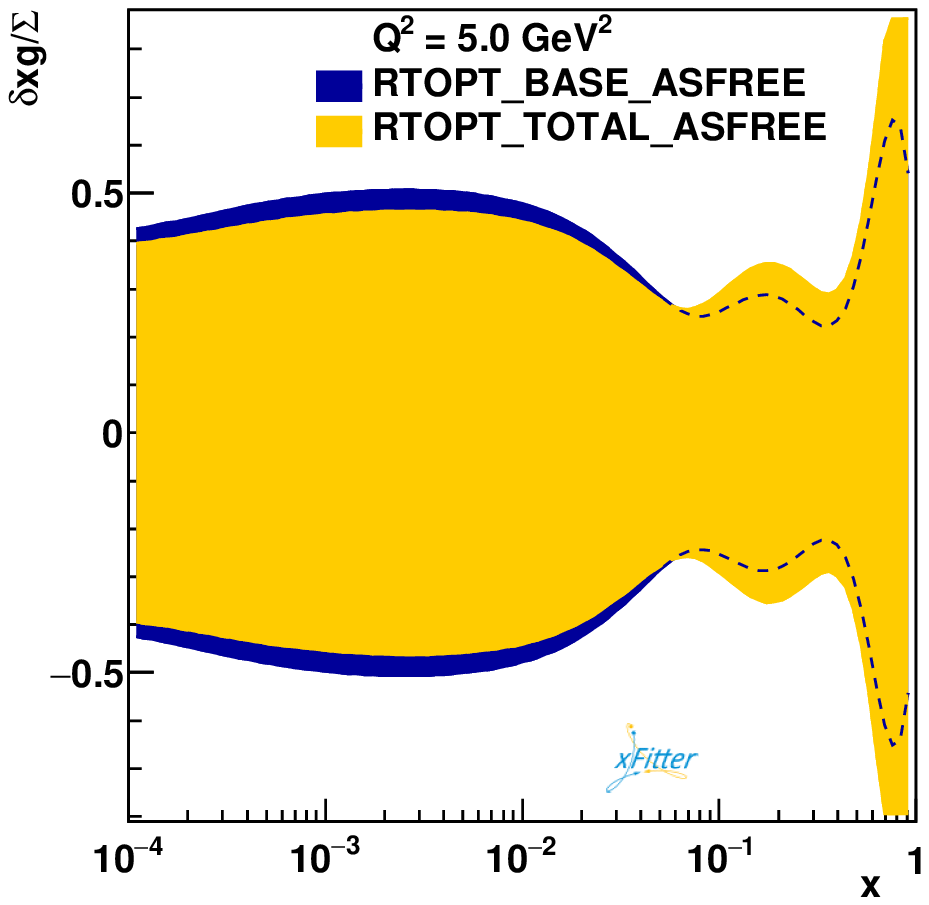}
\includegraphics[width=0.23\textwidth]{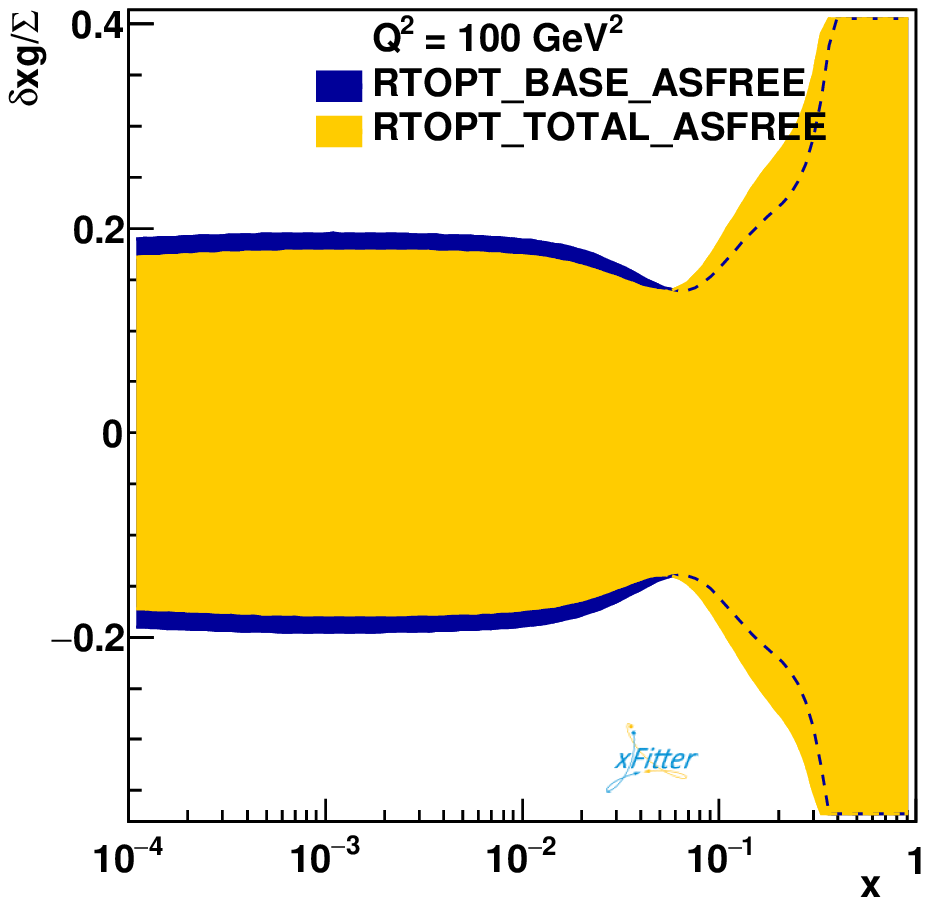}
\includegraphics[width=0.23\textwidth]{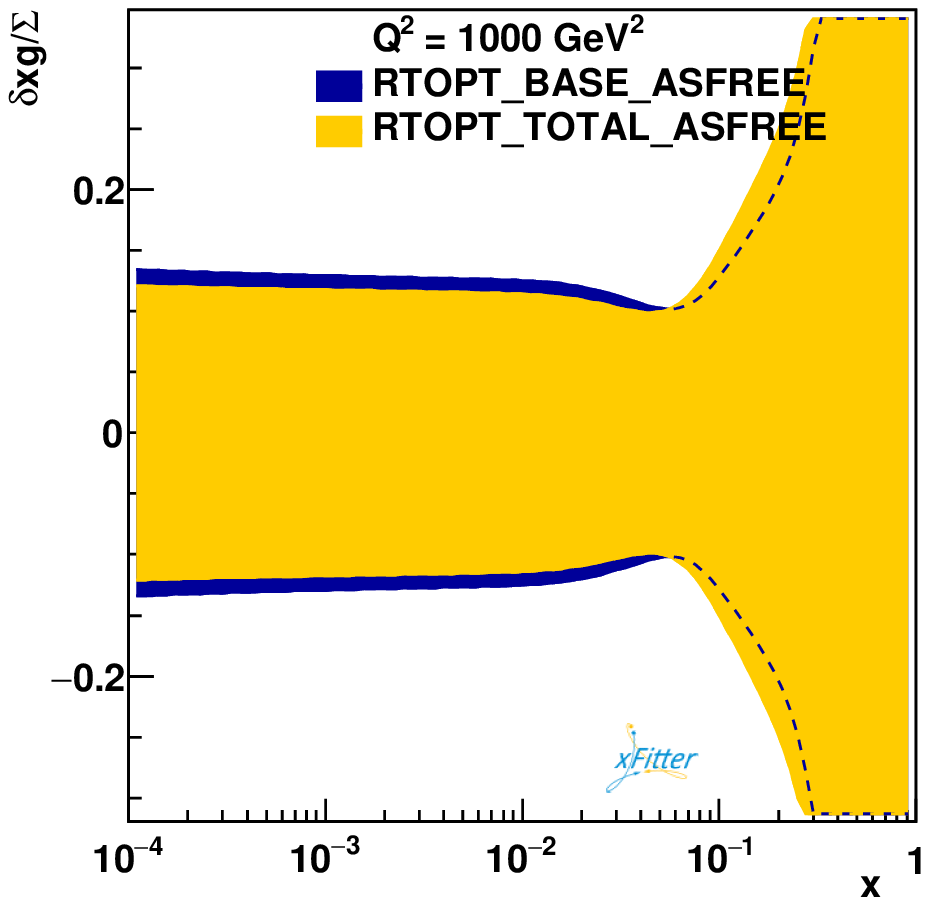}

\caption{The partial ratio of gluon distributions over $\Sigma$-PDFs, as extracted for the RTOPT scheme. The upper four diagrams correspond to the first scenario, while the lower four diagrams correspond to the second scenario and clearly show the impact of charm flavour data on the PDFs. }
\label{fig:7}
\end{figure*}

The total sea quark $\Sigma$-PDFs are defined by $\Sigma=2x(\bar u+\bar d+\bar s+\bar c)$. In Figs.~\ref{fig:6} and \ref{fig:7}  
we plot the partial ratio of gluon distributions over the $\Sigma$-PDF to show the impact of adding charm cross section H1-ZEUS combined  data to HERA I and II combined data, at $Q^2$ = 4, 5, 100 and 10000~GeV$^2$ in the RT and RTOPT schemes and for the two different scenarios. Clearly, these impacts can be seen only in the second scenario.

\clearpage
\section{\label{Summary}Summary}
Up to 36 percent of the cross sections at HERA originate from processes with charm quarks in the final state. In this QCD analysis we investigated the simultaneous impact of charm quark cross section H1-ZEUS combined data on the PDFs and on the determination of the strong coupling.

We chose the full HERA run I and II  DIS charged and neutral current data as a base data set and developed our QCD analysis at next-to-leading order in both RT and RTOPT schemes and for two separate scenarios using the HERAPDF parametrization form.

The sensitivity of PDF uncertainties to reduced charm quark cross section H1-ZEUS combined data at next-to-leading order, especially when in our second scenario we take the strong coupling, $\alpha_s(M^2_Z)$, as an extra free parameter, is reported in this QCD analysis.

 This analysis shows a dramatic reduction of some PDF uncertainties and good agreement of the strong coupling constant, $\alpha_s(M^2_Z)$, with the world average value, when the reduced charm quark cross section H1-ZEUS combined data are included.
 
 As we mentioned, the strong coupling, $\alpha_s(M^2_Z)$, plays a central role in the pQCD factorization theorem and the result of this QCD-analysis emphasis on its dramatic correlation with the PDFs reveals the impact of the charm flavour contribution.
 
 According our QCD analysis, in going from the RT scheme to the RTOPT scheme, we get $\sim 0.4$~\% and $\sim 0.9$~\%~improvement in the fit quality, without and with the charm flavour contribution, respectively.
 Also, we show that in going from the RT scheme to the RTOPT scheme, we get $\sim 0.9$~\% and $\sim 2.0$~\%~improvement in the strong coupling value, without and with the charm flavour contribution, respectively.

 A standard LHAPDF library file of this QCD analysis at next-to-leading order is available and can be obtained from the author via e-mail.
 


\end{document}